\documentclass{article}

\usepackage{microtype}
\usepackage{graphicx}
\usepackage{subfigure}
\usepackage{booktabs} 

\usepackage{hyperref}

\newcommand{\alglinelabel}{%
  \addtocounter{ALC@line}{-1}
  \refstepcounter{ALC@line}
  \label
}

\usepackage[accepted]{icml2024}

\usepackage{amsmath, amssymb, amsthm, thm-restate}
\usepackage{braket, physics, enumitem, relsize}
\usepackage{makecell, subfiles, multirow, wrapfig}
\usepackage[page]{appendix}
\usepackage[capitalize, nameinlink]{cleveref}
\usepackage{tabularx, array}
\setlist[itemize]{leftmargin=10pt}
\usepackage{float}
\usepackage{tikz, adjustbox}
\usetikzlibrary{quantikz}
\usepackage{xcolor}
\hypersetup{
  colorlinks   = true, 
  urlcolor     = blue, 
  linkcolor    = teal, 
  citecolor   = blue 
}
\usepackage{setspace}

\allowdisplaybreaks

\mathcode`l="8000
\begingroup
\makeatletter
\lccode`\~=`\l
\DeclareMathSymbol{\lsb@l}{\mathalpha}{letters}{`l}
\lowercase{\gdef~{\ifnum\the\mathgroup=\m@ne \ell \else \lsb@l \fi}}%
\endgroup

\theoremstyle{plain}
\newtheorem{thm}{Theorem}[section]
\newtheorem{prop}[thm]{Proposition}
\newtheorem{lem}[thm]{Lemma}
\newtheorem{corol}[thm]{Corollary}
\theoremstyle{defn}
\newtheorem{defn}[thm]{Definition}
\theoremstyle{rem}
\newtheorem{rem}[thm]{Remark}

\newtheorem{exm}{Example}
\newtheorem*{thm*}{Theorem}

\numberwithin{exm}{section}

\Crefname{prop}{Proposition}{Propositions}
\Crefname{alg-line}{Line}{Lines}
\crefname{alg}{Algorithm}{Algorithms}
\crefname{exm}{Example}{Examples}

\def\ptarget{p_{\text{data}}}
\def\KL{{\text{KL}}}

\DeclareMathOperator{\Var}{Var}
\DeclareMathOperator{\EE}{\mathbb{E}}
\DeclareMathOperator{\TV}{TV}
\DeclareMathOperator{\plog}{polylog}
\DeclareMathOperator{\poly}{poly}

\renewcommand{\vec}[1]{\overrightarrow{#1}}
\renewcommand{\epsilon}{\varepsilon}
\renewcommand{\tilde}{\widetilde}

\newcolumntype{x}[1]{>{\centering\hspace{0pt}\arraybackslash}m{#1}}

\icmltitlerunning{Gibbs Sampling of Continuous Potentials on a Quantum Computer}

\begin{document}

\twocolumn[
\icmltitle{Gibbs Sampling of Continuous Potentials on a Quantum Computer}




\begin{icmlauthorlist}
\icmlauthor{Arsalan Motamedi}{iqc,uw}
\icmlauthor{Pooya Ronagh}{iqc,uw,pi,irr}
\end{icmlauthorlist}

\icmlaffiliation{iqc}{Institute for Quantum Computing, University of Waterloo, Waterloo, ON, Canada}
\icmlaffiliation{uw}{Department of Physics \& Astronomy, University of Waterloo, Waterloo, ON, Canada}
\icmlaffiliation{pi}{Perimeter Institute for Theoretical Physics, Waterloo, ON, Canada}
\icmlaffiliation{irr}{Irr\'eversible, Vancouver, BC, Canada}

\icmlcorrespondingauthor{Arsalan Motamedi}{arsalan.motamedi@uwaterloo.ca}
\icmlcorrespondingauthor{Pooya Ronagh}{pooya.ronagh@uwaterloo.ca}

\icmlkeywords{Machine Learning, ICML}

\vskip 0.3in
]



\printAffiliationsAndNotice{}  

\begin{abstract}
Gibbs sampling from continuous real-valued functions is a challenging problem of
interest in machine learning. Here we leverage quantum Fourier transforms to
build a quantum algorithm for this task when the function is periodic. We use
the quantum algorithms for solving linear ordinary differential equations to
solve the Fokker--Planck equation and prepare a quantum state encoding the
Gibbs distribution. We show that the efficiency of interpolation and
differentiation of these functions on a quantum computer depends on the rate of
decay of the Fourier coefficients of the Fourier transform of the function. We
view this property as a concentration of measure in the Fourier domain, and
also provide functional analytic conditions for it. Our algorithm makes zeroeth
order queries to a quantum oracle of the function and achieves polynomial
quantum speedups in mean estimation in the Gibbs measure for generic
non-convex periodic functions. At high temperatures the algorithm also allows
for exponentially improved precision in sampling from Morse functions.
\end{abstract}

\section{Introduction}
\label{sec:intro}

In recent advances in machine learning (ML), a reincarnation of energy-based
models (EBM) has provided state-of-the-art performance in generative modeling
\cite{grathwohl2019your, song2021train, song2020score, ho2020denoising}.
Unlike the traditional EBMs such as Boltzmann machines and Hopfield neural
networks \cite{ackley1985learning, hopfield1982neural, hinton2006fast,
hinton2007learning} these models require Gibbs sampling from continuous
real-valued functions parameterized by large deep neural networks. However, both
the training of and inference from these models is extremely computationally
expensive and numerically unstable despite using state-of-the-art ML
accelerators such as graphical and tensor processing and streaming units
\cite{nijkamp2019learning}.

The computational challenge in training EBMs is sampling from the canonical
distribution represented by the model which is the Gibbs distribution
\begin{equation}
\label{eq:Gibbs-distribution}
p_\theta (x) = \exp(-E_\theta (x))/ Z_\theta
\end{equation}
of an associated $d$-dimensional \emph{energy potential}, $E_\theta: \mathbb
R^d \to \mathbb R$. Here $\theta \in \mathbb R^m$ denotes a vector of model
parameters and the normalizing constant $Z_\theta = \int e^{-E_\theta(x)}\, dx$
is the partition function of $p_\theta$. Gibbs sampling is typically done
using Monte Carlo integration of (the overdamped) Langevin dynamics
\cite{du2019implicit, nijkamp2020anatomy}, a stochastic differential equation
(SDE) governing diffusion processes.
Despite the limitations imposed by this computational bottleneck, EBMs have
been shown to provide improved representations of classical data. For
example, \citet{grathwohl2019your} and \citet{hill2020stochastic} overcome the
instabilities of the training process on particular datasets to provide
numerical evidence that EBMs can achieve more calibrated and adversarially
robust representations compared to conventional classifiers. We refer the
reader to \cref{sec:app-EBM} for more details on the usage of Gibbs sampling in
the training of and inference from EBMs.

In this paper, we investigate using quantum computation for Gibbs sampling from
continuous energy potentials. We use finite difference techniques for solving
differential equations on quantum computers \cite{berry2017quantum,
childs2020quantum, childs2021high, liu2021efficient, krovi2023improved} to solve
the Fokker--Planck equation (FPE), a second-order partial differential equation
(PDE) admitting the Gibbs distribution as its steady state solution. The FPE
and Langevin dynamics are associated with each other through It\^o integration
\cite{evans2023stochastic}. Interestingly, directly solving for the
steady state of the FPE requires solving linear systems with exponentially poor
condition numbers. We therefore also have to integrate the FPE for a long
enough time to asymptotically converge to the Gibbs state; as such, we do not
achieve a shortcut to the problem of long mixing time in equilibrium dynamics.
We do, however, achieve a high precision approximation to the Gibbs state in
total variation distance. 

\paragraph*{Technical setting and contributions.}

In the Euclidean space, the Gibbs measure is only well-defined for an unbounded
potential. This setting is not suitable for finite difference methods,
therefore we are interested in probability density functions with compact
support. This imposes boundary conditions on the FPE. In this paper we
consider periodic boundary conditions since they allow us to leverage quantum
Fourier transforms. In other words, we focus on potentials defined on high
dimensional tori. We assume that the energy function takes values in a real
interval of diameter $\Delta$ which we refer to as the diameter of the potential
hereon.\footnote{For simplicity, we use a constant thermodynamic $\beta= 1$
throughout (otherwise $\beta \Delta$ can be thought of as a single parameter).}
For a periodic function $f: \mathbb R^d \to \mathbb R$ the Fourier transform
coefficients correspond to points on the lattice $\mathbb Z^d$. We show
that these coefficients decay sub-exponentially away from the origin assuming
a bound on the growth of the derivatives of $f$ (\cref{def:semi-analytic}).
The latter is a functional analytic condition similar to analyticity, however
our definition is milder and therefore we call it \emph{semi-analyticity}.
We also show that if the Fourier coefficients on $\mathbb Z^d$ are viewed as
the density of a probability measure defined on this lattice, then
semi-analyticity is equivalent to the concentration of this measure
(\cref{thm:SAisBern}).

Many periodic functions are semi-analytic. For example any function with
finitely many non-zero Fourier coefficients is semi-analytic (\cref
{ex:Nyquist}). Semi-analyticity is determined using two parameters which we
denote as $C$ and $a$ in this paper. The first parameter represent the scale of
the function (i.e., scales linearly with scalar multiplication), but the second
parameter bears information about the geometry of the function and can be
viewed as an inverse radius of convergence of its Taylor expansion (see
\cref{sec:app-semianalytic}). We show some of the basic properties of analytic
and semi-analytic functions; i.e., how $C$ and $a$ change under arithmetic
operations and compositions (\cref{prop:semi-analytic-basic-prop}).
Consequently, we show that parameterized families such as deep neural networks
with analytic activation functions are analytic. However, activation functions
such as the sigmoid function creates sharp ramps in the energy landscape which
shrinks the radius of convergence (\cref{cor:neural-nets}). Beside
the algorithmic contributions in this paper, the above insights can shed light
on suitable alternatives to deep neural networks as parameterized models for
energy-based learning. We first show that real analytic periodic functions admit
efficient quantum algorithms for interpolation and differentiation provided
quantum states encoding very coarse discretizations of such functions (\cref
{thm:body-interpolation}). We also provide a lower bound result in \cref
{thm:body-lower-bound} showing the optimality of our discretization.

We refer the reader to \cref{ex:cosine-function,ex:inv-cos} for further
preliminary examples of semi-analytic functions. More systematically, in
\cref{sec:constructions}, we show several constructions of semi-analytic
functions from other ones (e.g., using compositions, or basic operations
between functions). We then continue with transcendental functions such as the
exponentiation and sigmoid functions (as they relates to machine learning
models in particular). As a corollary of these properties, we compute the
analytic parameters of neural networks constructed via sigmoid activation
functions in \cref{cor:neural-nets}. Additionally, as concrete numerical
demonstrations, \cref{fig:3,fig:Fourier} showcase the error of our
interpolation approach applied on the functions considered in \cref
{ex:cosine-function,ex:inv-cos}.

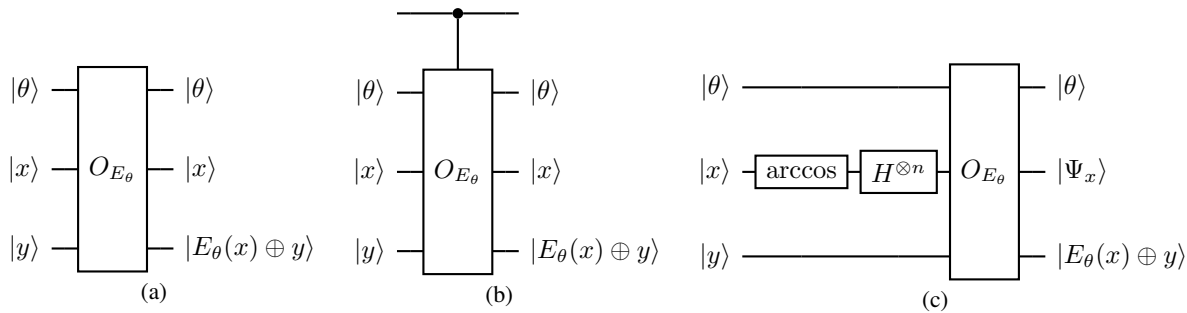
\begin{figure*}[t]
  \centering
  \begin{subfigure}[\label{fig:EBM-a}]{
	\begin{quantikz}[column sep=5pt, row sep={30pt,between origins}]
	\\
	\lstick{$\ket{\theta}$} &\qw & \gate[3]{O_{E_\theta}} &\qw &\qw  \rstick{$\ket{\theta}$} \\
	\lstick{$\ket{x}$} & \qw &  & \qw & \qw\rstick{$\ket{x}$} \\
	\lstick{$\ket{y}$} & \qw & & \qw & \qw \rstick{$\ket{E_\theta(x) \oplus y}$}
	\end{quantikz}}
  \end{subfigure}
  \begin{subfigure}[\label{fig:EBM-b}]{
  \begin{quantikz}[column sep=5pt, row sep={30pt,between origins}]
	\lstick{} & \qw & \ctrl{1} & \qw & \qw\rstick{} \\
	\lstick{$\ket{\theta}$} &\qw & \gate[3]{O_{E_\theta}} &\qw &\qw  \rstick{$\ket{\theta}$} \\
	\lstick{$\ket{x}$} & \qw &  & \qw & \qw\rstick{$\ket{x}$} \\
	\lstick{$\ket{y}$} & \qw & & \qw & \qw \rstick{$\ket{E_\theta(x) \oplus y}$}
	\end{quantikz}}
  \end{subfigure}
  \begin{subfigure}[\label{fig:EBM-c}]{
  \begin{quantikz}[column sep=5pt, row sep={32pt,between origins}]
  \\
	\lstick{$\ket{\theta}$} &\qw &\qw & \gate[3]{O_{E_\theta}} &\qw &\qw  \rstick{$\ket{\theta}$} \\
	\lstick{$\ket{x}$} & \gate{\arccos} &\gate{H^{\otimes n}} &  & \qw & \qw\rstick{$\ket{\Psi_x}$} \\
	\lstick{$\ket{y}$} & \qw &\qw & & \qw & \qw \rstick{$\ket{E_\theta(x) \oplus y}$}
	\end{quantikz}}
  \end{subfigure}
  \caption{
Schematics of the circuits of quantum oracles. All registers receive float-point
representations of real numbers. (a) The oracle for an energy function
$E_\theta$. The first register receives the model parameters
$\theta \in \mathbb R^m$, the second register receives a data sample
$x \in \mathbb R^d$. And, the last register is used to evaluate the energy
function. (b) A controlled variant of the same oracle, controlled on a single
qubit represented by the top wire. (c) The oracle of figure (a) receiving
an augmented sample in superposition as the state
$\ket{\Psi_x}= \sum_{b \in \{0, 1\}^d}\ket{(-1)^{b_i} \arccos x_i}$.}
\label{fig:EBM}
\end{figure*}

Given a parametrization $\theta$ of an energy potential\footnote{For example,
the weights and biases of a deep neural network, or any other parametrized
ansatz.} $E_\theta: \mathbb R^d \to \mathbb R$, one can construct a quantum
oracle
\begin{align}
\label{eq:energy-oracle}
\ket{\theta} \ket{x} \ket{y} \mapsto \ket{\theta} \ket{x} \ket{y \oplus E(x)}
\end{align}
realizing such a function with only a polylogarithmic overhead compared to the
gate complexity of the corresponding classical boolean circuit
\cite{nielsen2002quantum}. The quantum circuit of this oracle is depicted in
\cref{fig:EBM-a}. In what follows, we ignore the model parameter
register when the context is clear.

\begin{table*}[t]
\caption{
Summary of the query complexities of some of the classical and quantum
algorithms for sampling from a $d$-dimensional Gibbs distribution, highlighting
the cases of significant quantum advantage in boldface. $\epsilon$
denotes the error in the designated norm (TV for total variation distance, and
$W_2$ for 2-Wasserstein distance), and $\kappa_f$ denotes the Poincar\'e
constant of a function $f$. Our result (first row) corresponds to non-convex
periodic functions and the relevant Poincar\'e constant is that of the function
$E/2$ due to Born rule(see \cref{sec:algorithm}). Therefore, for families of
functions that the Poincar\'e constant is better than the Eyring-Kramers bound
we may achieve a quantum advantage in sampling. For example, for Morse
functions with unique global minima we achieve the query complexities in the
third row (\cref{cor:body-unique-optimum}). In comparison to the classical
counterpart (fourth row), our algorithm achieves an exponential advantage in
precision $\epsilon$ while only consuming zeroeth order queries to the
function. For estimating means of random variables of the Gibbs state, we
achieve a quadratic advantage for generic periodic functions (last column of
the first and second rows) and a quartic advantage in the case of Morse
functions with unique global minima. Here $\Delta_f$ is the diameter of the
range of values a function $f$ attains. The prior results (rows five to nine)
all require convexity assumptions on the potential. For $\mu$-strongly convex
functions the Poincar\'e constant $\kappa_E$ is $1/\mu$, famously known as the
Bakry-\'Emery criterion \cite{bakry2014analysis}.}
\label{tab:comparison}
\begin{center}
\footnotesize
\setlength\tabcolsep{1pt}
\setlength{\extrarowheight}{5pt}
\begin{tabular}{x{0.24\linewidth}x{0.19\linewidth}x{0.07\linewidth}x{0.2\linewidth}x{0.04\linewidth}x{0.24\linewidth}}
\toprule
Method & Potential type & Query order & \,\,Sampling\,\, complexity & \vspace{-5mm} Norm & Mean estimation complexity\\
\midrule
This paper & non-convex periodic & zeroeth
& $\tilde O\left(\kappa_{E/2} e^{\Delta / 2} d^7\right)$ & TV &
$\pmb{\tilde O\left(\kappa_{E/2} e^{\Delta / 2} d^7\Delta_f \epsilon^{-1}\right)}$ \\
Rejection sampling & non-convex & zeroeth
& $O\left(e^{\Delta} \right)$ & TV &
$O\left(e^{\Delta} \Delta_f^2 \epsilon^{-2}\right)$ \\
\midrule
This paper & Morse and periodic & zeroeth
& $\pmb{\tilde O\left(\lambda^{-2}e^{\Delta / 2} d^7\right)}$ & TV &
$\pmb{\tilde O\left(\lambda^{-2}e^{\Delta / 2} d^7\Delta_f \epsilon^{-1}\right)}$ \\
Cl. RLA~\cite{li2020riemannian} & Morse and periodic & first
& $\tilde O\left(\lambda^{-4} L^4 d^3\epsilon^{-2}\right)$ & TV &
$\tilde O\left(\lambda^{-4} L^4 d^3\Delta_f^2 \epsilon^{-4}\right)$ \\
\midrule
Cl. MRW~\cite{dwivedi2018log} & convex & first
& $\tilde O\left(L^2d^{3}\epsilon^{-2}\right)$ & TV
& $\tilde O\left(L^2d^{3}\Delta_f^2\epsilon^{-4}\right)$ \\
\midrule
Q. ULD~\cite{childs2022quantum} & strongly convex & zeroeth
& $\tilde O\left(\mu^{-2} L^2 d^{1/2}\epsilon^{-1}\right)$ & $W_{2}$ & -- \\
Cl. ULD~\cite{cheng2017underdamped} & strongly convex & first
& $\tilde O\left(\mu^{-2} L^2 d^{1/2}\epsilon^{-1}\right)$ & $W_{2}$ & -- \\
Q. MALA~\cite{childs2022quantum} & strongly convex & first
& $\tilde O\left(\mu^{-1/2} L^{1/2} d\right)$ & TV
& $\tilde O\left(\mu^{-1/2} L^{1/2} d\Delta_f \epsilon^{-1}\right)$ \\
Cl. MALA~\cite{lee2020logsmooth} & strongly convex & first
& $\tilde O\left(\mu^{-1} L d\right)$ & TV
& $\tilde O\left(\mu^{-1} L d \Delta_f^2 \epsilon^{-2}\right)$\\
\bottomrule
\end{tabular}
\end{center}
\end{table*}

We note that if $E$ has a positive minimum value
$E_* \geq 0$, a na\"ive classical rejection sampling routine, consisting of
drawing uniform random samples $x \in \mathbb R^d$ and accepting the sample
with probability $\exp(-E(x))$ will require $O(e^{\Delta + E_*})$ iterations
to generate an accepted sample from the Gibbs distribution. Similarly,
the mixing time of Langevin dynamics is $O(\kappa_E)$ where $\kappa_E$ is
the \emph{Poincar\'e constant} of the potential $E$ and for generic non-convex
functions this constant is in $O(e^\Delta)$. We prove the counterparts to these
two facts for periodic functions in \cref{prop:exp-decay-torus} and
\cref{prop:Poin}. The latter bound can be interpreted as a
Eyring-Kramers law for periodic functions \cite{berglund2011kramers}.

We use our Fourier interpolation techniques to achieve high precision
Gibbs sampling as summarized in \cref{tab:comparison}. Our algorithm uses
$\tilde O\left(\kappa_{E/2} e^{\Delta / 2} d^7 \plog(1/\epsilon)\right)$ queries
to the oracle $O_E$ for Gibbs sampling with an approximation error $\epsilon$ in
total variation distance (TV) (\cref{thm:main-result}). This provides an
opportunity for quantum speedup for families of functions with small Poincar\'e
constants. However, for a generic non-convex periodic function $E$ the relevant
Poincar\'e constant (i.e., that of $E/2$, due to Born rule) is in
$O(e^{\Delta/2})$. Therefore, our complexity factor
$\kappa_{E/2} e^{\Delta/2} = e^{\Delta}$ in the first row of
\cref{tab:comparison} saturates the complexity bound of classical rejection
sampling for periodic functions in absence of any additional structure. However
for families of functions with a Poincar\'e constant
better than the Eyring-Kramers bound we may achieve a quantum advantage. For
example, for Morse functions with unique global minima we achieve the query
complexities in the third row of the table (\cref{cor:body-unique-optimum}).
The classical counterpart for this result is
\citep[Theorem 2.4]{li2020riemannian} which results in the complexity bounds
of the fourth row. In
comparison, our algorithm achieves an exponential advantage in precision
$\epsilon$ while only consuming zeroeth order queries to the function.

The prior results, Metropolized random walk (MRW), the underdamped Langevin
dynamics (ULD) and the Metropolis-adjusted Langevin algorithm (MALA), together
with their accelerations via quantum random walk in \cite
{childs2022quantum} are also included in \cref{tab:comparison}. We note that
all these algorithms require convexity assumptions on the potential. While
convex potentials are of theoretical interest for analyzing and comparing
different sampling algorithm, they do not pertain to practical ML scenarios
as real-world distributions are not uni-modal.

The oracle of \cref{fig:EBM-a} is sufficient for sampling from the Gibbs
distribution of $E_\theta$ which pertains to inference from EBMs.
However, in practice we often collect samples in order to estimate the mean of
another quantity $f: \mathbb R^d \to \mathbb R$ with respect to the Gibbs
measure. For example, in training EBMs, the expectations of the gradients of
the energy function is desired (see \cref{sec:app-EBM}). In this case, a \emph
{white-box} access to the energy oracle will allow us to use amplitude
estimation techniques to achieve quantum advantage. By this we mean the ability
to construct a controlled variant (as in \cref{fig:EBM-b}) of the unitary in
\cref{fig:EBM-a} and its inverse. Given white-box access to both the energy and
random variable oracles (i.e., an oracle for evaluating $f$), our algorithm achieves a quadratic advantage in mean
estimation for generic periodic functions and a quartic advantage in the case
of Morse functions with unique global minima (\cref{cor:body-mean-estimation})
as reported in the last column of \cref{tab:comparison}.

\paragraph*{Practical applicability.}

Our algorithm outperforms other methods in the high temperature and
high precision regime. This is the sweet spot for generative modeling using
energy-based techniques since in ML the Gibbs sampling temperature is often set
to $\beta = 1$ or tuned as a finite hyperparameter, and the parameterized
models representing the energy potential are regularized to attain bounded
values (e.g., using $L_2$ regularization). We also note that our quantum
advantage in precision is at the cost of high order polynomial dependence on
dimension $d$. However, for applications processing classical data in
computational basis states, at least a linear dependence on $d$ for state
preparation and measurement is unavoidable. Moreover, in ML applications
Langevin dynamics is typically run on a latent space of much lower dimensions
than the input data \cite{pang2020learning,vahdat2021score,rombach2022high}.
Generated latent samples are then scaled up to high resolution results through
decoding techniques. Therefore, our exponential quantum advantage in
precision can be of critical significance in ML use-cases.

Many real-world datasets are naturally represented using periodic degrees of
freedom (e.g., rotation angles), particularly including problems within
physics. For example, many-body systems such as the $XY$
Hamiltonian \cite{araki1985ground}, and quantum field theories such as the
$\phi^4$ field theory \cite{jordan2012quantum} are of this kind. Nevertheless,
this is not typically the case for conventional ML training datasets (e.g., for
image and audio datasets represented by various channels of colour intensities
and acoustic amplitudes, respectively) \cite{ziyin2020neural}. In such cases,
we may assume that data is always normalized within a hypercube $[-1,1]^d$.
One solution is to extend our algorithm to functions
$f:[-1,1]^d\rightarrow \mathbb R$. We can do so by constructing the composite
function $g(x):=f(\cos x)$ and applying our algorithm to $g$. We show in \cref
{sec:app-chebyshev} that this procedure is not efficient unless data is
concentrated in the boundaries of the hypercube or possesses a dimension
independent variance. Since our algorithm is more suitable for periodic
functions, the alternative approach is to augment the dataset by inverting a
trigonometric function of its components as shown in \cref{fig:EBM-c}. Here
each dimension of data is lifted to two branches of the $\arccos$ function.
Therefore each sample $\ket{x}$ in the computational basis corresponds to $2^d$
inverse images, and an equal superposition $\ket{\Psi_x}$ of these
exponentially many inverse images is efficiently constructed using Hadamard
gates. By employing this approach, the resulting energy function learns the
correct function across all branches of $\arccos$.

We refer the reader to \cref{sec:app-EBM} for a concise introduction to training
and inference from EBMs. We also provide a detailed construction of the oracle
used for the simulation of the Fokker--Planck equation using the energy oracles
as building blocks (\cref{fig:Oracle4L}).

\paragraph*{Related works.}

To the best of our knowledge, this work and the independent
paper \citet{childs2022quantum} are the first efforts to analyze quantum
algorithms for the problem of Gibbs sampling from a continuous real-valued
function. \citet{childs2022quantum} achieve a quadratic speedup in
expediting a Monte Carlo simulation of Langevin dynamics using quantum random
walks but their result is restricted to strongly convex potentials. Similarly
classical algorithms that achieve high precision Gibbs sampling also make
assumptions about convexity or satisfaction of isoperimetric inequalities
\cite{roberts1996exponential, dwivedi2018log, chewi2021analysis}.
However, ML applications demand highly non-convex potentials that can capture
complex modes of data in a multimodal landscape. Fortunately, our algorithm
provides high precision Gibbs sampling of such potentials as long as they
satisfy periodic boundary conditions. In addition, the quantum speedup observed
by \citet{childs2022quantum} assumes accurate access to the gradients of the
potential whereas we only perform zeroeth order queries to the potential.

As another interesting point of comparison, \citet{ge2018simulated} focus on
Gibbs sampling from finite mixtures of log-concave distributions and incur very
high-order polynomial costs in dimension $d$ and inverse precision
$1/\epsilon$. Moreover, \citet{fan2023improved} consider smooth potentials that
satisfy Poincar\'e inequalities and achieve polylogarithmic dependence on
precision, similar to us. However, we have an exponential improvement with
respect to the Lipschitz constant, and more importantly, \citet
{fan2023improved} have an additional logarithmic dependence on the
$\chi^2$-divergence of a warm-start initial distribution and the desired
target distribution. As mentioned in their work, one has merely an
$\exp(d)$ bound on this divergence, which results in higher-order polynomial
overhead with respect to $d$. We also mention the results of
\citet{li2022quantum} and \citet{ozgul2023stochastic} who tackle quantum
optimization and sampling from non-convex functions with additional global
geometric assumptions.

Finally we note that prior quantum algorithms \cite
{terhal2000problem, poulin2009sampling, chowdhury2016quantum, van2017quantum}
apply amplitude amplification to achieve a Grover speedup in preparing the
Gibbs state of \emph{discrete} spin systems. However, a na\"ive application of
these techniques to (say, a discretization of) the continuous domain, will at
best result in a query complexity that scales with $\sqrt{\exp(d)}$.

\section{The algorithm}
\label{sec:algorithm}

Our goal is to obtain the steady state solutions to the Fokker--Planck equation
(FPE)
\begin{align}\label{eq:FP-background}
\partial_t \rho(x,t) &= \nabla \cdot \left(
e^{-E(x)} \,\nabla \left( e^{E(x)} \, \rho(x,t) \right) \right)
\end{align}
corresponding to a \emph{toroidal diffusion process}. That is, a diffusion
process obtained by projecting Langevin dynamics
\begin{align}\label{eq:Langevin}
\mathrm{d}Y_t = -\nabla E(Y_t) \mathrm{d}t+ \sqrt{2}\mathrm{d}W_t
\end{align}
on a high dimensional torus $\mathbb T= \mathbb R^d / \mathbb Z^d$. By this
notation we mean the topological quotient of $\mathbb R^d$ under the action of
$\mathbb Z^d$ via $g: x \mapsto g + x$ for all $g \in \mathbb Z^d$. We refer the
reader to \cref{sec:app-langevin} for more details on such stochastic
processes. Here $(W_t)_{t\geq 0}$ is a Wiener process and the drift term
$-\nabla E(Y_t)$ is along the
gradient of a periodic smooth function $E: \mathbb R^d \to \mathbb R$, which
is called the energy function, energy potential, or the potential, for short.
We assume that the fundamental domain of this quotient
is of length $l$ and more specifically $[-l/2, l/2]^d \subset \mathbb R^d$. The
unique steady state solution $\rho_s$ to \eqref{eq:FP-background} corresponds to
the Gibbs state $\rho_s(x) \propto e^{-E(x)}$ of the potential. We intend
to find this distribution by solving \eqref{eq:FP-background} using a uniform
distribution $\rho_0(x)\propto 1$ as an initial condition and accessing the
long time $T \gg 0$ asymptotes of the solution. We refer the reader to
\cref{alg:gibbs-sampling} for a pseudo-code of our algorithm.

By solving the FPE we mean preparing a quantum state that encodes the solution
of the PDE \eqref{eq:FP-background} on a discrete lattice. When discretizing an
$l$-periodic function, we first consider a lattice obtained from taking an odd
number, $2N + 1$, of equidistant points along each axis; $x_{n} = \frac{ln}
{2N+1}$ for all $n \in [-N..N]^d$. We denote this discrete lattice by $V_N$ and
the Hilbert space $\mathbb{C}^{V_N}$(i.e., the space of functions from $V_N$ to
$\mathbb{C}$) by $\mathcal V_N$. Our discretization scheme transforms the
generator
\begin{align}
\mathcal{L} (-)=\nabla \cdot \left(
e^{-E} \,\nabla \left( e^{E} \, - \right) \right)
\end{align}
of the FPE to a linear operator
$\mathbb L:\mathcal V_N \rightarrow \mathcal V_N$.
An explicit construction of
$\mathbb L$ is elaborated in \cref{sec:discrete-FP}. We then solve the
linear ordinary differential equation
\begin{align}\label{eq:linear-system}
    \frac{d}{dt} \vec{u(t)} = \mathbb L \vec{u(t)}
\end{align}
using the results of \citet{berry2017quantum} and \citet{krovi2023improved} to
find a high precision approximation of the solution, $u(T)$ (line
\ref{alg-line:3} of \cref{alg:gibbs-sampling}).

In solving this linear system we use the Fourier pseudo-spectral method to
achieve high precision finite difference approximations of the derivatives of
$u$ using merely a coarse lattice $V_N$. To this end we require a tameness
condition on the growth of the higher derivatives of $u$. We call this
condition \emph{semi-analyticity} for its close resemblance to the
notion of analyticity in real functional analysis. In \cref{sec:results-fourier}
we discuss the connections between semi-analyticity and the concentration
of measure for a random variable we define from the Fourier transform of
$u$. We also provide examples of semi-analytic functions. In particular,
we show that all periodic real analytic functions are semi-analytic as well.

We note that sampling from the discretization of the Gibbs distribution
\begin{align}
\ket{\rho_s}\propto \sum_x e^{-E(x)} \ket{x},
\end{align}
results in an ensemble at thermodynamic $\beta = 2$ instead of
at $\beta= 1$. Here the normalization constant of this state is
$1/\sqrt{\bar Z_{\beta = 2}}$, where the $\bar Z$ notation represents the
partition function of the discretized probability measure. To overcome this
problem, throughout we set the energy function of our interest to be
$\frac{1}2 E$. In the notation
$\ket x := \ket{x_1} \otimes \cdots \otimes \ket{x_d} \in \mathcal V_N$ for
addressing the points on the lattice, each $\ket{x_i}= \ket{n_i + N}$ is the
one-hot encoding of the index $n_i + N$ where $n_i \in \{-N, -N+1, \ldots, N\}$.

Moreover, the discretization of the Gibbs state will result in sampling from
each point of the lattice according to the discrete probability distribution
\begin{align}
p(x)= \frac{1}{\bar Z_{\beta = 2}} e^{-2 E(x)} \simeq
\frac{l^d}{(2N+1)^d Z_{\beta= 2}}e^{-2E(x)}.
\end{align}
This is the case since $\bar Z_{\beta = 2} \Delta x \simeq  Z_{\beta = 2}$.
Therefore our proposed algorithm is to draw samples $x \in V_n$ via
measurements in the computational basis states, and then, generate uniform
samples from the box
$\prod_{i=1}^d[ x_i - \frac{l}{4n+2}, x_i + \frac{l}{4n+2}]$
(line \ref{alg-line:5} of \cref{alg:gibbs-sampling}).

However, na\"ive usage of a small $N$ combined with this uniform sampling
strategy does not provide a good approximation to the Gibbs distribution.
This is the second step in our algorithm wherein the
semi-analyticity condition plays a critical role.
We show that given a real periodic function $u: \mathbb
R^d\rightarrow \mathbb R$, we can provide samples from a high-precision
approximation of the distribution proportional to $u^2$ by querying very few
points in the domain of the definition of $u$.
Using a very coarse lattice $V_N$ (i.e., with $N$ being small)
we achieve a sampling error of $O(e^{-N})$ by employing a technique from
classical signal processing involving representation of
$u$ in the Fourier domain, although we use quantum Fourier transforms (QFT) for
its implementation. We call this procedure \emph{upsampling} of $u$
\cite{oppenheim1975digital} (line \ref{alg-line:3} of \cref
{alg:gibbs-sampling}).

\begin{algorithm*}
\caption{Pseudocode of our Gibbs sampling algorithm.}
\label{alg:gibbs-sampling}
\begin{algorithmic}[1]
\INPUT Energy function oracle $O_E$, lattice parameters $N, M \in \mathbb N$,
solution time $T > 0$
\STATE Construct an oracle for the discretization $\mathbb L$
of the generator of the Fokker--Planck equation (see \cref{fig:Oracle4L}
in the appendix).\alglinelabel{alg-line:1}
\STATE Deploy the algorithm of \cite{berry2017quantum} to prepare
a quantum state approximating $\ket{u(T)}$ pertaining to the solution of
$\frac{d}{dt} \vec u = \mathbb L \vec u$, with $\vec u(0) = \mathbf 1$, at
time $t= T$.\alglinelabel{alg-line:2}
\STATE Apply the upsampling isometry $F_M^{-1} \iota F_N$ involving quantum
Fourier transformations on the prepared state (\cref{thm:body-interpolation}).
\alglinelabel{alg-line:3}
\STATE Measure the resulting state in the computational basis to obtain
a lattice point $x\in [-l/2,l/2]^d$.
\alglinelabel{alg-line:4}
\STATE Draw a sample $\tilde x$ uniformly at random from the box
$\prod_{i=1}^d \left[x_i - \frac{l}{4M+2}, x_i + \frac{l}{4M+2}\right]$
around $x$.
\OUTPUT Sample point $\tilde x$.
\alglinelabel{alg-line:5}
\end{algorithmic}
\end{algorithm*}

To this end, in \cref{sec:results-fourier} we introduce \emph
{Fourier interpolation} and use this technique to upsample our quantum state in
the Fourier domain (refer to \cref{sec:app-semianalytic} for further
information). However, our interpolation technique is useful beyond the
applications considered herein. For instance, quantum algorithms for solving
partial differential equations (e.g., \cite{childs2021high}) also prepare
quantum states that encode the solutions of the equations on coarse discrete
lattices. Our interpolation algorithm, applied as a post-processing quantum
circuit, allows one to find approximate solutions on finer lattices and even on
the continuous domain without discretization.

The quantum algorithm makes queries to oracles for the discrete generator
$\mathbb L$ which themselves require access to $O(dN)$ oracles of the energy
function at different points (see \cref{eq:fourier-1} in the appendix). In \cref
{sec:results-sampling} we show that assuming that the FPE generates a
semi-analytic one-parameter family of probability measures $\{e^{\mathcal
{L}t} \rho_0: t\geq 0\}$, \cref{alg:gibbs-sampling} samples from a distribution
$\epsilon$-close to the Gibbs distribution (in total variation distance) by
making $\tilde{\mathcal{O}}(d^7 \frac{e^{\Delta/2}}{l^2} \kappa_{E/2}
\plog(1/\epsilon))$ queries to the oracle \eqref{eq:energy-oracle} of the
energy function. Note that dilating the domain of the definition of the
energy function by a scalar $\alpha$ multiplies the Poincar\'e constant by
$\alpha^2$. This is why $\kappa_{E/2}$ is normalized by a factor of $l^2$
in this complexity.

\paragraph*{Extension to non-periodic domains.}

Let $f:[-1,1]^d\to\mathbb R$ be a $(C, a)$-analytic function with diameter
$\Delta$. Since $g(\theta):=f(\cos\theta)$ is $2\pi$-periodic, applying
\cref{alg:gibbs-sampling} to it allows us to sample from the distribution
$\Theta\sim e^{-g}/Z_g$. The random variable $Y=\cos(\Theta)$, with the cosine
function applied component-wise, follows the distribution
$Y\sim e^{-f}/\tilde Z_f \cdot \prod_{i=1}^d\frac{1}{\sqrt{1-x_i^2}}$. Note that
\begin{align}
\begin{split}
\tilde Z_f &:= \int \frac{e^{-f(x)}}{\prod_i \sqrt{1-x_i^2}} \mathrm dx \\
&=Z_f \cdot \EE_{X \sim e^{-f}/Z_f} \left[\prod_i (1-X_i^2)^{-1/2}\right].
\end{split}
\end{align}
In \cref{sec:app-chebyshev} we show that this procedure is equivalent to using
Chebyshev interpolation instead of Fourier interpolation.

Using rejection sampling, we may accept samples from $Y$ with probability
$\frac{\prod_i\sqrt{1-x_i^2}}{\EE\left[\prod_i (1-X_i^2)^{-1/2}\right]}$ to
prepare samples from the original Gibbs distribution $e^{-f}/Z_f$ on the
hypercube $[-1,1]^d$. This adds a prefactor overhead of
$\mathcal O(\EE\left[\prod_i (1-X_i^2)^{-1/2}\right])$ for sampling
(see \cref{cor:chebyshev-rejection}). Consequently, the rejection sampling
algorithm is efficient only if the Gibbs measure is concentrated in the interior
of the hypercube. For example, this includes quadratic potentials (i.e. Gaussian
Gibbs measures) with bounded variance. However, for such families of functions
$\Delta = \Omega(d)$ therefore the end-to-end algorithm is not efficient.
For more general potentials, this process samples efficiently from the boundary
of the hypercube as detailed in \cref{prop:boundary-sampling}.

\section{Interpolation results}
\label{sec:results-fourier}

We now introduce the notion of semi-analyticity for smooth functions and prove
several favourable properties of it using the Fourier spectral method. In what
follows, for a string of length $d$ of
non-negative integers $\alpha = (\alpha_1, \cdots, \alpha_d)\in\mathbb{Z}_+^d$
we define $\alpha! := \alpha_1!\cdots \alpha_d!$, and
$|\alpha| := \alpha_1 + \cdots + \alpha_d$, and use the following notation for
higher order derivatives:
\begin{align}
D^\alpha := \frac{\partial^{|\alpha|}}{\partial x_1^{\alpha_1}
\cdots \partial x_d^{\alpha_d}}.
\end{align}

\begin{defn}
\label{def:body-semi-analytic}
Let the function $u:\mathbb{R}^d\rightarrow \mathbb{R}$ be $l$-periodic along
all axes, and moreover, let $X\sim \text{Unif}\left( [-l/2,l/2)^d \right)$ be
a uniform random variable. We say $u$ is semi-analytic if there exists
$C, a \in \mathbb{R}_+$, such that for any $m\in\mathbb{N}$ we have
\begin{align}\label{eq:semi-analytic}
\left(\frac{l}{2\pi}\right)^m\,
\sqrt{\mathbb{E}\left( \sum_{\alpha:|\alpha|=m}
\left|D^\alpha u(X) \right|^2\right)}
\leq C\,a^m \, m!.
\end{align}
Furthermore, we refer to $C$ and $a$ as the semi-analyticity parameters.
\end{defn}

The Fourier transform of $u$
\begin{equation}
\label{eq:fourier-transform}
u(x) = \sum_{k\in\mathbb{Z}^d} \widehat{u}[k] \,
e^{i\frac{2\pi \langle k, x\rangle}{l}}
\end{equation}
has coefficients $\widehat{u}[k] = \frac{1}{l^d}\int_{\mathbb T}
u(x) e^{-i \frac{2\pi \langle k,x\rangle}{l}} \, dx\,$ assigned to the lattice
points on $\mathbb Z^d$. The values $|\widehat u[k]|^2$ form a probability
measure on this lattice. In \cref{thm:SAisBern} we show that semi-analyticity
is equivalent to the sub-exponential concentration of this measure.

We provide several examples of semi-analytic functions. Any function with
finitely many non-zero Fourier coefficients is semi-analytic
(\cref{ex:Nyquist}). Every periodic real-analytic function is semi-analytic
(\cref{prop:analytic-implies-semianalytic}).
We also show how semi-analyticity parameters change through basic operations
like addition, multiplication, and composition of functions
(\cref{prop:semi-analytic-basic-prop}). In \cref{cor:neural-nets}, we use these
results to find the analyticity parameters of deep neural networks, as the de
facto function approximators used in machine learning (which can act as the
parameterized oracles shown in \cref{fig:EBM-a} for our quantum algorithm).

We now present our main result regarding upsampling of a quantum state
represented on a discrete lattice to a target continuous distribution defined
in the continuous ambient space of the lattice. Recently,
\citet{ramos2022efficient} discussed the idea of upsampling in the context
of efficient representation of classical data on a quantum computer although
without a rigorous mathematical account. We, however, provide a rigorous
analysis of the upsampling technique and its precision with respect to the
target \emph{continuous} distribution, rather than only to a discretization of
it on a finer lattice. Given a tuple of indices
$n = (n_1, \cdots, n_d) \in \{0, \ldots, 2N\}^{\times d}$, we denote the
associated computational basis state in the Hilbert space
$\mathcal V_N \cong \left(\mathbb C^{2N+1}\right)^{\otimes d}$ by $\ket{n}$. We
further denote the discretization of $u$ by $\vec u \in \mathcal V_N$, and the
unit vector parallel to that by $\ket{u_N} \propto
\sum_{m\in\{0, \ldots, 2N\}^{\times d}} u(x_{m-N}) \ket{m}$. We now state our
main interpolation result.

\begin{thm}[Main interpolation result]\label{thm:body-interpolation}
Given an $L$-Lipschitz $(C, a)$-semi-analytic periodic function $u$, an integer
$N\geq 2ad$, and a (previously prepared) quantum state
$\ket{\psi}\in\mathcal V_N$, such that
$\norm{\ket \psi - \ket{u_N}}_2 \leq \delta$,
there exists a quantum algorithm with gate complexity $\mathcal O
\left( \frac{d N}{a} \plog\left(NdLl/C\right) \right)$
that returns samples from a distribution within at most $\epsilon$ total
variation distance from the distribution proportional to $u^2$, where
\begin{equation*}
\begin{split}
\epsilon \leq \delta +
\frac{16 \sqrt{2} e^3 \, C}{\mathcal{U}}e^{-0.6\, \frac{N}{a}},
\end{split}
\end{equation*}
and $\mathcal U = \sqrt{\EE u^2(X) }$,
with $X\sim\mathrm{Unif}([-\frac{l}{2},\frac{l}{2}]^d)$.
\end{thm}

In \cref{sec:app-chebyshev} we demonstrate that Chebyshev interpolation can also
be performed efficiently by means of QFT and an analogous interpolation result
holds as stated in \cref{prop:chebyshev-interpolation}. Moreover,
in \cref{ex:inv-cos} we show a family of functions that provide
adversary witnesses obstructing the usage of very coarse discretization
for upsampling to arbitrarily small target errors.

\begin{thm}
\label{thm:body-lower-bound}
Let $u$ be a $(C,a)$-semi-analytic function. Consider any exact discretization
$\ket {u_N}$ on the discrete lattice with $N\leq \theta a/16$, where
$\theta\in(0,1)$. There is no algorithm that can return samples close to the
actual distribution (proportional to $u^2$) with a guaranteed error of less
than $(1-\theta)^2\,\frac{1}{1024e}$.
\end{thm}

We conclude this section by noting that even the first and second order
derivatives of a semi-analytic function $u$ can be approximated with high
precision (\cref{prop:high-precision-derivatives}). This is instrumental in
constructing high precision approximations to the generator $\mathcal L$ of the
FPE.

\section{Algorithm complexity}
\label{sec:main-results}

We now state the computational complexity of our Gibbs sampler
(\cref{alg:gibbs-sampling}).

\begin{thm}[Main sampling result]\label{thm:body-main-result}
Given an $L$-Lipschitz periodic potential $E$,
suppose that the one-parameter family of all probability measures
$\{e^{\mathcal{L}t} \rho_0: t\geq 0\}$ consists of semi-analytic functions with
parameters $C$ and $a$. \cref{alg:gibbs-sampling} samples from
a distribution $\epsilon$-close to the Gibbs distribution (in total variation
distance), by making
$\tilde {O}\left(a^4 d^7 \kappa_{E/2} e^{\Delta/2} l^{-2}\right)$
queries to the oracle of the energy function. The algorithm succeeds with
bounded probability of failure and returns a flag indicating its success.
In addition, the gate complexity of the algorithm is larger only by a factor of
$\plog(Cad e^{\Delta} (1+lL))/\epsilon)$.
\end{thm}

As a corollary, we show that the complexity of our algorithm improves under
simplifying assumptions on the geometry of the saddle points of the energy
function. Recall that a function $E$ is called a Morse function if all its
critical points are non-degenerate; i.e., if $\nabla^2 E(x)$ is non-singular
whenever $\nabla E(x)= 0$. \citet{mei2016landscape} quantify this condition
with additional parameters. We use a simplified definition
compared to this reference, and call $E$ to be a $\lambda$-strongly Morse
function if the spectrum of $\nabla^2 E(x)$ is bounded below by $\lambda > 0$
in absolute value at every critical point; equivalently, if $\norm{\left
(\nabla^2 E(x)\right)(v)} \geq \lambda \norm{v}$ for all critical points $x$
and all vectors $v$.

The strong Morse condition allows every saddle point in the energy landscape
to have \emph{steep enough} escape directions. Therefore, intuitively, the
dynamics opposite the gradient flow is not obstructed. \citet{li2020riemannian}
generalize this definition by allowing flat eigendirections in the saddle
points as long as the exponentiation map along these directions leaves us inside
the critical loci. We also note that for Morse function on compact domains,
the strong Morse criteria is always satisfied for some parameter $\lambda$.
Applying \citep[Proposition 9.14]{li2020riemannian} to weak Morse functions on
the products of spheres results in a better Poincar\'e constant than the
general bound in \cref{prop:Poin}. This is a generalization of the Bakry-Emery
criterion \citep[Proposition 4.8.1]{bakry2014analysis} well beyond strong
convexity. We have,

\begin{corol}
\label{cor:body-unique-optimum}
Let $E$ be a $\lambda$-strongly Morse potential with a unique global minimum.
Furthermore, assume that $E$, $\nabla E$, and $\nabla^2 E$ are Lipschitz
continuous with respective parameters $L_1$, $L_2$, and $L_3$. Letting
$\mathcal C$ denote the set of critical points of the energy potential $E$, we
also make an additional technical assumption as in
\citep[Proposition 9.14]{li2020riemannian}, namely that
\begin{align*}
C_F = \min\left\{ 1, \frac{\lambda}2,
\underset{x: d(x,\mathcal C) > \frac{\lambda}{L_3}}{\inf}
\frac{\norm{ \nabla E(x)}}{d(x,\mathcal C)} \right\} \in (0, 1]
\end{align*}
satisfies
$\max\left( \frac{4}{\lambda^2},
\frac{6L_2 d}{C_F^2} \right) \leq \frac{C_F^2}{12L_2 L_3^2d}$.
Then the query complexity of our algorithm reduces to
$\tilde O\left(a^4 d^7 \lambda^{-2} e^{\Delta/2}\right)$.
\end{corol}

This follows from the observation that $E$ satisfies a Poincar\'e inequality
with $\kappa_E = \mathcal O\left(\frac{1}{\lambda^2}\right)$
as per \citep[Proposition 9.14]{li2020riemannian}. We note that this proposition
is stated for $S^n \times \cdots \times S^n$ where $n \geq 2$. However, in
presence of a unique global minimum the result remains valid for $n=1$ as well;
i.e., in the case of high-dimensional tori.\footnote{We thank Mufan (Bill) Li
for confirming this fact.}

We now investigate how the Gibbs sampler discussed earlier can be employed to
calculate the expected values of random variables with bounded variance.
Specifically, we consider a periodic function $f:[-\frac{l}{2},\frac{l}
{2}]^d\rightarrow \mathbb R$ that belongs to $L^2(\rho)$, and aim at
estimating $\mathbb E[f(X)]$, where $X$ is a random variable with distribution
$\rho$. To this end we use the state-of-the-art mean estimation algorithm
presented in \citet{kothari2023mean}.

\begin{corol}[Mean estimation]
\label{cor:body-mean-estimation}
Let $E$ be an energy function, satisfying the assumptions made in \cref
{thm:body-main-result}. Furthermore, let $f$ be an $L_f$-Lipschitz $l$-periodic
function with diameter $\Delta_f$. There is a quantum algorithm that returns an
estimate $\widehat \mu$ to $\mathbb E[f(X)]$, with additive error at most
$\epsilon >0$ and success probability at least $1-\delta$, making
\begin{align}
\tilde O \left(a^4 d^7 e^{\Delta/2} \frac{\kappa_{E/2}}{l^2}
\frac{\Delta_f}{\epsilon} \log(\frac{1}{\delta})\right)
\end{align}
queries to the controlled and standalone oracles of the energy function $E$ and
the function $f$.
\end{corol}

Therefore a quantum computer can prepare a distribution $\epsilon$-close in TV
distance to the Gibbs distribution of Morse functions defined on tori using
$\tilde O\left(\lambda^{-2} e^{\Delta / 2} d^7\right)$
queries to the energy oracle, while the Riemannian Langevin diffusion of
\citet{li2020riemannian} uses
$\tilde O\left(\lambda^{-4} L^4 d^3\epsilon^{-2}\right)$
classical queries to the energy function.
For mean estimation on the same potentials, quantum computation requires
$\tilde O\left(\lambda^{-2}e^{\Delta / 2} d^7\Delta_f \epsilon^{-1}\right)$
queries to the controlled and standalone energy oracles, while classically
one requires
$\tilde O\left(\lambda^{-4} L^4 d^3\Delta_f^2 \epsilon^{-4}\right)$ queries
to the energy function. This suggests an exponential quantum speedup in the
sampling precision, and a quartic speedup in the precision of mean estimation.

For mean estimation we also obtain a quadratic speedup in the range $\Delta_f$
of the quantity $f$. However, since $\Delta = \mathcal O (Ll\sqrt d)$ we
require to sample at temperatures in $\Omega(Ll\sqrt d)$ in order to avoid an
exponentially poor performance in the dimension of the energy potential and its
Lipschitz constant. Nevertheless, even at low temperatures this algorithm
retains a quadratic advantage in comparison to classical rejection sampling.

\section{Conclusion}
\label{sec:conc}

In this paper, we propose a quantum algorithm for Gibbs sampling from a
continuous potential defined on a $d$-dimensional torus. Our algorithm queries
the quantum oracle of the energy potential
$\tilde O(d^7 \kappa_{E/2} e^{\Delta/2})$ times in the most notable factors,
with only polylogarithmic scaling with respect to the approximation error of the
collected samples from the Gibbs distribution in total variation distance.
Here $\Delta$ is the diameter of the range
of the potential or alternatively the thermodynamic $\beta$ if the potential
was considered to be normalized in the range. We also provide examples of
conditions under which at high enough temperatures our algorithm is suggestive
of exponential quantum advantage at this task.

Our motivation for this research is to use quantum computation as a building
block of learning schemes. For instance, the frontiers of research in
energy-based learning can take advantage of improved Gibbs samplers from
continuous potentials in order to both achieve a better representation
of knowledge, and require significantly lower power consumption. Our algorithm
achieves this end by solving the second-order PDE known as the Fokker--Planck
equation (FPE). When incorporated into energy-based learning
(\cref{sec:app-EBM}), the quantum algorithm does not use coherent queries to
classical data, but rather use Hamiltonian simulation techniques to solve a
PDE. Therefore, classical data does not need to be prepared in quantum random
access memory (QRAM) as typically assumed in the literature on quantum
algorithms.

This indicates that, more broadly, investigating steady states of PDEs other
than the FPE can also be instrumental in designing classical and quantum machine
learning algorithms. Our analysis made it apparent that except for the problem
of long mixing time in equilibrium dynamics, the exponential hardness in Gibbs
sampling at low temperatures exhibits itself when the eigendirections of the
generator of the FPE are far from perpendicular. We believe that this
technical constraint may be ramified for special families of potentials which
ideally exhibit sufficient expressivity for learning tasks (or for other
applications).

In order to obtain these results we take advantage of the efficiency of quantum
Fourier transforms in manipulating functions in their Fourier representations.
We show that this performance requires sub-exponential concentration of the
Fourier components. We also show that this is equivalent to a condition milder
than analyticity which we name semi-analyticity. We quantify analyticity and
semi-analyticity of functions using parameters we introduce and track how
these parameters change under arithmetic operations and compositions. However,
many similar properties remain open to be investigated. We also generalize
our upsampling results to non-periodic functions using Chebyshev polynomials.

Finally, we mention that our method makes queries directly to the oracle of
the energy potential, and therefore is a zeroeth order method. This is unlike
typical classical algorithms for Gibbs sampling, specially ones that use
the stochastic integration of Langevin dynamics, the SDE associated to the
FPE. It therefore remains open to investigate the opportunity
for improving our results using quantum queries to the first order oracles
of the potential.

\section*{Acknowledgments}

AM and PR acknowledge the support of NSERC Discovery grant RGPIN-2022-03339. PR
further acknowledges the support of Mike and Ophelia Lazaridis, Innovation,
Science and Economic Development Canada (ISED), Irr\'eversible Inc., and the
Perimeter Institute for Theoretical Physics. Research at the Perimeter
Institute is supported in part by the Government of Canada through ISED, and by
the Province of Ontario through the Ministry of Colleges and Universities.

\section*{Impact statement}

This paper presents work whose goal is to advance the fields of machine learning
and qunatum computation. There are many potential societal consequences of our
work, none which we feel must be specifically highlighted here.

\bibliography{main}
\bibliographystyle{icml2024}

\newpage
\appendix
\onecolumn

\section{Semi-analytic functions}
\label{sec:app-semianalytic}

In this section we analyze the effect of discretization on estimating
distributions and derivatives of differentiable functions. Previous works
(e.g., \cite{childs2021high}) assume an upper bound for all derivatives of
the function. This, however, is a restrictive assumption as it excludes simple
functions such as $\cos(2x)$. We show that a milder condition such as
analyticity (or an even a weaker condition we call semi-analyticity) is enough
for such results to hold.

We now introduce the notion of semi-analyticity for smooth functions and prove
several favourable properties of it using the Fourier spectral method. We borrow
some of the ideas presented in \citet[Section 2.2]{shen2011spectral}, although
ibid is only concerned with functions of a single variable and focused on
interpolation errors. In what follows, for a string of length $d$ of
non-negative integers $\alpha = (\alpha_1, \cdots, \alpha_d)
\in\mathbb{Z}_{\geq 0}^d$ we define $\alpha! := \alpha_1!\cdots \alpha_d!$, and
$|\alpha| := \alpha_1 + \cdots + \alpha_d$, and use the following notation for
higher order derivatives:
\begin{align}
D^\alpha := \frac{\partial^{|\alpha|}}{\partial x_1^{\alpha_1}
\cdots \partial x_d^{\alpha_d}}.
\end{align}

\begin{defn}[\cref{def:body-semi-analytic} in the manuscript]
\label{def:semi-analytic}
Let $u:\mathbb{R}^d\rightarrow \mathbb{R}$ be $l$-periodic along all axes, and
moreover, let $X\sim \text{Unif}\left( [-l/2,l/2)^d \right)$ be a uniform
random variable. We say $u$ is semi-analytic if there exists
$C, a \in \mathbb{R}_+$, such that for any $m\in\mathbb{N}$ we have
\begin{align}\label{eq:semi-analytic}
\left(\frac{l}{2\pi}\right)^m\,
\sqrt{\mathbb{E}\left( \sum_{\alpha:|\alpha|=m}
\left|D^\alpha u(X) \right|^2\right)}
\leq C\,a^m \, m!
\end{align}
Furthermore, we refer to $(C, a)$ as the semi-analyticity parameters.
\end{defn}

Note that for a semi-analytic function $(C,a)$ are scale invariant; i.e.,
replacing $u(\cdot)$ by $u(\frac{\cdot}{\alpha})$ and at the same time changing
the fundamental domain to $[-\frac{\alpha\, l}{2}, \frac{\alpha\, l}{2}]^d$,
for any $\alpha>0$, would result in another $(C,a)$-semi-analytic function.
One could absorb the coefficient $\left(\frac{l}{2\pi}\right)$ into $a$,
however, we find our current formulation more convenient.

For simplicity, consider the case of having a univariate function $f$. We recall
that the Taylor expansion around the point $x_0$ is $f(x) = f(x_0) + \sum_
{m=1}^\infty \frac{f^{(m)}}{m!} (x-x_0)^m$. Hence, imposing the condition
$\left|f^{(m)}(x_0)\right| \leq a^m \, m!$ on the growth of the derivatives
guarantees convergence of this series for all $x\in(x-a^{-1},x+a^{-1})$.
Similarly, in the multi-variate case, imposing the condition $\left|D^{\alpha}f
(x_0)\right| \leq \alpha! \, a^{|\alpha|}$ guarantees the convergence of the
Taylor expansion in the box $\prod_{i=1}^d (x_{0,i}-a^{-1},x_{0,i}+a^
{-1})$, where $x_{0,i}$ denotes the $i$-th component of $x_0$. Although a
rigorous connection between analyticity and semi-analyticity is provided below,
we emphasize that we can understand the parameter $a$ as an inverse convergence
radius.

Recall that for a periodic function $u$ over $[-\frac{l}{2},\frac{l}{2}]^d$,
the discretization on $2N + 1$ points along each axis results in a vector
$\vec{u_{N}} \in \mathcal{V}_N$. We also use the notation
$u_{N}[n] := u\left( \frac{l\, n} {2N+1} \right), \forall n\in[-N .. N]^d$.
We now define the Fourier transform of an $l$-periodic function
$u:\mathbb{R}^d\rightarrow \mathbb{R}$ via
\begin{equation}
\begin{split}
u(x)
&= \sum_{k\in\mathbb{Z}^d} \widehat{u}[k] \,
e^{i\frac{2\pi \langle k, x\rangle}{l}}, \quad \text{ where } \\
\widehat{u}[k] &= \frac{1}{l^d}\int_{\mathbb T}
u(x) e^{-i \frac{2\pi \langle k,x\rangle}{l}} \, dx\,.
\end{split}
\end{equation}

Note that the Fourier transform of $D^\alpha u$ is
$\left(\frac{2\pi}{l}\right)^{|\alpha|}(ik_1)^{\alpha_1} \cdots
(ik_d)^{\alpha_d} \widehat{u}[k]$, and hence by Parseval's theorem we have
$\left(\frac{l}{2\pi}\right)^{|\alpha|}\EE\left[\left(D^\alpha u\right)^2\right]
= \sum_{k\in\mathbb Z^d}k_1^{2\alpha_1}k_2^{2\alpha_2}
\cdots k_d^{2\alpha_d} |\widehat{u}[k]|^2$. Moreover, since
$\sum_{\alpha:|\alpha|=m} k_1^{2\alpha_1}\cdots k_d^{2\alpha_d}
= \left(k_1^2 + \cdots + k_d^2\right)^m$, we conclude that
\cref{def:semi-analytic} is equivalent to
\begin{align}
|u|_m:=\sqrt{\sum_{k\in\mathbb{Z}^d} \norm{k}^{2m}
\left| \widehat{u}[k] \right|^2} \leq C\, a^m\, m!.
\end{align}
It is straightforward to check that the quantity introduced above is
a semi-norm.

We now provide examples of semi-analytic functions and show that this condition
is quite mild. In particular, in \cref{prop:analytic-implies-semianalytic} we
prove that every analytic function is semi-analytic (hence, the naming).

\begin{exm}\label{ex:Nyquist}
Any function with finitely many non-zero Fourier coefficients is semi-analytic
with $C = \sqrt{\Var[ u(X) ]}$ and $a=k_0$, where
$k_0 := \max\{\norm{k}:\, \widehat{u}[k] \ne 0\}$.
\end{exm}

\begin{exm}\label{ex:cosine-function}
For any $z > 0$, the function $u(x) = e^{z\cos(x)}$ with domain $[-\pi,\pi]^d$
is $\left(I_0(z)\,e^{z/2}, \max\{\frac{z}2,1\}\right)$-semi-analytic with $I_0$
being the modified Bessel function of the first kind. To see this, note that
the Fourier coefficients of $u$ are described by the modified Bessel function
of the first kind \citep[page 376]{abramowitz1964handbook}:
\begin{equation}
\begin{split}
\widehat{u}[k] &= \frac{1}{2\pi}
\int_{x=-\pi}^{\pi} e^{-ikx} \, e^{z\cos(x)} \, dx = I_k(z).
\end{split}
\end{equation}
Furthermore,
\begin{equation}\label{eq:semi-norm}
\begin{split}
I_k(z) &= \left(\frac{z}{2}\right)^k
\sum_{l\geq 0} \frac{\left(\frac{z}{2}\right)^{2l}}{l! (k+l)!}
\leq \left(\frac{z}{2}\right)^k \frac{1}{k!}
\sum_{l\geq 0} \frac{\left(\frac{z}{2}\right)^{2l}}{(l!)^2}
= \frac{1}{k!} \, \left(\frac{z}{2}\right)^k\, I_0(z).
\end{split}
\end{equation}
Therefore, $\left|\widehat{u}[k] \right|
\leq \left(\frac{z}{2}\right)^k\,\frac{I_0(z)}{k!}$, which allows us to write
\begin{equation}
\begin{split}
\sum_{k\in\mathbb{Z}} |k|^{2m} \left| \widehat{u}[k] \right|^2
&\leq \left(I_0(z)\right)^2\,
\sum_{k\in\mathbb{Z}} \left(\frac{z}{2}\right)^{2k}
\frac{k^{2m}}{\left(k!\right)^2}.
\end{split}
\end{equation}
Hence, using the fact that the $l_1$-norm is larger in value than the
$l_2$-norm, we conclude that
\begin{align}
|u|_{m} \leq I_0(z) \,
\sum_{k\in\mathbb{Z}} \left(\frac{z}{2}\right)^k \frac{k^{m}}{k!}
\leq I_0(z)\, e^{z/2} \, \left(\max\{\frac{z}2,1\}\right)^m\, m!
\end{align}
where the last inequality follows from \cref{lem:cool}.
\end{exm}

\begin{prop}\label{prop:analytic-implies-semianalytic}
Every periodic real-analytic function is semi-analytic.
\end{prop}

\begin{proof}
Let $f$ be a real analytic and periodic function. It follows from
\citet[Lemma 1]{komatsu1960characterization} that there exist $C, a>0$ such that
\begin{align}
\sup_{x\in\mathbb T} \abs{D^\alpha f(x)} \leq C \, a^{\abs{\alpha}}\, \alpha!.
\end{align}
From this we conclude that
\begin{equation}
\begin{split}
\sqrt{\sum_{\alpha: |\alpha| = m } \EE\left[ \left(D^\alpha f\right)^2 \right]}
\leq C \, a^m \, \sqrt{\sum_{\alpha:|\alpha|=m} (\alpha!)^2 }
\leq C \, a^m \, \sum_{\alpha:|\alpha|=m} \alpha!
\leq 3^{d-1} C\, a^m \, m!
\end{split}
\end{equation}
where the last inequality follows from \cref{lem:fact2}.
\end{proof}

We let $F_{N}$ denote the unitary representing the $d$-dimensional discrete
Fourier transform and adopt the notation
$\tilde{u}_{N} := F_{N} \vec{u_{N}}$. We drop the subscript $N$
when it is clear from the context. Moreover, consider
$\Sigma \subseteq \Gamma$ as an embedding of a finite alphabet $\Sigma$
in $\Gamma$, $\Gamma$ being either a larger finite alphabet or $\mathbb N$.
We also consider the natural embedding of spaces of functions
$\iota: l^2 (\Sigma )\hookrightarrow l^2 (\Gamma)$ induced by the inclusion
$\Sigma \subseteq \Gamma$ and the usual $l^2$ norm
$\|a\|= \sqrt{\sum_{n \in \Gamma} |a[n]|^2}$ and the induced metric
$d(a, b)= \|a- b\|$.

We observe that when dealing with a periodic function having a Fourier spectrum
with bounded support, such that $\widehat u[k] = 0$ for $k\notin
[-k_0 .. k_0]^d$, Nyquist's well-known theorem guarantees exact recovery of the
function \cite{oppenheim1975digital}. Specifically, given a discretization on a
lattice $V_N$ with $N\geq k_0$, there exists a classical algorithm to
reconstruct the entire continuous function $u(x)$. In the following, we show
that under the milder condition of semi-analyticity, one can still
achieve approximate reconstructions. Even though the reconstructions will not be
entirely accurate, we can limit the errors to a poly-logarithmic overhead by
exploiting the fact that the values of $\widehat u[k]$ are exponentially small
for sufficiently large $k$ (as shown in \cref{lem:tail} below). Furthermore, it
is worth noting that by measuring quantum states in the computational basis, we
obtain a sample drawn from a distribution corresponding to the squared
amplitudes. This feature enables us to develop a sampler in the continuum.

\begin{lem}
\label{lem:tail}
For a $(C,a)$-semi-analytic periodic function $u$, with $N$ an integer
satisfying $N \geq 2a$, we have
\begin{align}
\sqrt{\sum_{k:\norm{k}_2\geq N }
\left| \widehat{u}[k] \right|^2 }
\leq 2 e^{3}\, C\, e^{-\frac{N}{a} \left( 1-\frac{1}{2e}\right)}.
\end{align}
\end{lem}

\begin{proof}
We have
\begin{align}
\sum_{k:\norm{k}_2\geq N } \left| \widehat{u}[k] \right|^2
\leq N^{-2m} \sum_{k:\norm{k}_2\geq N }
\norm{k}^{2m} \left| \widehat{u}[k] \right|^2
\leq N^{-2m}\, C^2 \, a^{2m} \, \left(m!\right)^2.
\end{align}
Now using $m! \leq \frac{m^{m+1}}{e^{m-1}}$
we obtain
\begin{align}
N^{-m} \, a^m \, m!\leq \frac{e^2}{a} N^{-m} \,
\left(\frac{a(m+1)}{e}\right)^{m+1}.
\end{align}
Setting $m+1 = \lfloor N/a \rfloor$ yields the bound
\begin{align}
N^{-m} \, a^m \, m!\leq \frac{e^3}{a} N \, e^{-N/a}.
\end{align}
We can now use the inequality $x \leq \alpha e^{\frac{x}{\alpha \, e}}$ for all
$x\in\mathbb{R}$ and all $\alpha >0$ by setting $\alpha = 2$ and $x= N/a$ to
complete the proof.
\end{proof}

The reader may notice analogies between the result of \cref{lem:tail} and the
sub-exponential decay bounds in the literature of concentration of measure. We
discuss this connection in \cref{sec:Meas-Consenteration}.

\begin{lem}
\label{lem:distance-bound}
Let $u$ be a $(C,a)$-semi-analytic periodic function with period
$[-\frac{l}{2}, \frac{l}{2}]^d$, and let $N$ be an integer such that
$N\geq 2a d$. We have
\begin{align}\label{eq:error-big}
d\left( \frac{1}{(2N+1)^{d/2}}\,  \tilde{u}_{N} ,\widehat{u} \right)
\leq 2 \sqrt{2} e^3\, C e^{-\frac{3N}{5a}}.
\end{align}
\end{lem}

\begin{proof}
We start by noting that
\begin{equation}
\begin{split}
\frac{1}{(2N+1)^{d/2}}\,\tilde{u}_{N}[k]
&= \frac{1}{(2N+1)^d} \sum_{n \in [-N..N]^d } u_{N}[n] \,
e^{-i \frac{2\pi \langle k,n \rangle}{2N+1} } \\
&= \frac{1}{(2N+1)^d} \sum_{n \in [-N..N]^d }
\sum_{k'\in\mathbb{Z}^d} \widehat{u}[k'] \,
e^{-i \frac{2\pi \langle k - k',n \rangle}{2N+1} }\\
&= \widehat{u}[k]
+ \sum_{p\in \mathbb{Z}^d \setminus \{0\}} \widehat{u}[k+(2N+1)p].
\label{eq:wrap}
\end{split}
\end{equation}
Therefore,
\begin{align}
d\left( \frac{1}{(2N+1)^{d/2}}\,  \tilde{u}_{N} ,\widehat{u} \right)^2
&= \sum_{k\in[-N..N]^d} \left| \sum_{p\in\mathbb{Z}^d\setminus \{0\}}
\widehat{u}[k+(2N+1)p] \right|^2
+ \sum_{k\in\mathbb{Z}^d \setminus [-N..N]^d}
\left| \widehat{u}[k] \right|^2
\end{align}
where the second term was upper-bounded in \cref{lem:tail}. As for the first
term
\begin{equation}
\begin{split}
\sum_{k\in[-N..N]^d}  \left| \sum_{p\in\mathbb{Z}^d\setminus \{0\}}
\widehat{u}[k+(2N+1)p] \right|^2 &\leq \sum_{k\in[-N..N]^d}
\bigg(\sum_{p\in\mathbb{Z}^d\setminus \{0\}} \norm{k+(2N+1)p}^{-2m} \\
& \qquad \times \sum_{p\in\mathbb{Z}^d\setminus \{0\}}
\norm{k+(2N+1)p}^{2m} \left| \widehat{u}[k+(2N+1)p] \right|^2 \bigg) \\
& \leq a^{2m} C^2\, (m!)^2 \, \max_{k\in[-N..N]^d}
\sum_{p\in\mathbb{Z}^d\setminus \{0\}} \norm{k+(2N+1)p}^{-2m}.
\end{split}
\end{equation}
We note that
\begin{equation}
\begin{split}
\max_{k\in[-N..N]^d} \sum_{p\in\mathbb{Z}^d\setminus \{0\}}
\norm{k+(2N+1)p}^{-2m}
& \leq \max_{k\in[-N..N]^d} \sum_{p\in\mathbb{Z}^d\setminus \{0\}}
\norm{k+2N\,p}^{-2m}\\
& \leq N^{-2m} \, \max_{x\in[-1,1]^d} \sum_{p\in\mathbb{Z}^d\setminus \{0\}}
\norm{x+2\,p}^{-2m},
\label{eq:sefuleq}
\end{split}
\end{equation}
where the first inequality follows from term-by-term comparison of the sums. Using \cref{lem:useful}, if $m \geq d$, we get
\begin{equation}
\begin{split}
d\left( \frac{1}{(2N+1)^{d/2}}\,
\tilde{u}_{N} ,\widehat{u} \right)
\leq 2\sqrt{2} a^m  C\, N^{-m} \, 2^{d/2}\, m!
\leq 2\sqrt{2} a^m  \, C\,N^{-m} \, 2^{n/4a}\, m!
\end{split}
\end{equation}
and setting $m= \lfloor N/a \rfloor - 1$ (which guarantees $m\geq d$) yields
\begin{equation}
\begin{split}
d\left( \frac{1}{(2N+1)^{d/2}}\, \tilde{u}_{N}, \widehat{u} \right)
\leq 2 e^3 \sqrt{2}\,
C e^{-\frac{n}{a}\left(1 - \frac{1}{2e}-\frac{\ln(2)}{4} \right)}
\end{split}
\end{equation}
which concludes the proof.
\end{proof}

For our purposes, we will be applying \cref{lem:distance-bound} to normalized
vectors. Here we highlight the following distance bound as a corollary
following immediately from \cref{lem:distance-bound} and \cref{lem:unitnorm}.
\begin{corol}
\label{corol:er}
Let $N\geq 2ad$ and $u$ be a $(C,a)$-semi-analytic periodic function. We have
\begin{align}
d\left(F_{N}\ket{u_{N}} ,\frac{\widehat{u}}{\mathcal{U}} \right)
\leq \frac{4\sqrt{2} e^3\, C }{\mathcal{U}} e^{-0.6 N/a}
\end{align}
where $\mathcal{U}=
\left(\mathbb{E}
\left[|u(X)|^2\right]\right)^{1/2}$, with $X\sim \mathrm{Unif}\left([-l/2,l/2]^d\right)$
\end{corol}

Now we show that upsampling a semi-analytic function is useful in achieving
minor aliasing effects.\footnote{In signal processing, aliasing effects refer
to the errors caused by Fourier interpolation, specially when the tails of the
Fourier transformation (that is, the very high and very low frequency
components) have non-negligible amplitudes \cite{oppenheim1975digital}.} Recall
the sampling procedure introduced in \cref{sec:algorithm}. That is, we measure
the output state of the algorithm, say $\ket{\psi}$, in the computational basis
to obtain $x\in V_{N}$. We then sample uniformly at random from the box
$\prod_{i=1}^d[x_i - \frac{l}{4N+2}, x_i+ \frac{l}{4N+2}]$. We call this
procedure \textit{continuous sampling} from $\ket{\psi}$.

\begin{rem}\label{rem:cont-sampling}
The total variation distance of continuous sampling from two quantum states is
upper bounded by the $l_2$-norm of their difference. To see this, let
$\ket{\psi}$ and $\ket{\phi}$ be two quantum states. We denote the probability
density associated with the random variable obtained from continuous sampling
from $\ket{\psi}$ by $\mu_\psi$ and we note that
\begin{align}
\mu_\psi(x) = \sum_{n\in[-N..N]^d}
\mathbf{1}_{ \{x\in\mathbb{B}_n \}}
\left| \psi[n] \right|^2 \left(\frac{2N+1}l\right)^d
\end{align}
where $\mathbf 1_{\{x\in\mathbb{B}_n\}}$ is the identifier function; i.e., it is
$1$ if $x\in\mathbb B_n =  \prod_{i=1}^d \left[ \frac{n_i \ell}{2 N+ 1} - \frac{l}{4N+2},
\frac{n_i \ell}{2 N+ 1} + \frac{l}{4N+2} \right]$ and $0$ otherwise. One can then write
\begin{align}
\frac12\int dx \, \left| \mu_\psi(x) - \mu_{\phi} (x) \right|
= \frac12\sum_{n\in[-N..N]^d} \left| |\psi[n]|^2 - |\phi[n]|^2 \right|
\leq \norm{\ket\psi - \ket\phi}
\end{align}
where the inequality follows from \cref{lem:TV-bound}.
\end{rem}

\begin{prop}\label{prop:upsampling}
Given a $(C, a)$-semi-analytic periodic function $u$, an integer $N\geq 2ad$,
and a quantum state $\ket{\psi_{N}} \in \mathcal{V}_{N}$ satisfying
$\norm{\ket{\psi_{N}} - \ket{u_{N}}} \leq \delta$, there exists
$M \in \mathbb N$ such that continuous sampling from
$F_{M}^{-1} \iota F_{N} \ket{\psi_{N}}$ results in an $\epsilon$-approximation
to the continuous distribution proportional to $u^2$ in total variation
distance, where
\begin{equation}
\begin{split}
\epsilon \leq \delta +
\frac{8 \sqrt{2} e^3 \, C}{\mathcal{U}}e^{-0.6\, \frac{N}{a}}.
\end{split}
\end{equation}
\end{prop}

\begin{proof}
Firstly, note that for any integer $r$, since sampling from $\ket{u_
{r}}$ reaches the actual $|u|^2$ distribution as $r\rightarrow \infty$, there
exists an integer $M^\ast$ such that $\ket{u_{r}}$ for all $r\geq M^\ast$ gives
$\delta$-approximation of the continuous distribution. If $N \geq M^\ast$, then
the statement is trivially satisfied after setting $M = N$. Otherwise, let
$M = M^\ast$, and note that due to \cref{corol:er} and by an application of the
triangle inequality
\begin{align}
\norm{\iota F_{N} \ket{u_{N}} - F_{M} \ket{u_{M}}}
\leq \frac{8\sqrt{2}e^3\, C}{\mathcal{U}} e^{-0.6\,\frac{N}{a}}.
\end{align}
Since isometries preserve the $l_2$-norm, by another application of the triangle
inequality we get
\begin{align}
\norm{F_{M}^{-1} \iota F_{N}\, \ket{\psi_{N}} - \ket{u_{M}}}
\leq \frac{8\sqrt{2}e^3\, C}{\mathcal{U}} e^{-0.6\, \frac{N}{a}}.
\end{align}
Finally, using another triangle inequality for the total-variation distance and
\cref{rem:cont-sampling} we obtain the result.
\end{proof}

We now investigate the gate complexity of interpolating a semi-analytic
function.

\begin{thm}[\cref{thm:body-interpolation} in the manuscript]
\label{thm:interpolation}
Given an $L$-Lipschitz $(C, a)$-semi-analytic periodic function $u$, an integer
$N\geq 2ad$, and a quantum state $\ket{\psi}\in\mathcal V_N$, such that
$\norm{\ket \psi - \ket{u_N}} \leq \delta$,
there exists a quantum algorithm with gate complexity $\mathcal O
\left( \frac{d N}{a} \plog\left(NdLl/C\right) \right)$
that returns samples from a distribution within at most $\epsilon$ total
variation distance from the distribution proportional to $u^2$, where
\begin{equation*}
\begin{split}
\epsilon \leq \delta +
\frac{16 \sqrt{2} e^3 \, C}{\mathcal{U}}e^{-0.6\, \frac{N}{a}},
\end{split}
\end{equation*}
and $\mathcal U = \sqrt{\EE u^2(X) }$, with $X\sim\mathrm{Unif}([-\frac{l}{2},\frac{l}{2}]^d)$.
\end{thm}

Notice that by `given a state $\ket\psi$' in the theorem above, we mean that we are given one copy of $\ket\psi$.

\begin{proof}
In this proof, for an integer $\alpha \in \mathbb N$
the notation $[\alpha]$ stands for $\{0, 1, \ldots, \alpha\}$.
Using \cref{lem:choose-M}, we set $M = \left\lceil\frac1{2\epsilon'}
\frac{L l d/2 + 10/3\sqrt 2\,ae^4 C}{\mathcal{U}}\right\rceil$ with
$\epsilon' = \frac{8 \sqrt{2} e^3 \, C}{\mathcal{U}}e^{-0.6\, \frac{N}{a}}$.
We now show that we can implement $F_M^{-1} \iota F_N$ with
$\tilde{\mathcal O}( d\log M)$ gates. Firstly, note that
\begin{equation}
F_N = \bigotimes_{i=1}^d f_N^{(i)}
\end{equation}
where $f_N = \sum_{m,k\in[2N]} e^{-i \frac{2\pi (k-N) (m-N)}{2N+1}}
\ket{k}\bra{m}$, and the superscript $i$ means that it acts non-trivially on
the $i$-th register. We notice that $f_N = T_N \hat f_N T_N$, where
$T_N = \sum_{k\in[2N]} e^{-i \frac{2\pi N k}{2N+1}} \ket k \bra k$, and
$\hat f_N = \sum_{m,k \in [2N]}e^{-i \frac{2\pi k m}{2N+1}} \ket{k}\bra{m}$
is the usual quantum Fourier transform and thus can be implemented using
$\mathcal{O}(\log N \log \log N)$ gates. Furthermore, it is straightforward to
implement $T_N$ in $\mathcal O(\log N)$. Overall, the gate complexity of $F_N$
is $\tilde{\mathcal{O}}(d \log N)$, since it can be implemented via
$d$ applications of $f_N$ in parallel.

It remains to show that $\iota$ itself can also be implemented using
$\mathcal O(d \log M)$ gates. Consider the isometry
$\hat \iota: \mathcal V_N \rightarrow \mathcal V_M$ defined as
$\hat \iota = \bigotimes \widehat{\iota'}^{(i)}$ via
$\widehat{\iota}': \ket{n} \mapsto \ket{n}$ (for $n\in[2N]$). We note that
$\widehat{\iota}'$ can be performed by adding auxiliary
qubits prepared in the $\ket 0$ state. Also, from
$\iota = \bigotimes_{i=1}^d \iota'^{(i)}$ with
$\iota': \ket{k+N} \mapsto \ket{k+M}$ for $k\in[-N..N]$,
we conclude that $\iota' = S \widehat{\iota'}$,
where $S \in U\left(\mathcal{V_M}\right)$ is a shift operator for integers
represented in the computational basis states,
$S:\ket{m} \mapsto \ket{m+M-N \text{ mod } 2M+1}$.
Finally, note that $S = \hat f_N S'\hat f_N$, given
$S' = \sum_{k\in[2N]} e^{-i \frac{2\pi (M-N)k}{2M+1}} \ket{k} \bra {k}$,
and the latter operator has gate complexity $\mathcal O(\log M)$.
We therefore conclude that the complexity of implementing $\iota$ is
$\mathcal{O}\left(d\log M \right)$.
\end{proof}

\subsection{Concentration of measure}\label{sec:Meas-Consenteration}

Sub-exponential distributions are studied in the context of high-dimensional
probability theory. Intuitively, a random variable is considered
sub-exponential if its probability distribution function has a tail that
vanishes exponentially or faster \cite{vershynin2018high}. We make a connection
between this concept and our notion of semi-analyticity, which will later allow
us to better understand the latter class of functions. Let us recall the
Bernstein random variables, which will appear to be useful later in this section.

\begin{defn}\label{lem:BernRV}
$X$ is a Bernstein random variable, if $X \geq 0$ almost surely, and
for some $A,b>0$ its moments are upper bounded as
\begin{align}
\EE X^m \leq A\, b^{m}\, m!,
\end{align}
for all positive integers $m$.
\end{defn}

Following \citet{boucheron2013concentration} we prove concentration bounds on a
Bernstein random variable.
\begin{lem}\label{lem:BernTail}
Let $X$ be the Bernstein random variable defined in \cref{lem:BernRV}. It is the
case that
\begin{align}\label{eq:BernsteinTail}
\mathbb P[X\geq t] \leq \begin{cases}
\max(A,1)\, e^{-\frac{(t-b)^2}{8b^2}}, \quad &\text{if } t\leq3b,\\
e\,\max(A,1)\, e^{-\frac{t}{2b}}, \quad &\text{if } t>3b.
\end{cases}
\end{align}
\end{lem}

\begin{proof}
Let us first upper bound the generating function corresponding to $X$. Let
$0 \leq\lambda < b^{-1}$, then
\begin{align}
\EE e^{\lambda X}
= 1 + \sum_{m\in \mathbb N} \frac{\lambda^m \EE X^m}{m!}
\leq 1 + A \sum_{m\in \mathbb N}(b\lambda)^m
= 1 + A \frac{\lambda b}{1-\lambda b}.
\end{align}
Moreover, if $0\leq\lambda\leq\frac{1}{2b}$, we have
$\frac{1}{1-\lambda b} \leq 1+ 2\lambda b$, which together with the identity
$1+x\leq e^x$ yield
\begin{align}
\EE e^{\lambda X} \leq \max(A,1)\,\exp{\lambda b + 2\lambda^2 b^2},
\quad \forall \lambda\in[0,\frac{1}{2b}].
\end{align}
We may now upper bound the tail probability via Chernoff's bound
\begin{align}
\mathbb P [X\geq t]
\leq \inf_{0\leq\lambda} e^{-\lambda t} \EE[e^{\lambda X}]
\leq \max(A,1) \, \min_{0\leq\lambda\leq \frac{1}{2b}} \,
\exp{-\lambda( t-b-2\lambda b^2)}.
\end{align}
For $t\leq 3b$, we make the choice $\lambda = \frac{t-b}{4b^2}$, and otherwise,
we choose $\lambda = \frac{1}{2b}$ to conclude the result.
\end{proof}

Note that the tail of a Bernstein random variable shows a sub-exponential
behavior eventually, as described by \cref{eq:BernsteinTail}. Indeed, the set
of Bernstein random variables coincides with the set of positive
sub-exponential distributions as stated bellow.

\begin{prop}
The set of Bernstein random variables is the set of almost surely positive
random variables that are sub-exponential.
\end{prop}

\begin{proof}
This follows from the characterization of sub-exponential random variables
in \citep[Proposition 2.7.1]{vershynin2018high}, according to which, the
positive random variable $X$ is sub-exponential if and only if
$\EE X^m \leq Q^m\, m^m$ for some $Q\geq0$.\footnote
{\citet{vershynin2018high} take the exponent $m$ to be any real number
larger than $1$, but one can readily observe that $m\in\mathbb N$ is also a
sufficient condition in their proofs.} Now, let $X$ be a sub-exponential
distribution. Using $\frac{m^m}{e^{m-1}} \leq m!$, we have
\begin{align}
\EE X^m \leq Q^m \, m^m \leq e^{-1} (Qe)^m \, m!,
\end{align}
which concludes that $X$ has a Bernstein property.

Conversely, assume $X$ has a Bernstein property. From $m! \leq m^m$, one
concludes that
\begin{align}
\EE X^m \leq A b^m m! \leq (\max(A,1)\, b)^m \, m^m,
\end{align}
which provides that $X$ is a sub-exponential random variable.
\end{proof}

Now we show the connection between the notions of concentration of measure
and the semi-analyticity condition in \cref{def:semi-analytic}. The next
definition allows us to make this connection clear.

\begin{defn}
Consider the Fourier transform of an $l$-periodic function $u:[-\frac{l}
{2},\frac{l}{2}]^d\rightarrow \mathbb R$, denoted by $(\hat u[k])_k$. Note that
$(\abs{\hat u[k]}^2/\mathcal U^2)_k$ defines a probability distribution on the
sample space $\mathbb Z^d$. We call the random variable $K_u$ corresponding to
this distribution the Fourier random variable of $u$.
\end{defn}

With this definition at hand, we make the following connection between
semi-analyticity and the Bernstein random variables.

\begin{thm}\label{thm:SAisBern}
A periodic function $u$ is semi-analytic if and only if $\norm{K_u}$ has the
Bernstein property. In particular
\begin{itemize}
    \item if $u$ is $(C,a)$-semi-analytic, then $\norm{K_u}$ has a Bernstein
    property with parameters $(C\mathcal U^{-1}, a)$; and
    \item if $\norm{K_u}$ has a Bernstein property with parameters $(A,b)$,
    then $u$ is $(\sqrt{2Ae}, 4b)$-semi-analytic.
\end{itemize}
\end{thm}

\begin{proof}
$u$ is semi-analytic $\Rightarrow$ $\norm{K_u}$ has a Bernstein property:
From \eqref{eq:semi-norm}, we have

\begin{align}
\sqrt{\EE \norm{K_u}^{2m}} \leq \mathcal U^{-1} \, C\, a^m \, m!.
\end{align}

Putting this together with the Jensen inequality
$\EE \norm{K_u}^m \leq \sqrt{\EE \norm{K_u}^{2m}}$, proves that $\norm{K_u}$ is
a Bernstein random variable with parameters $(C\mathcal U^{-1}, a)$.

$\norm{K_u}$ has a Bernstein property $\Rightarrow$ $u$ is semi-analytic: By
definition we have $\EE\norm{K_u}^{2m} \leq A \, b^{2m}\, (2m)!$. Note that
\begin{align}
(2m)! \leq \frac{(2m)^{2m+1}}{e^{2m-1}}
\leq 2e\, 4^{2m}\, \frac{m^{2m-2}}{e^{2m-2}}
\leq 2e\, 4^{2m} (m!)^2
\end{align}
where the second inequality uses $4^m \geq m^3$. This implies
$\sqrt{\EE\norm{K_u}^{2m}} \leq \sqrt{2Ae}\, (4b)^m \, m!$.
\end{proof}

Making this connection allows us to obtain a Fourier concentration result
similar to \cref{lem:tail}.

\begin{corol}\label{corol:l2Concenteraion}
Let $u$ be $(C,a)$-semi-analytic. It is the case that
\begin{align}
\sum_{k:\norm{k}\geq t} \abs{\hat{u}[k]}^2 \leq \begin{cases}
    \max(C,\mathcal{U}) \, e^{-\frac{(t-a)^2}{8a^2}}, \quad &\text{if }t\leq 3a,\\
    e\, \max(C,\mathcal U) \, e^{-\frac{t}{2a}}, \quad &\text{if } t>3a.
\end{cases}
\end{align}
\end{corol}

\begin{proof}
This follows directly from the first implication in \cref{thm:SAisBern} and
\cref{lem:BernTail}.
\end{proof}

Note that the result of \cref{lem:tail} can also be proven using the Markov
inequality $\mathbb P\left[\norm{K_u}>t \right] = \mathbb P\left[\norm
{K_u}^m>t^m \right]\leq \frac{\EE \norm{K_u}^m}{t^m}$ by a suitable choice of
$m$. This correspondence further lets us find functions that saturate the
semi-analyticity condition, as in the following example.

\begin{figure}[t]
\centering
\begin{subfigure}[]{}
{\includegraphics[width=0.4\textwidth]{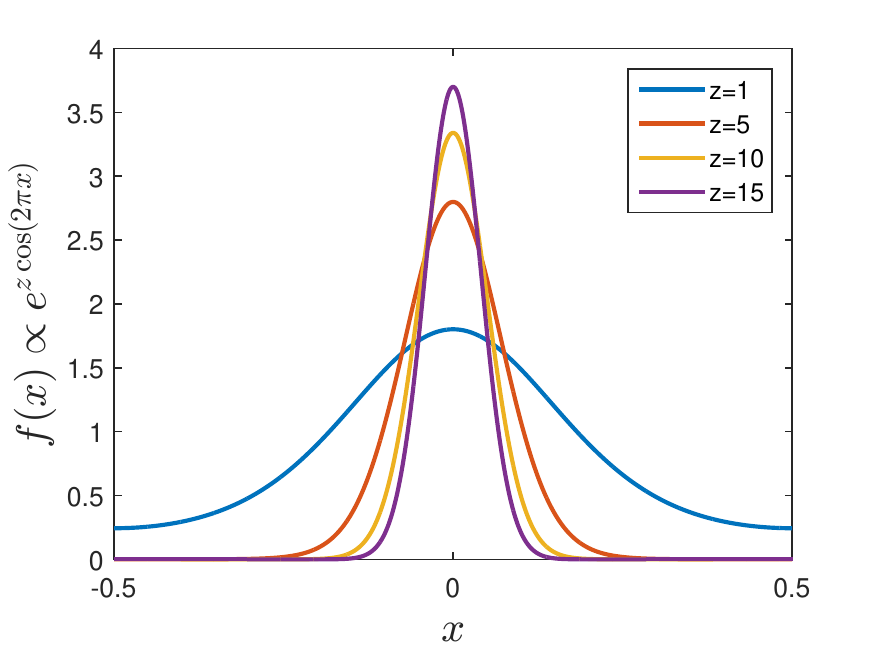}}
\end{subfigure}
\begin{subfigure}[]{}
{\includegraphics[width=0.4\textwidth]{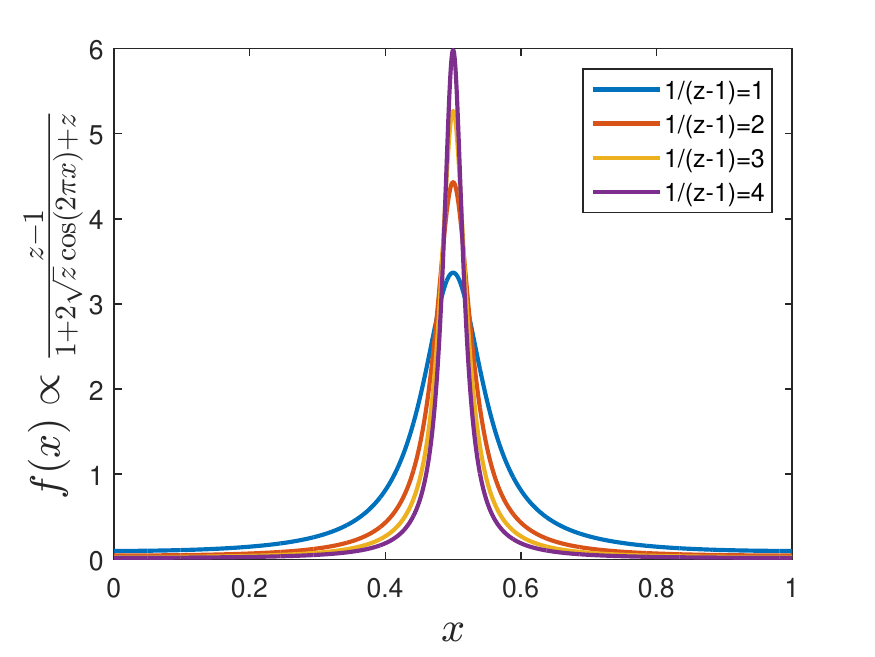}}
\end{subfigure}\\
\begin{subfigure}[]{}
{\includegraphics[width=0.4\textwidth]{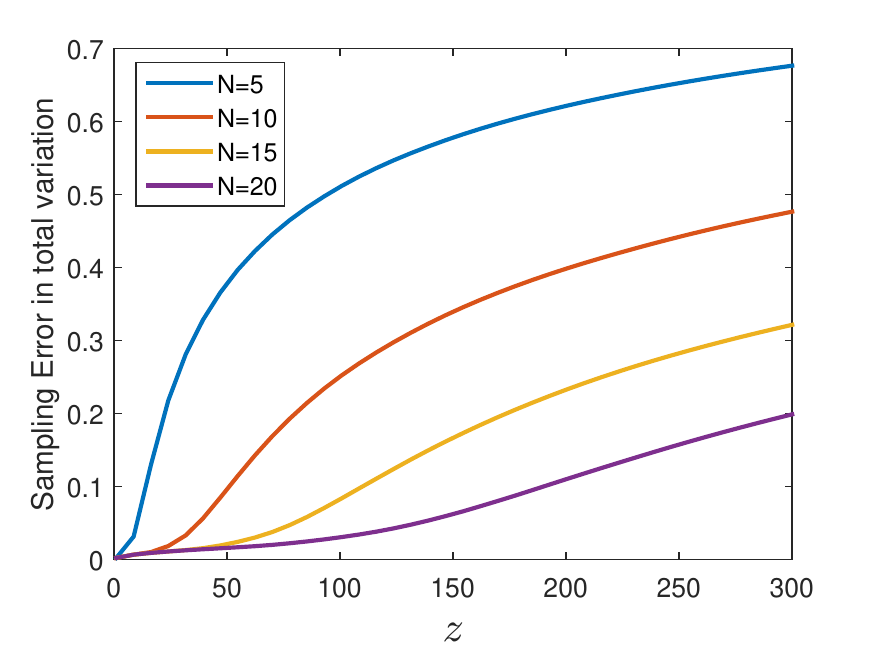}}
\end{subfigure}
\begin{subfigure}[]{}
{\includegraphics[width=0.4\textwidth]{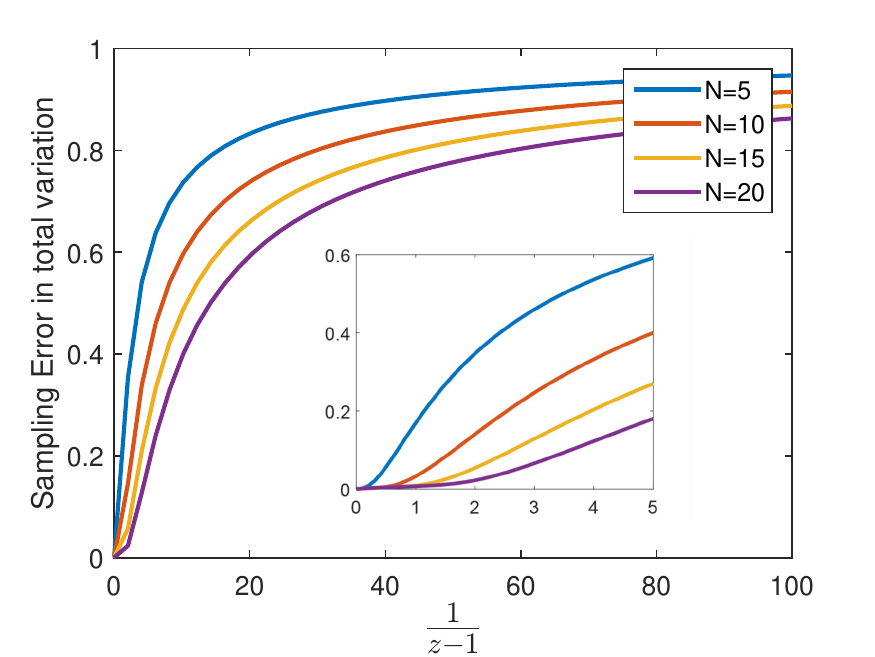}}
\end{subfigure}
\caption{
(a) and (b) show two families of functions considered respectively in
\cref{ex:cosine-function} and \cref{ex:inv-cos}. The functions are normalized so
that $\int_{x\in[0,1]} \mathrm dx\, \left(f(x)\right)^2 = 1$. That is, $f^2$
represents a distribution over one period. Note how in (a) the smoothness of the
functions is controlled by the parameter $z$ and in (b) it is controlled by
$(z-1)^{-1}$.
(c) and (d) show the Fourier interpolation accuracy on the two respective
families of functions considered in (a) and (b). We demonstrate the
interpolation error given the state $\ket{f_N}$ for different $N$. Note that
in both cases having $N$ larger than our upper bounds on $a$ results in a
sampling error less than $0.1$. The sampling error is shown with respect to the
smoothness parameters $z$ and $(z-1)^{-1}$, obtained by the application of the
upsampling algorithm using $M=200$.
Recall from \cref{ex:cosine-function} and \cref{ex:inv-cos} that we may think of
$\max(1,\frac{z}{2})$ and $\max(8,\frac{8}{z-1})$ as upper bounds on the
(average) inverse convergence radius of the respective functions in panels (a)
and (b).}
\label{fig:3}
\end{figure}

\begin{exm}
\label{ex:inv-cos}
Let $z>1$. The $2\pi$-periodic function $u(x) = \frac{z-1}{1-2\sqrt{z}\cos
(x)+z}$ satisfies the following inequality.
\begin{align}
\label{eq:lower-tail}
\frac{1}{(1+z^{-1})^{1/2}}\frac{m!}{(1-z^{-1})^{m}}
\leq \sqrt{\EE \norm{K_u}^{2m}} \leq
\sqrt{\frac{2e}{1+z^{-1}}} \, \max\left( 8,\frac{8}{z-1}\right)^{m} m!
\end{align}
Therefore $u$ is both upper bounded and lower bounded by growth rates in the
definition of semi-analyticity, although for different choices of parameters.
We refer the reader to \cref{fig:3} for visual demonstrations.

To obtain \eqref{eq:lower-tail}, note that the Fourier transform of $u(x)$ is
$\widehat u[k] = z^{-\frac{|k|}2}$ since
\begin{align}
1+2\sum_{k=1}^{\infty} \cos(kx) z^{-\frac k 2}
= \frac{z-1}{1+z-2\sqrt{z} \cos x}.
\end{align}
This implies $\mathcal U^2 = \frac{1+z^{-1}}{1-z^{-1}}$, and moreover the
moments of $\norm{K_u}$ can be lower bounded as follows
\begin{align}
\frac{1+z^{-1}}{1-z^{-1}}\EE \norm{K_u}^m
&= \sum_{k=0}^\infty k^m\, z^{-k}\\
&\geq \frac{\partial^m}{\partial (z^{-1})^m} \left( \sum_{k\geq0} z^{-k}\right)
= \frac{\partial^m}{\partial (z^{-1})^m} \left( \frac{1}{1-z^{-1}} \right)
= \frac{m!}{(1-z^{-1})^{m+1}}.
\end{align}
We, therefore, have
\begin{align}
\frac{1}{1+z^{-1}} \frac{m!}{(1-z^{-1})^m}
\leq \EE \norm{K_u}^m \leq \frac{1}{1+z^{-1}} \max\{2, \frac{2}{z-1} \}^m \, m!
\end{align}
where the upper bound follows from \cref{lem:example-3}.
\end{exm}

We now use another result from probability theory, namely the Paley--Zygmund
lower bound \cite{paley1932note, petrov2007lower}, to prove that taking $\Omega
(a)$ points is necessary to produce samples from a distribution arbitrarily
close to the one generated by the underlying distribution.

\begin{lem}[Paley--Zygmund]
Let $X$ be a non-negative random variable (that is, $X\geq 0$ almost surely).
For any $\theta \in(0,1)$, it is the case that
\begin{align}
\mathbb P\left[X>\theta \EE [X]\right]
> (1-\theta)^2\frac{\left(\EE [X]\right)^2}{\EE[X^2]}.
\end{align}
\end{lem}

\begin{proof}
Note that $X = X \mathbf{1}_{X\leq \theta \EE[X]} +
X \mathbf{1}_{X> \theta \EE[X]}$, from which it is straightforward to conclude
\begin{align}
\EE [X] \leq \theta \EE[X] +
\sqrt{\EE[X^2] \mathbb P\left[X > \theta \EE[X]\right]},
\end{align}
where the second term is due to the Cauchy--Schwartz inequality.
\end{proof}

\begin{thm}[\cref{thm:body-lower-bound} in the manuscript]
Let $u$ be a $(C,a)$-semi-analytic function. Consider any exact discretization
$\ket {u_N}$ on the discrete lattice with $N\leq \theta a/16$, where
$\theta\in(0,1)$. There is no algorithm that can return samples close to the
actual distribution (proportional to $u^2$) with a guaranteed error of less
than $(1-\theta)^2\,\frac{1}{1024e}$.
\end{thm}

\begin{proof}
Note that $f(x) = \frac{C}{\sqrt{e}} \frac{z-1}{1-2\sqrt{z}\cos(2\pi x/l) +z}$
is $(C,a)$-semi-analytic for $z=1+\frac{8}{a}$. From the example
above, we have $\EE \norm{K_f} \geq \frac{a}{16}$, therefore
\begin{align}\label{eq:PZ-thm}
\mathbb P \left[\norm{K_u} > \theta \frac{a}{16}\right]
> \mathbb{P} \left[\norm{K_f} > \theta \EE \norm{K_f}\right]
> (1-\theta)^2 \frac{(a/16)^2}{2e a^2} = (1-\theta)^2 \frac{1}{512 e}.
\end{align}
Hence, $\norm{K_f}$ is large with a considerable probability. Let $g:
[-l/2,l/2]\rightarrow \mathbb R$ be a function with the Fourier transform
\begin{align}
\widehat{g}[k] = \alpha \begin{cases}
\sum_{p\in\mathbb{Z}^d} \widehat{f}[k+p(2N+1)], &\text{if } k\in[-N..N]^d,\\
0, &\text{otherwise.}
\end{cases}
\end{align}
Here $\alpha$ is a normalization constant chosen such that
$\EE_{X\sim \text{Unif}}[g(X)^2]=C^2$. One can readily verify that $g\left(\frac
{nl}{2N+1}\right) = \alpha f\left(\frac{nl}{2N+1}\right)$. Therefore, $\ket
{f_N} = \ket{g_N}$. Moreover, note that $g$ is also $(C,a)$-semi-analytic due
to \cref{ex:Nyquist} and that the total variation distance between the
distributions whose densities are proportional to $\abs{f}^2$ and $\abs
{g}^2$ is at least $\mathbb P\left[ \norm{K_f}> N \right]$, which is itself
lower bounded by $(1-\theta)^2\frac{1}{512e}$ due to $N<\theta\frac{a}
{16}$ and \eqref{eq:PZ-thm}. Let $\ket{\psi_N}\in\mathcal V_N$ denote the
discretization of $f$ and $g$ (so $\ket{\psi_N} = \ket{f_N} = \ket{g_N}$). Given
the promises and the state $\ket{\psi_N}$, any algorithm will sample from a
distribution, say $\mathcal P$, which is at least $\frac{1-\theta^2}
{1024e}$ away from at least one of $P_f$ and $P_g$. Hence, the algorithm fails
as stated upon processing either $g$ or $f$ as the underlying functions.
\end{proof}

\subsection{The Fourier differentiation method}
\label{sec:app}

Here we describe the Fourier pseudo-spectral method used in our work and prove
several useful properties of it. Let $u:\mathbb{R}^d\rightarrow \mathbb{R}$ be
$l$-periodic in all dimensions. We define the Fourier derivatives on the
discretized lattice as follows:
\begin{align}
\tilde{\partial}_j u_{N}[n]
:= F^{-1}_{N} \left( \frac{i2\pi k_j}{l}
\left(F_{N} \vec{u_{N}}\right)[k] \right).
\end{align}
A straightforward calculation yields the following convolution relation
\begin{align}
\label{eq:fourier-1}
\tilde{\partial}_j u_{N}[n]
= \sum_{m\in[-N..N]} u_{N} [n_1,
\cdots, n_{j-1}, m, n_{j+1}, \cdots, n_{d}] \, a[n_j - m]
\end{align}
where
\begin{align}
\label{eq:fourier-a}
a[m] =\begin{cases}
0,& \text{if } m=0, \\
\frac{\pi\, (-1)^{m+1}}{l\,\sin\left( \frac{\pi\,m}{2N+1} \right)},&
\text{otherwise.}
\end{cases}
\end{align}
Higher order derivatives can then be defined as consecutive applications of the
first order operators:
\begin{equation}
\begin{split}
\tilde{\partial}^{r}_j u_{N}[n]
:=& F^{-1}_{N} \left[ \left(
\frac{i2\pi k_j}{l}\right)^r \left(F_{N} \vec{u_{N}}\right)[k] \right]\\
=& \sum_{m\in[-N..N]} u_{N} [n_1,
\cdots, n_{j-1}, m, n_{j+1}, \cdots, n_{d}] \, a^{(r)}[n_j - m],
\end{split}
\end{equation}
where $a^{(r)} = a \ast a \ast \cdots \ast a$ is the $r$-fold convolution. This
means that taking the $r$-th derivatives in the $j$-th dimension is identical to
$r$ consecutive applications of the first derivative in direction $j$. However,
if the number of discretization points is even the Fourier derivatives may be
define differently (as in \citet{shen2011spectral}) in which case this
composability property may not hold. Note that $\norm{\tilde
{\partial_j} u}_2 \leq 2\pi\,N/l\, \norm{u}_2$ since each derivative is an
operator with eigenvectors being the Fourier basis with eigenvalues
$\frac{i2\pi k_j}{l}$. In what follows, we discuss some properties of this
differentiation operation. Most notably, we show that it respects the Leibniz
product rule and that the maximum derivative is at most $\mathcal O (N\log N)$
bigger than the largest value the function attains. Note that analogously
when using $Df:= \frac{f(x+h)-f(x)}{h}$ for finite difference approximation
of conventional derivatives, using $h = \frac{1}{2N+1}$, the approximation is
at most $\mathcal O(N)$ larger than the maximum value of $f(x)$.

\begin{prop}
\label{prop:FourierProperties}
Let $u$ and $v$ be two $l$-periodic functions in all dimensions. The Fourier
derivatives $\tilde{\partial_j}$ have the following properties:
\begin{enumerate}[label=(\alph*), itemsep= -1pt,
ref=\cref{prop:FourierProperties}\alph*]
\item The product rule:
$\tilde{\partial_j}(u\cdot v)
= (\tilde{\partial_j} u) \cdot v + u \cdot (\tilde{\partial_j} v)$.
\label{prop:Fourier1}
\item $ \norm{\tilde{\partial_j} u}_{\infty}
\leq \frac{2\pi}{l} \, \norm{u}_{\infty}
(2N+1)\left[ \frac{1}{\pi}\ln\left( \frac{4N+2}{\pi} \right)
+ \frac{1}{2} \right]$. Also, if $N>3$, one obtains a simpler (but
worse) upper bound $\norm{ \tilde{\partial}_j u}_\infty
\leq \frac{48}{l} \norm{u}_\infty N\ln N$.
\label{prop:Fourier2}
\item $\sum_{n\in[-N..N]^d} \left(\tilde{\partial_j} u\right)_{[n]} = 0$.
\label{prop:Fourier3}
\item $\tilde{\partial_j}^2$ is a symmetric operator.
\label{prop:Fourier4}
\end{enumerate}
\end{prop}

\begin{proof} (a) It suffices to show that the Fourier transforms of the two
sides coincide.
\begin{equation}
\begin{split}
(2N+1)^{d/2}
&\biggl\{F_{N}\left(\tilde{\partial}_j (u\cdot v) \right)\biggr\}[k]
\overset{(1)}{=} \sum_{q\in[-N..N]^d}
\frac{i2\pi k_j}{L} \, \widehat{u}[q] \, \widehat{v}[k-q] \\
&= \sum_{q\in[-N..N]^d} \frac{i2\pi (q_j+k_j-q_j)}{L}\,
\widehat{u}[q] \, \widehat{v}[k-q]\\
&= \sum_{q\in[-N..N]^d} \left(\frac{i2\pi q_j}{L}
\widehat{u}[q] \right) \widehat{v}[k-q] +
\left(\frac{i2\pi (k_j - q_j)}{L}
\widehat{v}[k - q] \right)\widehat u[q]\\
&\overset{(2)}{=} (2N+1)^{d/2}
\left[F_{N}\left((\tilde{\partial}_j u) \cdot v\right)
+ F_{N}\left((\tilde{\partial}_j v) \cdot u\right)\right]
\end{split}
\end{equation}

Here (1) and (2) follow from the fact that the Fourier transform of the
pointwise multiplication of two functions is the convolution of their Fourier
transforms (up to the normalization factor $(2N+1)^{d/2}$).

(b) To show this, we make use of equation \eqref{eq:fourier-1}:
\begin{equation}
\begin{split}
\left| (\tilde{\partial}_j u)[n] \right|
&= \left| \sum_{m\in[-N..N]}
u[n_1, n_2, \cdots, m, \cdots, n_d] \, a[n_j-m] \right|\\
&\overset{(1)}{\leq} \frac{2\pi}{l} \, \norm{u}_{\infty} \,
\sum_{m=1}^{N} \frac{1}{\sin\left(  \frac{\pi\,m}{2N+1} \right)}\\
&\leq \frac{2\pi}{l} \, \norm{u}_{\infty} \,
\left( \int_{x=1}^{N} \frac{dx}{\sin\left( \frac{\pi \, x}{2N+1}\right)}
+ \frac{1}{\sin \left(\pi/(2N+1)\right)} \right)\\
&\overset{(2)}{\leq} \frac{2\pi}{l} \, \norm{u}_{\infty} \,
\left( -\frac{(2N+1)}{\pi} \ln\left( \tan\left( \frac{\pi}{4N+2} \right) \right)
+ \frac{2N+1}{2} \right)\\
&\overset{(3)}{\leq} \frac{2\pi}{l} \, \norm{u}_{\infty}
(2N+1)\left[ \frac{1}{\pi}\ln\left( \frac{4N+2}{\pi} \right)
+ \frac{1}{2} \right]
\end{split}
\end{equation}
where (1) follows from H\"older's inequality, and (2) follows from noting that
$\sin(\pi/(2N+1)) \geq \frac{2}{2N+1}$. Finally, (3) follows from
the fact that $\tan(x) \geq x$ for $0 \leq x < \pi/2$. The claim follows since
these inequalities hold for any $n\in[-N..N]^d$. For $N > 3$, in order to
simplify the right hand side of (3) we use the fact that $1+x\leq 2x$ if
$x\geq 1$. This implies that
\begin{align}
\left| (\tilde{\partial}_j u)[n] \right|
\leq \frac{48}{l} \norm{u}_\infty N\ln N.
\end{align}

(c) Note that for any vector $v$ defined on the discrete lattice, one has
\begin{align}
\sum_{n\in[-N..N]^d} v[n] = \left(\sqrt{2N+1}\right)^d \, \hat{v}[0]
\end{align}
where $\hat{v}$ represents the Fourier transform of $v$. Noting that
$\left(F_{N} \tilde{\partial}_j v\right)_{[0]} =
\left( \frac{2\pi k_j}{l} \, \hat{u}[k] \right)_{[0]} = 0$ completes the proof.

(d) \eqref{eq:fourier-1} and \eqref{eq:fourier-a} show that
$\tilde{\partial}_j$ is anti-symmetric. And since composition of
an anti-symmetric operator with itself is symmetric the result follows.
\end{proof}

\begin{figure}[t]
\centering
\includegraphics[width = 0.5\textwidth]{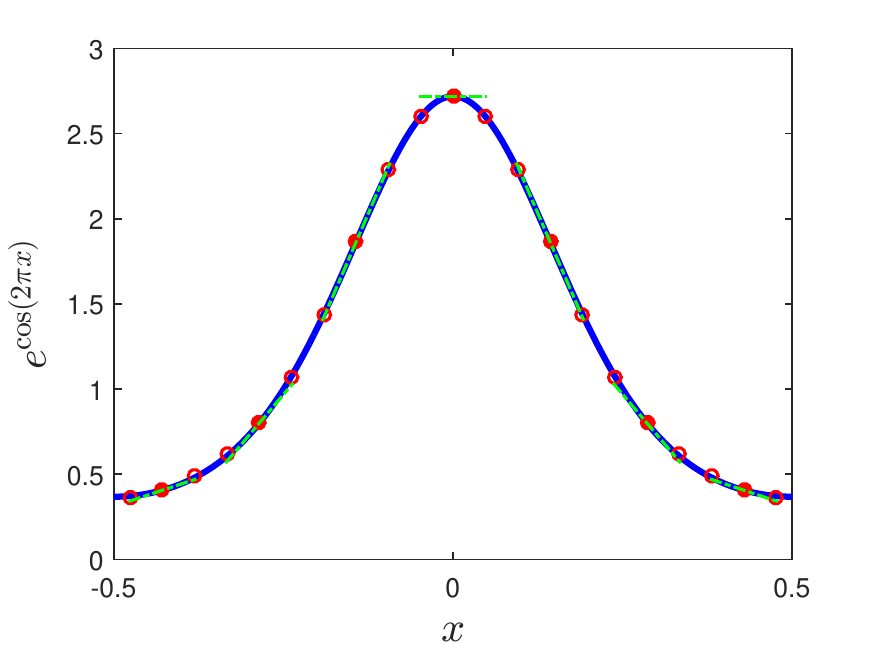}
\caption{Applying the Fourier interpolation of \cref{thm:interpolation} on the
input function $u(x) = e^{\cos 2\pi x}$ of \cref{ex:cosine-function}. The plot
shows the interpolation results with $N=3$ and $M=10$. Filled circles correspond
to the initial samples, and the hollow circles represent the interpolation
output. The solid blue line represents the graph of the underlying function $u$.
And the dashed green lines show the Fourier derivative estimations.}
\label{fig:Fourier}
\end{figure}

So far we talked about the interpolation results for semi-analytic functions. We
may now show that the Fourier differentiation technique is able to estimate the
first and second order differentiation with high accuracy. Note that this is
non-trivial as the Fourier differentiation operator is not bounded. The
proof of this result borrows ideas from \cite{shen2011spectral}.
\cref{fig:Fourier} depicts an example of the Fourier interpolation and this
derivative estimation method.

\begin{prop}
\label{prop:high-precision-derivatives}
Let $u$ be $(C,a)$-semi-analytic and periodic, and let $N\geq 4 ad$. It is
the case that
\begin{align}
\sqrt{\sum_{j=1}^d
\norm{\overrightarrow{\partial_j u}_{N}
- \overrightarrow{\tilde{\partial_j} u }_{N}}^2}
&\leq \frac{40\sqrt{2}\pi e^3\, a}{l}\,
C\left(2N+1\right)^{d/2}\, e^{- \frac{N}{2a}}, \text{ and }
\label{eq:first}\\
\norm{\overrightarrow{\nabla^2 u}_{N}
- \overrightarrow{\tilde{\nabla^2} u}_{N}}
&\leq \frac{200\sqrt{2}\pi^2 e^3\, a^2}{l}\, C^2
\left(2N+1\right)^{d/2}\, e^{-0.4 \frac{N}{a}}.
\label{eq:second}
\end{align}
\end{prop}

\begin{proof}
We first prove \eqref{eq:first}. We have
\begin{align}
\partial_j u[n]
&= \sum_{k\in \mathbb{Z}^d} \frac{i 2\pi k_j}{l}
\widehat{u}[k]\, e^{i\frac{2\pi \langle k,n\rangle}{2N+1}} \\
&= \sum_{k\in[-N..N]^d} e^{i\frac{2\pi \langle k,n\rangle}{2N+1}}
\sum_{p\in\mathbb{Z}^d} \frac{i2\pi (k_j+p_j(2N+1))}{l}
\widehat{u}[k + (2N+1)p].
\end{align}
And similarly,
\begin{align}
\tilde{\partial_j} u[n]
&= \frac{1}{(2N+1)^{d/2}}\sum_{k\in [-N..N]^d} \frac{i 2\pi k_j}{l}
\tilde{u}[k]\, e^{i\frac{2\pi \langle k,n\rangle}{2N+1}} \\
&= \sum_{k\in[-N..N]^d} e^{i\frac{2\pi \langle k,n\rangle}{2N+1}} \,
\sum_{p\in\mathbb{Z}^d} \frac{i2\pi k_j}{l} \widehat{u}[k+(2N+1)p]
\label{eq:wrap-ap}
\end{align}
where \eqref{eq:wrap-ap} follows from \eqref{eq:wrap}. Hence
\begin{align}
\partial_j u[n] - \tilde{\partial_j} u[n]
= \sum_{k\in[-N..N]^d} e^{i\frac{2\pi \langle k,n\rangle}{2N+1}} \,
\sum_{p\in\mathbb{Z}^d\setminus \{0\}} \frac{i2\pi p_j(2N+1)}{l}
\widehat{u}[k+(2N+1)p]
\end{align}
which using Parseval's theorem gives
\begin{equation}
\begin{split}
\sum_{j=1}^d \norm{ \overrightarrow{\partial_j u}_{N}
- \overrightarrow{\tilde{\partial}_j u }_{N}}^2
&= (2N+1)^{d} \sum_{j=1}^d \sum_{k\in[-N..N]^d}
\left| \sum_{p\in\mathbb{Z}^d\setminus \{0\}}
\frac{i 2\pi (2N+1) p_j}{l} \widehat{u}[k+(2N+1)p] \right|^2 \\
& \leq \frac{4\pi^2}{l^2} (2N+1)^d \sum_{j=1}^d
\sum_{k\in[-N..N]^d}
\Biggl\{\left( \sum_{p\in\mathbb{Z}^d \setminus\{0\}}
\norm{k+(2N+1)p}^{-2m} \right) \\
& \,\,\,\,
\times \left(\sum_{p\in\mathbb{Z}^d\setminus\{0\}}
\norm{k+(2N+1)p}^{2m} |(2N+1)p_j|^2 \, \left|
\widehat{u}[k+(2N+1)p] \right|^2\right) \Biggr\}.
\end{split}
\end{equation}
Then, using inequality \eqref{eq:sefuleq} together with \cref{lem:useful}, along
with the fact that for each $j\in[d]$ we have
$|(2N+1)p_j| \leq 2|(2N+1)p_j+k_j|$, we get
\begin{align}
\sum_{j=1}^d \norm{ \overrightarrow{\partial_j u}_{N}
- \overrightarrow{\tilde{\partial}_j u }_{N}}^2
&\leq \frac{32\pi^2}{l^2} C^2\left(2N+1\right)^d 2^d\, N^{-2m} \,
[(m+1)!]^2 a^{2m+ 2}\\
&\leq \frac{32\pi^2}{l^2}\, C^2
\left(2N+1\right)^d 2^{N/2a}\, N^{-2m} \, [(m+1)!]^2 \, a^{2m+ 2}
\end{align}
for $N\geq 2ad$. Hence, by choosing $m = \lfloor N/a \rfloor -2$ one achieves
the upper bound
\begin{align}
\sqrt{\sum_{j=1}^d \norm{ \overrightarrow{\partial_j u}_{N}
- \overrightarrow{\tilde{\partial}_j u }_{N}}^2}
\leq \frac{8\sqrt 2 \pi e^3}{l} C \left(2N+1\right)^{d/2}
N\, e^{-0.6 \frac{N}{a}}.
\end{align}
Again, using the inequality $x \leq \alpha \, e^{\frac{x}{e\alpha}}$ for all
$x$ and all positive $\alpha$, and setting $\alpha = 5$ completes the proof.
Now we prove \eqref{eq:second}. As in above, we start by writing the Fourier
transform of the Laplacians:
\begin{align}
\tilde{\nabla^2} u_{N}[n]
&= \sum_{k\in[-N..N]^d} \frac{-4\pi^2}{l^2}\,
\norm{k}^2 \sum_{p\in\mathbb{Z}^d} \widehat{u}[k+(2N+1)p] \\
\nabla^2 u_{N}[n]
&= \sum_{k\in[-N..N]^d} \frac{-4\pi^2}{l^2}
\sum_{p \in \mathbb{Z}^d} \norm{k+(2N+1)p}^2 \, \widehat{u}[k].
\end{align}
Therefore,
\begin{equation}
\begin{split}
\norm{ \overrightarrow{\tilde{\nabla^2} u}_{N}
- \overrightarrow{\nabla^2 u}_{N}}^2
&= \frac{16\pi^4}{l^4} (2N+1)^d\sum_{k\in[-N..N]^d}
\left| \sum_{p\in\mathbb{Z}^d\setminus\{0\}}
\left(\norm{k+(2N+1)p}^2 - \norm{k}^2 \right)\,
\widehat{u}_{N}[k] \right|^2 \\
&\overset{(a)}{\leq}
\frac{16\pi^4}{l^4}(2N+1)^d \sum_{k\in[-N..N]^d}
\Bigg\{ \left( \sum_{p\in\mathbb{Z}^d\setminus\{0\}}
\norm{k+p(2N+1)}^{-2m}\right) \\
&\quad \quad \quad \quad \times
\left(\sum_{p\in\mathbb{Z}^d\setminus\{0\}}
\norm{k+(2N+1)p}^4 \norm{k+(2N+1)p}^{2m}
\left| \widehat{u}_{N}[k] \right|^2 \right)\Bigg\}
\end{split}
\end{equation}
where (a) uses the Cauchy-Schwartz inequality along with the fact that
$\norm{k} \leq \norm{k+p(2N+1)}$ for all $k\in[-N..N]^d$. Again, we use
\cref{lem:useful} and set $m= \lfloor N/a \rfloor - 3$ (which is guaranteed to
be a natural number since $N\geq 4ad$) to conclude the proof.
\end{proof}

\subsection{Construction of semi-analytic functions}
\label{sec:constructions}

One question that arises in our study of semi-analyticity is the behavior of the
semi-analyticity parameters $C$ and $a$ under composition rules. For example,
we may be interested in the semi-analyticity parameters of the function
approximators represented by deep neural networks. Recall that the definition
of semi-analyticity involves taking high order derivatives. Therefore, we make
multiple uses of the Fa\`a~di~Bruno formula \cite
{roman1980formula, krantz2002primer}, according to which we have
\begin{align}
\label{eq:FaadiB}
\frac{d^m}{dx^m} &f\left(g(x)\right) \nonumber\\
&= \sum_{\stackrel{i_1,i_2,\cdots,i_m\in\{0,\cdots,m\}}{i_1+2i_2+\cdots+mi_m=m}}
\frac{m!}{i_1!\cdots i_m!} \, f^{(i_1+\cdots+i_m)}(g(x))
\left(\frac{dg/dx}{1!}\right)^{i_1} \cdots
\left(\frac{d^m g/dx^m}{m!}\right)^{i_m}
\end{align}
for any pair of smooth functions $f,g:\mathbb R \to \mathbb R$. Working in
$d$ dimensions, we need to apply the multivariate Fa\`a~di~Bruno formula, which
is provided below.

\begin{prop}\label{prop:Multi-FB}
Let $g:\mathbb R^d\rightarrow \mathbb R$ and $f:\mathbb R\rightarrow \mathbb R$
be smooth functions and let $\alpha \in\mathbb Z_{\geq 0}^d$. We have
\begin{align}\label{eq:Multi-FB}
D^{\alpha}(f \circ g)(x)
= \alpha! \sum_{\lambda=1}^{|\alpha|} f^{(\lambda)}(g(x)) \,
\sum_{s=1}^{|\alpha|} \sum_{p_s(\lambda,\alpha)}
\prod_{j=1}^s \frac{1}{k_j!}\left(\frac{D^{l_j} g}{l_j!}\right)^{k_j},
\end{align}
where
\begin{align*}
p_s(\lambda,\alpha) :=
\left\{ (k_1,\cdots, k_s,l_1,\cdots,l_s):
k_i>0,\, 0 \prec l_1 \prec \cdots \prec l_s,\,
\sum_{j=1}^s k_j l_j = \alpha, \, \sum_{j=1}^s k_j = \lambda \right\}.
\end{align*}
Here, for $\mu, \nu \in \mathbb Z_{\geq 0}^d$, we say $\mu \prec \nu$ if the
1-norms compare as (i) $|\mu| < |\nu|$, or (ii) if $|\mu| = |\nu|$ then use
lexicographic ordering.
\end{prop}

\begin{proof}
This proposition is obtained by setting $m=1$ in Theorem 2.1 of
\citet{constantine1996multivariate}.
\end{proof}

Reference \citep[Lemma 1.4.1]{krantz2002primer} proves that the coefficients in
\eqref{eq:FaadiB} follow
\begin{align}
\sum_{\stackrel{i_1,i_2,\cdots,i_m\in\{0,\cdots,m\}}{i_1+2i_2+\cdots+mi_m=m}}
\frac{(i_1+\cdots+i_m)!}{i_1!\cdots i_m!}
R^{i_1+\cdots+i_m}= \frac{R}{R+1} \, (R+1)^m,
\end{align}
for any $R>0$. We use similar ideas to extend this result to the multivariate
case.

\begin{lem}\label{lem:comp-FB}
Let $\alpha$ be a $d$-dimensional vector of non-negative integers. We have
\begin{align}
\sum_{\lambda=1}^{|\alpha|} \lambda! \, R^\lambda \, \sum_{s=1}^{|\alpha|}
\sum_{p_s(\lambda,\alpha)} \prod_{j=1}^s \frac{1}{k_j!}
= \frac{R}{R+1} \left(R+1\right)^{|\alpha|}.
\end{align}
\end{lem}

\begin{proof}
The proof is a generalization of \citep[Lemma 1.4.1]{krantz2002primer}. Let
$g(x) = \frac{1}{1-\sum_{i=1}^d x_i}$ and $f(x) = \frac
{1}{1-R(x-1)}$. In what follows, we consider $f(g(x))$, and its Taylor
expansion, and subsequently, will apply \eqref{eq:Multi-FB} to get the desired
relation.

To begin with, we observe the following. Let $\alpha =
(\alpha_1, \cdots, \alpha_d)\in \mathbb Z_{\geq 0}^d$. One can readily verify
that
\begin{align}\label{eq:Dsum}
D^\alpha (x_1+\cdots+x_d)^n\big|_{x=0}
= \alpha! \, \mathbf{1}_{\{|\alpha| = n\}}
\end{align}
where $\mathbf 1$ is the identifier function (i.e., it is $1$ if the condition
inside the brackets is satisfied, and $0$ otherwise). Therefore, as $g
(x)= \sum_{n=0}^\infty \left( \sum_{i=1}^d x_i \right)^n$ in a neighbourhood
of $x=0$, we conclude
\begin{align}
\left(D^{\alpha} g\right)(0) = \alpha!.
\end{align}

Additionally, it is transparent that $(f \circ g)(x) = \frac{1-\sum_{i=1}^d x_i}
{1-(R+1)\sum_{i=1}^d x_i}$, which provides the following expansion on a
neighbourhood of $x=0$.
\begin{align}\label{eq:fog}
(f\circ g)(x) = 1 + \frac{R}{R+1} \sum_{n=1}^{\infty} (R+1)^{n}
\left( \sum_{i=1}^d x_i \right)^n
\end{align}
Combining \eqref{eq:Dsum} and \eqref{eq:fog} provides $D^\alpha(f\circ g)
(0) = \frac{R}{R+1} (R+1)^{|\alpha|} \alpha!$. Furthermore, it is
straightforward to find $f^{(\lambda)}(g(0)) = \lambda!\, R^{\lambda}$, from
which the lemma follows by substitutions into \eqref{eq:Multi-FB}.
\end{proof}

We are now ready to study the composition of analytic functions of many
variables. To make our claims easier to state and comprehend, let us first
introduce some notation.

\begin{defn}
\label{dfn:analytic-semi-analytic-class}
Let $f$ be analytic on an open set $U$. We say $f\in A_U(C,a)$, if
\begin{align}
\sup_{x\in U} D^{\alpha} f(x) \leq C\, a^{|\alpha|}\, \alpha!,
\end{align}
and we say that $g\in S_U(C,a)$ if
\begin{align}
\sqrt{\underset{X\sim \mathrm{Unif}(U)}{\mathbb{E}}\left(
\sum_{\alpha:|\alpha|=m}
\left|D^\alpha g(X) \right|^2\right)} \leq C\, a^m\, m!.
\end{align}
\end{defn}

Note that if $f$ is periodic and analytic in its fundamental domain, then it is
analytic on the entire domain $\mathbb R^d$. We use the notation $B_\infty
(M)$ for the open $l_\infty$-ball $(-M, M)^d \subset \mathbb R^d$, and the
notation $A_M$ as a shorthand for $A_{B_\infty (M)}$.

\begin{rem}
Using the same argument as in \cref{prop:analytic-implies-semianalytic}, for any
open set $U$, we have $A_U(C,a) \subseteq S_U(3^{d-1}C,a)$.
\end{rem}

\begin{prop}
\label{prop:semi-analytic-basic-prop}
Let $f_1\in A_U(C_1,a_1)$ and $f_2\in A_V(C_2,a_2)$ for two open domains
$U$ and $V$. The following statements hold.
\begin{enumerate}[label=(\alph*), itemsep= -1pt,
ref=\cref{prop:semi-analytic-basic-prop}\alph*]
\item $f_1 + f_2 \in A_{U \cap V}(C_1+C_2, \max\{a_1,a_2 \})$ if
$U, V \subseteq \mathbb R^d$ are in the same Euclidean domains. Also, the same
property holds for the semi-analytic families.
\item $f_1\cdot f_2\in A_{U \cap V}(C_1 C_2, a_1+a_2)$.
\item Let $V$ contain the image of $f_1$, that is $f_2 \in A_{f_1(U)}(C_2,a_2)$.
Then $f_2 \circ f_1\in
A_U\left(\frac{C_1 a_2 C_2}{1+C_1  a_2 }, a_1 (1+C_1 a_2)\right)$.
\end{enumerate}
\end{prop}

\begin{proof}
(a) Note that for any $\alpha\in\mathbb Z_+^d$ the quantity $\sup_
{x\in U \cap V} \abs{ D^\alpha f(x)}$ defines a semi-norm. Using the triangle
inequality of this semi-norm, we can obtain the result. For the
semi-analyticity part, let $|u|_{m} := \sqrt{ \EE \left(\sum_
{\alpha:|\alpha|=m} D^{\alpha} u(x) \right)^2}$ and note that
$|\cdot|_m$ is also a semi-norm, and in particular,
$|f_1 + f_2|_m \leq |f_1|_m + |f_2|_m$.

(b) We note that
\begin{align}
D^\alpha (f_1\cdot f_2)(x)
= \sum_{\beta\in\prod_{i=1}^d\{0,\cdots,\alpha_i\}}
{\alpha\choose\beta} D^{\alpha-\beta} f_1(x) \cdot D^{\beta} f_2(x),
\end{align}
where we have used the convention ${\alpha\choose\beta} := \prod_{i=1}^d
{\alpha_i\choose\beta_i}$. Using the upper bounds on the derivatives of $f_1$
and $f_2$, we get
\begin{align}
\begin{split}
\sup_{x\in U \cap V} \big|D^\alpha (f_1\cdot f_2)(x)\big|
&\leq C_1C_2 a_1 \prod_{i=1}^d
\sum_{\beta_i=0}^{\alpha_i} a_1^{\alpha_i-\beta_i} a_2^{\beta_i}\\
&\leq C_1C_2 \prod_{i=1}^d  (a_1+a_2)^{\alpha_i}
= C_1C_2 (a_1+a_2)^{|\alpha|}.
\end{split}
\end{align}

(c) Using \cref{prop:Multi-FB}, we have
\begin{align}
\sup_{x\in U} D^{\alpha}(f_2\circ f_1)
&= \sup_{x\in U} \alpha! \sum_{\lambda=1}^{|\alpha|} f_2^{(\lambda)}
\sum_{s=1}^{\alpha} \sum_{p_s(\lambda,\alpha)}
\prod_{j=1}^s \frac1{k_j!} \left(\frac{D^{l_j} f_1}{lj!} \right)^{k_j}\\
&\leq \alpha! C_2 a_1^{|\alpha|}
\sum_{\lambda=1}^{\alpha} \lambda! a_2^{\lambda } C_1^\lambda
\sum_{s=1}^{|\alpha|} \sum_{p_s(\lambda,\alpha)} \prod_{j=1}^s \frac1{k_j!}\\
&\overset{(1)}{=} \alpha! \frac{C_2C_1a_2}{1+a_2C_1}
\left[ a_1(1+a_2C_1) \right]^{|\alpha|},
\end{align}
where (1) follows from \cref{lem:comp-FB}.
\end{proof}

\begin{corol}\label{corol:exp-comp}
Let $f\in A_U(C,a)$. Then, $e^f \in A_U(\frac{C}{1+C} e^{\Delta}, (1+C) a)$,
where $\Delta = \sup_{x\in U} f(x)$ (compare this with \cref
{ex:cosine-function}).
\end{corol}

\begin{proof}
This follows from \cref{prop:semi-analytic-basic-prop}(c) and the fact that
$g(x) = e^x$ is in $A_\Delta(e^\Delta,1)$.
\end{proof}

\begin{corol}\label{corol:sigmoid}
Let $f\in A_M(C,a)$. Then,
$\sigma(f(x)):=\frac{1}{1+e^{-f(x)}} \in
A_{M}\left(1,a (1+C (1+e^\Delta))\right)$,
where $\Delta = \sup_{x\in(-M, M)^d} \abs{f(x)}$.
\end{corol}

\begin{proof}
This follows \cref{prop:semi-analytic-basic-prop}(c) and noting that the
function $g(x)=\frac{1}{1+x}$ is in $A_{\mathbb R_+}(1,1)$ because
$g^{(m)}(x) = (-1)^m \frac{m!}{(1+x)^m}$.
\end{proof}

In the final corollary of this section we find the analyticity parameters of
deep neural networks which are the de facto function approximators in deep
learning.

\begin{corol}
\label{cor:neural-nets}
Let $f: \mathbb R^d \to \mathbb R$ be the function represented by a deep neural
network consisting of $D$ fully-connected layers with sigmoid activation
functions. We denote the $i$-th layer weights matrix with $W^{(i)}$ and the bias
vector with $b^{(i)}$. Then $f \in A_1(\tilde C, \tilde a)$ is with
$\tilde C = 1$ and
\begin{align}
\tilde a \leq 2^D \exp\left(\sum_{k=1}^D 2
\norm{W^{(k)}}_\infty+\norm{b^{(k)}}_\infty\right)
\end{align}
\end{corol}

Here the norm $\norm{\cdot}_\infty$ is the maximum absolute row sum i.e.,
$\norm{X}_\infty=\max_{i} \sum_{j} \abs {X_{ij}}$ for any matrix $X$. Also,
note that $A_1$ in the statement above could be generalized to $A_M$ for
arbitrary $M>0$, by rescaling the weights and biases of the first layer.

\begin{proof}
Let us denote the input and output of the $i$-th neuron of the $k$-th layer
be denoted by $f^{(k)}_i$ and $g^{(k)}_i= \sigma (f^{(k)})_i$, respectively.
We prove the result by induction on $D$. As for the base case,
note that the input functions to the neurons of the first layer are $f^{
(1)}_i:=\langle w^{(1)}_i, x\rangle + b_i^{(1)}$, where $w^{(1)}_i:=(W^{(1)}_
{ij})_j$ is the $j$-th row of the weight matrix of the first layer. Hence,
by \cref{prop:semi-analytic-basic-prop}(a) we get
$f^{(1)}_i\in A(\norm{w_i^{(1)}}_1,1)$. Also, note that
$\sup_{x\in[-1,1]^d} \abs{f^{(1)}_i(x)} \leq
\norm{w_i^{(1)}}_{1} + \abs {b_i^{(1)}}$. Applying \cref{corol:sigmoid} yields
$g^{(1)}_i\in A_1(1, a_i)$ with $a_i=2(1+\norm{w_i}_{1})e^{\norm{w_i^{(1)}}_{1}
+ \abs {b_i^{(1)}}} \leq 2 e^{2\norm{w_i^{(1)}}_{1} + \abs{b_i^{(1)}}}$.
Taking a maximum over $i$ proves the base case.

Now, assuming the bounds are valid for the neural network consisting of only the
first $k$ layers, we prove the bound for the first $k+1$ layers. By assumption
$g^{(k)}_j\in A_1(1,\tilde a_k)$, where
\begin{align*}
\tilde a_k \leq
2^k \exp \left(\sum_{l=1}^k 2\norm{W^{(l)}}_\infty
+ \norm{b^{(l)}}_\infty\right),
\end{align*}
and that $f^{(k+1)}_i = \langle w^{(k+1)}_i, g\rangle + b_i^{(k+1)}$. This,
together with \cref{prop:semi-analytic-basic-prop}(a) implies
$f_i^{(k+1)} \in A_1\left(\norm{w_i^{(k+1)}}_\infty, \tilde a_k\right)$. As
$g^{(k+1)}_i = \frac{1}{1+e^{-f_i^{(k+1)}}}$, we may use \cref{corol:sigmoid}
once more to complete our induction.
\end{proof}

\subsection{Extension to infinite lattices}
\label{sec:infinite-extension}

In this subsection, we extend the definition of semi-analyticity and the
interpolation results to non-periodic functions. Consider a function $u:\mathbb
{R}^d \to \mathbb{R}$, such that $u$ and all its derivatives are in $L^2
(\mathbb{R}^d)$. We are given the following quantum state, which encodes the
values of $u$ at certain points
\begin{align*}
\ket{u_{H}} \propto \sum_{j\in\mathbb Z^d}
u(jH) \ket{j_1}\otimes \ket{j_2}\otimes \cdots \otimes \ket{j_d},
\end{align*}
where $H>0$ is the discretization parameter, and $j=(j_1,j_2,\cdots,j_d)$ is a
vector of integers. Note that the Fourier transform in this case is a function
$\widehat u\in L^2(\mathbb R^d)$ that satisfies the following equations.
\begin{align}
\label{eq:inverse-cont-Fourier}
u(x) &= \frac{1}{(2\pi)^{d/2}} \int_{\omega\in\mathbb R^d}
e^{i\langle \omega, x\rangle} \widehat u(\omega) \, \mathrm d\omega\\
\label{eq:cont-Fourier}
\widehat u(\omega) &= \frac{1}{(2\pi)^{d/2}} \int_{x\in\mathbb R^d}
e^{-i\langle \omega, x\rangle} u(x) \, \mathrm dx
\end{align}
Moreover, associated to $\vec u\in l^2(\mathbb Z^d)$ is a Fourier transform
$\tilde u:\mathbb R^d \to \mathbb R$ that satisfies
\begin{align}\label{eq:approximate-Fourier}
\tilde u(\omega)
:= \left(\frac{H}{\sqrt{2\pi}}\right)^{d}
\sum_{j\in\mathbb Z^d} e^{-i\langle \omega, jH\rangle} u(jH).
\end{align}
The coefficients of \eqref{eq:inverse-cont-Fourier}, \eqref{eq:cont-Fourier},
and \eqref{eq:approximate-Fourier} are chosen so that
\begin{align}
\int_{x\in\mathbb R^d} \abs{u(x)}^2 \, \mathrm dx
&= \int_{\omega\in\mathbb R^d} \abs{\widehat u(\omega)}^2 \, \mathrm d\omega,\\
H^d \, \sum_{j\in\mathbb Z^d} \abs{u(jH)}^2
&= \int_{\omega\in[-\frac{\pi}{H}, \frac{\pi}{H}]^d}
\abs{\tilde u(\omega)}^2 \, \mathrm d\omega.
\end{align}

Note the similarities between \eqref{eq:cont-Fourier} and \eqref
{eq:approximate-Fourier}, and that $\tilde u \rightarrow \widehat u$ as
$H\rightarrow 0$ pointwise. Indeed, it is transparent that $\tilde u
(\omega)$ is periodic with period $\frac{2\pi}{H}$ along each axis, and that
\begin{align}\label{eq:cont-fourier-wrapping}
\tilde u(\omega) = \sum_{k\in \mathbb Z^d}
\widehat u\left(\omega + \frac{2\pi k}{H}\right).
\end{align}
Note that $\tilde u$ depends only on the values of $u$ at the lattice points
$H \mathbb Z^d$. Moreover, if $\hat u$ has a bounded support circumscribed
within a fundamental domain of $\tilde u$ then \eqref
{eq:cont-fourier-wrapping} implies that $\tilde u(\omega) = \widehat u
(\omega)$ for all $\omega\in[-\frac{\pi}{H},\frac{\pi}{H}]^d$, and therefore
one can exactly recover the function $u$ (i.e., $u(x)$ can be found within
arbitrarily small error at any $x\in\mathbb R^d$). This is indeed a restatement
of the Nyquist theorem.

In what follows, we focus on the case where the support of $\widehat u$ is
possibly the entire domain $\mathbb R^d$, but an interpolation with
exponentially small error is still feasible. The arguments closely follow those
of the previous subsections regarding periodic functions, and hence, we keep
our proofs brief. From hereon, we use $\norm{\cdot}$ to refer to the $2$-norm
for functions in $L^2(\mathbb R^d)$.

\begin{defn}\label{defn:cont-semi-analyticity}
A function $u\in L^2(\mathbb R^d)$ is said to be semi-analytic if
\begin{align}
\norm{ \sum_{\alpha:|\alpha|=m} D^\alpha u} \leq C \, a^m \, m!
\end{align}
for some $C,a\geq 0$. As before, we refer to $C$ and $a$ as the
semi-analyticity parameters.
\end{defn}
We note that \cref{defn:cont-semi-analyticity} is equivalent to
\begin{align}
\sqrt{\int_{\omega\in\mathbb{R}^d}\norm{\omega}^{2m}
\abs{\widehat u(\omega)}^2 \, \mathrm d\omega} \leq C\, a^m \, m!.
\end{align}
While establishing a connection between this notion of semi-analyticity and
analyticity is challenging, we provide several examples of such functions.

\begin{exm}
Any function which has a Fourier transform with bounded support is $(\norm
{u}, \omega_0)$-semi-analytic, where $\omega_0=\sup\{ \norm{\omega}_2: \widehat
{u}(\omega)\ne0\}$.
\end{exm}

\begin{exm}
The Gaussian function $u(x) := \sqrt{\frac{2a}{\pi}} e^{-\frac{a}2 x^2}$
with inverse variance $a$ is $(1,a)$-semi-analytic. This is due to the fact that
$|\widehat u (\omega)|^2$ corresponds to the probability distribution function
of $\mathcal N(0,a)$, and hence
\begin{align}
\sqrt{\int_{\omega\in\mathbb{R}}\abs{\omega}^{2m}
\abs{\widehat u(\omega)}^2 \, \mathrm d\omega}
= \sqrt{ \underset{X\sim\mathcal N(0,a)}{\EE}[X^{2m}] }
= \left(\frac{a}{2}\right)^{m} \sqrt{(2m-1)!!} \leq a^m \, m!.
\end{align}
\end{exm}

\begin{exm}
The function $u(x) = \frac{2\lambda}{\lambda^2 + x^2}$ is
$(\sqrt{e/\lambda}, 2/\lambda)$-semi-analytic. That is due to the fact that
$\widehat u(\omega) = e^{-\lambda |\omega|}$. Hence,
\begin{align*}
\sqrt{\int_{\omega\in\mathbb{R}}\abs{\omega}^{2m}
\abs{\widehat u(\omega)}^2 \, \mathrm d\omega}
= \sqrt{ \int_{\omega} \omega^{2m} e^{-2\lambda \abs{\omega}} \,
\mathrm d\omega} = \frac{1}{(2\lambda)^{m+\frac12}} \sqrt{(2m)!}
\leq \sqrt{\frac{e}{\lambda}} \, \left(\frac{2}{\lambda}\right)^m  m!.
\end{align*}
\end{exm}

We may now show the interpolation result of this section.

\begin{thm}
Let $0<h<H$. Further, let $V:l^2(\mathbb Z^d) \to l^2(\mathbb Z^d)$ be an
isometry defined by its action on the computational basis
\begin{align}
V: \ket{j} \mapsto \sum_{k\in\mathbb Z} V_{jk} \ket{k}
\end{align}
where $V_{jk} := \left(\frac{\sqrt{Hh}}{\pi}\right)\frac{(-1)^k \sin\left( \frac
{\pi  hk}{H}\right)}{hk - Hj}$. Then, for a $(C,a)$-semi-analytic function
$u\in L^2( \mathbb R^d)$, it is the case that
\begin{align}
\norm{ V^{\otimes d} \ket{u_H} - \ket{u_h}}
\leq \frac{8\sqrt{2}e^3\, C}{\norm{u}} e^{-0.6\,\frac{\pi}{aH}}
\end{align}
if $\frac{\pi}{H} \geq 2ad$.
\end{thm}

\begin{proof}
To better understand the proof, it is helpful to provide some intuition
beforehand. Although the technical details closely resemble the previous
results, we discuss the reasoning behind it. Note that our aim is to
approximate the function values at the lattice points $h\mathbb Z^d$, given the
values on $H \mathbb Z^d$, for $h<H$. To do so, we note that
\begin{align}\label{eq:cont-thm-1}
\begin{split}
u(hj) &= \frac{1}{(2\pi)^{d/2}}
\int_{\omega \in \mathbb R^d} e^{i\langle jh, \omega\rangle }
\widehat{u}(\omega) \, \mathrm d\omega\\
&\approx \frac{1}{(2\pi)^{d/2}}
\int_{\omega \in [-\frac{\pi}{H}, \frac{\pi}{H}]^d}
e^{i\langle jh, \omega\rangle} \tilde{u}(\omega) \, \mathrm d\omega\\
&= \left(\frac{H}{2\pi} \right)^d
\int_{\omega \in [-\frac{\pi}{H}, \frac{\pi}{H}]^d}
e^{i\langle jh, \omega\rangle} \sum_{k\in\mathbb Z^d}
e^{-i\langle Hk, \omega\rangle} u(kH)\, \mathrm d\omega\\
&= \sum_{k\in \mathbb Z^d} u(kH) \prod_{a=1}^d W_{j_ak_a},
\end{split}
\end{align}
where $W_{jk}:= \left(\frac{H}{2\pi} \right) \int e^{i (jh - Hk) \omega} = \left
(\frac{H}{\pi}\right) \frac{(-1)^{k}\sin(\frac{jh\pi}{H})}{jh-kH}$. A
straightforward computation reveals that $\sum_{j\in\mathbb Z}W_{jk} \overline
{W}_{jl} = \sqrt{\frac{H}{h}} \delta_{kl}$, and hence, we conclude that $V$ is
an isometry.

It remains to prove that the approximation error in \eqref{eq:cont-thm-1} is
small. With a similar argument as the one used in \cref{lem:tail}, we can show
that
\begin{align}
\sqrt{\underset{{\omega\notin [-\frac{\pi}{H},\frac{\pi}{H}]^d}}
\int \abs{\widehat u(\omega)} \, \mathrm d\omega }
\leq 2e^3 C \, e^{-\frac{0.6\pi}{Ha}},
\end{align}
and also, following the proof of \cref{lem:distance-bound}, we obtain
\begin{align}
\sqrt{\int_{\omega\in[-\frac{\pi}{H},\frac{\pi}{H}]^d} \abs{\widehat u(\omega)
- \tilde u(\omega)}^2 \, \mathrm d\omega+
\int_{\omega\notin [-\frac{\pi}{H},\frac{\pi}{H}]^d}
\abs{\widehat u(\omega)}^2 \, \mathrm d\omega}
\leq 2 \sqrt{2} e^3\, C e^{-\frac{3\pi}{5aH}}.
\end{align}
The rest of the proof follows directly from the arguments made in \cref
{corol:l2Concenteraion}, and \cref{prop:upsampling}.
\end{proof}

We also note that the concentration results of
\cref{sec:Meas-Consenteration} are readily extendable to the non-periodic cases
studied here. In particular, one can show that a function $u$ is semi-analytic if
and only if $|\widehat u|^2/\norm{u}$ defines a sub-exponential distribution.
We leave it open to investigate whether the foundations provided
in the subsection can be used as the building blocks of quantum algorithms using
registers of quantum modes with infinitely many levels (such as bosonic quantum
computers made from quantum harmonic oscillators).

\subsection{Chebyshev interpolation}
\label{sec:app-chebyshev}

In polynomial interpolation, the question of interest is the following. Given
values of a function $f$ at  $x_1<x_2<\cdots<x_N$, find a polynomial that
passes through all point $(x_i, f(x_i))$ and approximates $f$ everywhere else.
It is by Weierstrass approximation theorem that we know such an interpolation is
feasible with arbitrary precision. Moreover, Legendre's interpolating approach
provides the generic solution
\begin{align}\label{eq:interpolation}
p(x) = \sum_{i} f(x_i) \prod_{j\neq i} \frac{x-x_j}{x_i-x_j}.
\end{align}
The error of the above interpolation at a point $x$, for an $N$ times
differentiable function $f$ is given by
\begin{align}\label{eq:error-exp}
e(x) = \frac{f^{(N)}(x)}{N!}\prod_i(x-x_i).
\end{align}
It is easy to observe that $\max_{x\in[-1,1]}|e(x)|
\sim f^{(N)} \frac{l^N}{N}$. However, since the $N$-th derivative of $f$
can grow rapidly (e.g., as fast as $N!$), this interpolation scheme may fail to
work on analytic functions. A well-known example is the Runge function.
However, one can choose the points $x_i$ more wisely to avoid such divergence
in approximation, which is the topic of the rest of this section.

\paragraph{Chebyshev polynomials.}
Chebyshev polynomials are the functions
\begin{align}\label{eq:def-chebyshev}
T_k(x) = \cos(k\arccos(x)),
\end{align}
defined on the domain $[-1, 1]$ and have the following recursive property
\begin{align}
T_k(x) = 2x T_{k-1}(x) - T_{k-2}(x),
\end{align}
which readily shows that $T_k$ is of degree $k$. Moreover, they satisfy the
following orthogonality condition
\begin{align}
\int_{-1}^1 T_n(x) T_m(x) \frac{\mathrm dx}{\sqrt{1-x^2}} = \begin{cases}
0 & \text{if }n\neq m,\\
\pi & \text{if } n=m=0,\\
\frac{\pi}{2} & \text{if }m=n\neq 0,
\end{cases}
\end{align}
and form a complete basis for the function space $L^2(\mu)$, where
$\mathrm d\mu= \frac{\mathrm dx}{\sqrt{1-x^2}}$. In the multi-variate case,
this measure is of the form $\frac{\mathrm dx}{\prod_i \sqrt{1-x_i^2}}$. As a
result, for any $f\in L^2(\mu)$, we have a decomposition
\begin{align}
f(x) = \sum_{n\geq 0} c_n T_n(x),
\end{align}
where
\begin{align}
c_n = \kappa_n \int_{-1}^1 f(x) T_n(x) \frac{\mathrm dx}{\sqrt{1-x^2}},
\end{align}
with $\kappa_0 = \pi$ and $\kappa_n=\pi/2$ for all $n>0$. Note that with the
change of variable $\cos \theta = x$, one may write the Chebyshev coefficients
as
\begin{align}
c_n = \frac{\kappa_n}2 \int_{-\pi}^{\pi}
f(\cos \theta) T_n(\cos\theta) \mathrm d\theta
&= \frac{\kappa_n}2 \int_{-\pi}^{\pi}
f(\cos \theta) \cos(n\theta) \mathrm d\theta \\
&= \pi \int_{-\pi}^{\pi} f(\cos\theta) e^{in\theta} \mathrm d\theta.
\end{align}
Here the last equality is due to the fact that $g(\cdot) = f(\cos(\cdot))$
is an even function. Hence, with the notation of this appendix, we have
$c_n = \pi\mathcal F\left\{ g \right\}_{[n]}$.

\paragraph{Chebyshev grid.}

We consider the lattice defined by the $N$ points
$x_k = \cos(\frac{\pi}{2}+\frac{k\pi}{2n+1})$ where $k \in \{1, \cdots, n\}$.
Note that these points correspond to the roots of $T_{N+1}(x)$ and satisfy
\begin{align}
\prod_{i=1}^N (x-x_i) = 2^{-N} T_{N+1}(x).
\end{align}
Hence, using this grid together with Lagrange interpolation, we obtain that the
remainder formula reads as $e(x)\sim \frac{f^{(N)}}{2^N N!} T_{N+1}(x)$. Note
that $|T_{N+1}(x)|\leq 1$ by definition from \eqref{eq:def-chebyshev}.

A similar interpolation to that of \cref{thm:body-interpolation} can be done
by means of QFT on the Chebyshev grid. We let $\ket{v_N}$ to be the $N$-point
discretization of $f$ on this grid. In other words, we have
$\ket{f_N}:=\sum_{k=1}^N f(x_k) \ket{k}$. The proposition below immediately
follows from \cref{thm:body-interpolation} and
\cref{prop:semi-analytic-basic-prop}.

\begin{prop}[Chebyshev interpolation]
\label{prop:chebyshev-interpolation}
Let $f:[-1,1]^d\rightarrow \mathbb R$ be a $(C,a)$-analytic function and
$\ket{\psi}$ be $\epsilon$-close to $\ket{f_N}$. Then, for any $M > N$ there is
a quantum curcuit that consists of $\tilde{\mathcal O}(d)$ elementary gates and
prepares $\ket{f_M}$, with error at most
$\epsilon + \frac{16e^{3}}{\mathcal U}e^{-0.6N/(1+a)}$.
\end{prop}

Applying the Chebyshev interpolation to the Gibbs state $e^{-f}/ Z_f$ results
in approximate samples from $\frac{e^{-f}}{\prod_{i=1}^d\sqrt{1-x_i^2}}$ on the
hypercube $[-1,1]^d$. We now show that most of this mass is concentrated in the
neighbourhood of the boundary of co-dimension $k= O(1)$ if $\Delta = o(d)$.
Note that we cannot have a concentration with $\Delta=\Omega(d)$, as the measure
defined by $f(x) = \beta\norm{x}^2$ is concentrated at the origin if $\beta >1$.

\begin{prop}
\label{prop:boundary-sampling}
The measure defined by
\begin{align}
\mathrm d\mu \propto \frac{e^{-f(x)}}{\prod_{i=1}^d \sqrt{1-x_i^2}} \,\mathrm dx
\end{align}
in high dimension is concentrated at around any boundary of co-dimension
$k=O(1)$, if $\Delta =o(d)$. More precisely, let $\mathcal C_k^{(\epsilon)}
:=\{x\in[-1,1]^d: |\{i:|1-x_i|\leq \epsilon \}|\geq k\}$. We have
\begin{align}
\int_{x\in\mathcal C^{(\epsilon)}_k} \mathrm d\mu = \Omega(1),
\end{align}
for any $k=O(1)$, $\epsilon=\omega(1/d^2)$, and $\Delta=o(d)$.
\end{prop}

\begin{proof}
We observe that
\begin{align}\label{eq:induction-step-chebyshev-measure}
\frac{\int_{x\notin\mathcal C^{(\epsilon)}_{k+1}}
\mathrm d\mu}{\int_{x\in\mathcal C^{(\epsilon)}_k} \mathrm d\mu}
\leq e^\Delta \cdot \left( \frac{\pi-\sqrt{2\epsilon}}{\pi} \right)^{d-k-1}
= \exp(\Delta - (d-k-1)\log(1-\frac{\sqrt{2\epsilon}}{\pi})).
\end{align}
It is also straightforward to show that
\begin{align}\label{eq:base-chebyshev-measure}
\int_{x\notin\mathcal C^{(\epsilon)}_1} \mathrm d\mu
\leq \exp(\Delta - d\log(1-\frac{\sqrt{2\epsilon}}{\pi})).
\end{align}
The result now follows from \eqref{eq:base-chebyshev-measure} and
\eqref{eq:induction-step-chebyshev-measure}.
\end{proof}

\begin{corol}
\label{cor:chebyshev-rejection}
Let $E:[-1,1]^d\to\mathbb R$ be an L-Lipschitz potential function, and suppose
that the one-parameter family of all measures
$\{e^{\mathcal L t}\rho_0: t\geq 0 \}$ consists of analytic functions with
parameters $C$ and $a$. There is a quantum algorithm that samples from a
distribution $\epsilon$-close to the Gibbs measure (in total variation
distance), by making
\begin{align}
\begin{split}
\mathcal O\Bigg( d^3\kappa'_{E/2} e^{\Delta/2}
\EE_{X\sim e^{-E}/Z}[e^{\norm{x}^2}] &
\max \Bigg\{ (1+a)^4 d^4, \log^4 \!\left(
\frac{\sqrt{d} e^{\frac{5\Delta}4} C (1+a)^3 (1+l\,L)}{\epsilon} \right)\!
\Bigg\}\\
& \plog\left( \frac{(1+a)d e^{\Delta}\log(C(1+lL))}{\epsilon} \right)\! \Bigg).
\end{split}
\end{align}
queries to the energy oracle.
Note that $\kappa'_{E/2}$ here is the Poincar\'e constant of the toric diffusion
obtained by the drift $G(\theta):=E(\cos\theta)$. Moreover, one can estimate the
mean of a random variable, say $f(X)$, where $X\sim e^{-E}/Z$, by making
\begin{align}
\tilde{\mathcal O}\left( d^7(1+a)^4 e^{\Delta/2}\kappa'_{E/2}
\EE_{X\sim e^{-E}/Z}[e^{\norm{x}^2}]\frac{\Delta_f}{\epsilon} \right)
\end{align}
queries to the controlled and standalone oracles of the energy function $E$ and
the function $f$. The algorithm succeeds with high probability with a flag
indicating its success.
\end{corol}

\section{Langevin diffusion on a torus}
\label{sec:app-langevin}

Let $\{Y_t\}_{t\geq 0}$ be a continuous time stochastic process in $\mathbb R^d$
satisfying the overdamped Langevin dynamics at thermodynamic $\beta = 1$,
\begin{align}
\label{eq:langevin-sde}
dY_t = -\nabla E(Y_t)\, dt+ \sqrt{2}\, dW_t
\end{align}
where $(W_t)_{t\geq 0}$ is a Wiener process. This equation is well-studied
in the literature. In particular, it is known that for confining energy
potentials the process is time reversible, ergodic, with a unique stationary
distribution proportional to $e^{-E}$
\citep[Proposition 4.2]{pavliotis2014stochastic}. Note that the condition for
being confining imposes the function to be non-periodic. However, we are
interested in a counterpart to the same results on a torus. We refer the reader
to \citet{garcia2019langevin} for notions of toroidal diffusions and wrapping
of a diffusion process in the Euclidean domain on a torus (that is, the
pushforward of the original process under the quotient map of the torus).
More generally, \citet{hsu2002stochastic} discusses stochastic calculus on
manifolds.

We start with a periodic energy function with period $l$ in all dimensions. It
is shown \citep[Proposition 2]{garcia2019langevin} that the corresponding
wrapped Langevin dynamics is Markovian, ergodic, time-reversible, and admitting
a unique stationary distribution, if the second derivatives of $e^{-E}$ are
H\"older continuous. This condition is satisfied in our case since compactness
of the torus implies that the third partial derivatives of $e^{-E}$ attain their
maxima. Therefore we can derive a Lipschitz property for the second derivatives,
resulting in their H\"older continuity. For further clarity, we will denote the
wrapped process by $\{X_t\}_{t\geq 0}$ and write the overdamped toroidal
Langevin diffusion in the same form as \eqref{eq:langevin-sde} given by
\begin{align}
\label{eq:langevin-sde-torus}
dX_t = -\nabla E(X_t)\, dt+ \sqrt{2}\, dW_t.
\end{align}

The Fokker--Planck equation associated to \eqref{eq:langevin-sde} viewed as an
It\^o stochastic differential equation
\begin{align}
\label{eq:FP}
\partial_t \sigma(y,t) = \nabla \cdot \left(
e^{-E} \,\nabla \left( e^{E} \, \sigma(y,t) \right) \right)
\end{align}
describes the evolution of the probability density function $\sigma(-, t)$ of
$Y_t$ (see \cite{pavliotis2014stochastic} for more details). The probability
density $\rho_t= \rho(-, t)$ of the wrapped process $X_t$ satisfies
\begin{align}
\rho(x,t) = \sum_{k\in\mathbb{Z}^d} \sigma(y + k\, l)
\end{align}
and hence, one gets the following parabolic differential equation with periodic
boundary conditions as the Fokker--Planck equation corresponding to the toroidal
diffusion process $X_t$. We will call this the toroidal Fokker--Planck equation.
\begin{align}
\label{eq:toroidal-FP}
\partial_t \rho(x,t) &= \nabla \cdot \left(
e^{-E} \,\nabla \left( e^{E} \, \rho(x,t) \right) \right)
\end{align}
The corresponding generator $\mathcal{L}$ is also a well-defined operator
\begin{align}
\label{eq:prettyFP}
\mathcal{L}(-) = \nabla \cdot \left(
e^{-E} \,\nabla \left( e^{E} \, - \right) \right).
\end{align}

\begin{rem}\label{rem:prettyFP}
It is worth mentioning that the Fokker--Planck generator is usually written
in the form
\begin{align}
\mathcal{L}(-) = \nabla^2 E (-) + \nabla E \cdot \nabla (-) + \nabla^2 (-)
\end{align}
however as we will see in \cref{sec:discrete-FP}, the generator is better
behaved under discretization when it is written in the form \eqref{eq:prettyFP}.
Nevertheless, discretizing all the derivatives in the usual Fokker--Planck
equation results in an operator of the same form
\begin{align}\label{eq:FP-equality}
\tilde{\nabla^2} E (-) + \tilde{\nabla} E \cdot \tilde{\nabla} (-)
+ \tilde{\nabla^2} (-) = \tilde{\nabla} \cdot \left(
e^{-E} \,\tilde{\nabla} \left( e^{E} \, - \right) \right)
\end{align}
where the tilde on top represents Fourier differentiation operators. This is due
to the fact that discrete Fourier differentiation obeys Leibniz's rules
(see \cref{sec:app} and \cref{sec:discrete-FP}).
\end{rem}

\begin{rem}\label{rem:stationary}
We can derive the uniqueness of the stationary state of the toroidal
Fokker--Planck equation by considering the operator
\begin{align}
\mathcal{L}' := e^{E/2} \circ \mathcal{L} \circ e^{-E/2}.
\end{align}
Let $\pi$ be a density function satisfying $\mathcal L \pi= 0$. It is
straightforward to see that for any periodic density function $\rho$
\begin{align}
\langle \rho, \mathcal{L}' \rho \rangle = - \int_{x\in[-L/2,L/2]^d} e^
{-E} \norm{ \nabla\left(e^{E/2} \rho \right) }^2 \leq 0
\end{align}
with equality happening if and only if $\rho(x) \propto e^{-E(x)/2}$. We
apply this inequality to $e^{E/2} \pi$ and conclude that $\pi = \rho_s$.
\end{rem}

The trend to equilibrium for this stochastic process is studied in the
literature \cite{markowich2000trend, berglund2011kramers, bakry2014analysis,
vanhandel2016probability}. A functional inequality known as the
\emph{Poincar\`e inequality} is equivalent to exponentially fast convergence
of Langevin diffusion \citep[Theorem 4.2.5]{bakry2014analysis} with a rate
known as the \emph{Poincar\`e constant}. Here we show that an analogous
inequality holds for potentials on a torus and we find a corresponding
Poincar\`e constant. But first, we will argue that the aforementioned
exponential decay property for diffusions in the Euclidean space translates
to a counterpart on the torus for toroidal diffusions.

\begin{prop}
\label{prop:exp-decay-torus}
Given the toroidal Fokker--Planck equation \eqref{eq:toroidal-FP}, let $\rho_s$
be the corresponding steady distribution. Further, assume that for a constant
$\lambda > 0 $ any differentiable function $f \in L^2 (\rho_s)$ satisfies
\begin{align}
\Var_{\rho_s} \left[ f \right]
\leq \lambda \, \EE_{\rho_s} \left[ \norm{ \nabla f }^2 \right].
\end{align}
Then the following decay in the distance of $\rho_t$ and $\rho_s$ is satisfied:
\begin{align}
\label{eq:poincare-tv}
\| \rho_t / \rho_s - 1\|_{L^2(\rho_s)}
\leq e^{-t/\lambda } \| \rho_0 / \rho_s - 1\|_{L^2 (\rho_s)}.
\end{align}
\end{prop}

\begin{proof}
Let us denote the ratio of the distributions by $h_t= h(-, t) = \rho_t/\rho_s$.
Using the Fokker--Planck equation \eqref{eq:FP} one has
$\partial_t h = \rho_s^{-1} \nabla\cdot \left( \rho_s \, \nabla h \right)$,
which implies equality (a)
\begin{align}
\label{eq:Poin1}
\frac{d}{dt} \int_{x\in[-l/2,l/2]^d} \rho_s\, (h-1)^2
&= 2 \int_{x\in[-l/2,l/2]^d} \rho_s\,(h-1) \partial_t h\\
&\overset{(a)}{=} -2 \int_{x\in[-l/2,l/2]^d} \rho_s \norm{\nabla h}^2.
\label{eq:Poin2}
\end{align}
Note that $\EE_{\rho_s}\left[h\right] = 1$ and hence, the
left hand side of \eqref{eq:Poin1} is the time derivative of
$\Var_{\rho_s}\left[ h_t \right]$, while the right hand side of \eqref{eq:Poin2}
is $-2 \EE_{\rho_s}\left[ \norm{ \nabla h }^2 \right] $. We conclude that
$\Var_{\rho_s}\left[ h_t \right] \leq
e^{-2\lambda t} \, \Var_{\rho_s}\left[ h_0 \right]$ and therefore the result
follows.
\end{proof}

\begin{rem} We note that by Jensen's inequality
\footnote{The quantity $\norm{p/q-1}^2_{L^2(p)}$ is referred to as the
$\chi^2$ divergence of distributions $p$ and $q$.}
\begin{align}
\EE_{\rho_s} [|\rho_t/ \rho_s - 1|]
\leq \sqrt{\EE_{\rho_s} (\rho_t/ \rho_s - 1)^2}
\end{align}
which yields
\begin{align}
\int_{x\in[-l/2,l/2]^d} \left| \rho_t (s) - \rho_s(x) \right| & \leq
\sqrt{ \int_{x\in[-l/2,l/2]} \rho_s(x)
\left( \frac{\rho_t}{\rho_s} -1 \right)^2}.
\end{align}
Note that the left hand side is twice the total-variation distance between the
two distributions, hence
\begin{align}
\TV(\rho_t, \rho_s) \leq \frac{1}2\,
\sqrt{\Var_{\rho_s}\left[ \rho_t / \rho_s \right]}.
\end{align}
\end{rem}

We can now show that for all bounded energy functions on tori there exists a
universal Poincar\`e constant exists.

\begin{prop}\label{prop:Poin}
Let $E$ be an $l$-periodic energy potential with a bounded range $\Delta$.
Then for all $l$-periodic $f \in L^2(\rho_s)$
\begin{align}
\Var_{\rho_s}\left[ f(X) \right] \leq
\frac{l^2\,e^{\Delta}}{4\pi^2} \,
\EE_{ \rho_s}\left[ \norm{ \nabla f }^2_2 \right].
\end{align}
\end{prop}

\begin{proof}
We have
\begin{equation}
\begin{split}
\Var_{\rho_s}\left[ f(X) \right]
&= \Var_{\rho_s}\left[ f(X) - \int_{x\in[-\frac{l}{2},\frac{l}{2}]^d}
\frac{1}{l^d} f(x) \right]\\
& \leq \EE_{\rho_s}\left[ \left(f(X) - \int_{x\in[-\frac{l}{2},\frac{l}{2}]^d}
\frac{1}{l^d} f(x) \right)^2 \right]\\
& \leq \frac{e^{-\min_{x} E(x)}}{Z} \, \int_{x\in[-\frac{l}{2},\frac{l}{2}]^d}
\left(f(x) - \int_{x\in[-\frac{l}{2},\frac{l}{2}]^d} \frac{1}{l^d} f(x) \right)^2
\label{eq:poin4}
\end{split}
\end{equation}
and also, due to Parseval's theorem
\begin{align}\label{eq:Poin3}
\int_{x\in[-\frac{l}{2},\frac{l}{2}]^d}
\left(f(x) - \int_{x\in[-\frac{l}{2},\frac{l}{2}]^d} \frac{1}{l^d} f(x)\right)^2
&= l^d \sum_{k\in\mathbb{Z}^d\setminus\{0\}} \left|\widehat{f}[k]\right|^2,
\end{align}
where $\widehat{f}$ is the Fourier transform of $f$. Note that the $k=0$ term is
excluded in \eqref{eq:Poin3} since we have subtracted the average of
$f$ on the left hand side. On the other hand, since the Fourier transform of
$\nabla f$ is $\frac{2i\pi}{l ^2}k\, \widehat{f}[k]$, one could again use
Parseval's theorem to write
\begin{equation}
\begin{split}
\int_{x\in[-\frac{l}{2},\frac{l}{2}]^d} \norm{\nabla f}^2
= l^d \sum_{k\in\mathbb{Z}^d}
\frac{4\pi^2}{l^2}\norm{k}^2\,\left|\widehat{f}[k]\right|^2
\geq l^d
\sum_{k\in\mathbb{Z}^d\setminus\{0\}}
\frac{4\pi^2}{l^2}\,\left|\widehat{f}[k]\right|^2.
\label{eq:poin6}
\end{split}
\end{equation}
Now, the inequality $\frac{e^{-\max_x E(x)}}{Z}\,
\int_{x\in[-\frac{l}{2},\frac{l}{2}]^d} \norm{\nabla f}^2
\leq \EE_{\rho_s}\left[ \norm{\nabla f}^2 \right]$ together with
\eqref{eq:poin4}, \eqref{eq:Poin3}, and \eqref{eq:poin6} prove the claim.
\end{proof}

\begin{corol}\label{corol:PI}
In \cref{prop:exp-decay-torus} we have
$\lambda = \frac{l^2\, e^\Delta}{4\pi^2}$ as the universal
Poincar\'e constant.
\end{corol}

\subsection{\bf Discretization of the Fokker--Planck equation}
\label{sec:discrete-FP}

We now introduce the discrete operator $\mathbb L$ obtained from the generator
$\mathcal{L}$ of the diffusion process:
\begin{equation}\label{eq:def-dis-L}
\begin{split}
\mathbb L: \mathcal V_N &\to \mathcal V_N \\
\vec f &\mapsto \tilde{\nabla}
\cdot\left( e^{-E} \tilde{\nabla}(e^E \vec f)\right)
\end{split}
\end{equation}
where tilde on the top of derivative is used to represent Fourier derivative
operators (see \cref{sec:app}). Note that by the product rule of the Fourier
derivative operator (\ref{prop:Fourier1}) we can rewrite $\mathbb L$ in terms of
derivatives of $e^{-E}$ and the function that $\mathbb L$ acts on as follows:
\begin{align}
\mathbb L(-) =
e^E \, \left(\norm{ \tilde{\nabla} e^{-E} }^2
+ \tilde{\nabla^2} e^{-E} \right) (-)
- e^E\, \tilde{\nabla } e^{-E}\cdot \tilde{\nabla} (-)
+ \tilde{\nabla^2} (-).
\end{align}
In what follows, we denote the condition number of a matrix $A$, by $\kappa_A$.
We also use the shorthand notation $[-N..N]:=\{-N,-N+1, \cdots, N\}$.
\begin{lem}
\label{lem:discL}
The discrete operator \eqref{eq:def-dis-L} has the following properties.
\begin{enumerate}[label=(\alph*), ref=\cref{lem:discL}\alph*, itemsep=0pt, partopsep=0pt]
\item It is diagonalizable as $\mathbb L = V^{-1}\,D\,V$ with
$\kappa_V \leq e^{\Delta/2}$, where $D$ is a negative semi-definite diagonal
matrix (i.e., $D\leq 0$).
\label{lem:discL1}
\item The kernel of $\mathbb L$ is one dimensional and is spanned by the
discretized Gibbs distribution $\vec{\rho_s}$.
\label{lem:discL2}
\item The operator norm of $\mathbb L$ is bounded above via
$\norm{\mathbb L} \leq \frac{dN^2}{l^2} \min \left\{ 4\pi^2
+ 2606 \Delta (\ln N)^2, 4\pi^2\,e^{\Delta}\right\}$ for $N>3$.
\label{lem:discL3}
\end{enumerate}
\end{lem}

\begin{proof}
\textit{Claims (a) and (b):} Let $U = e^{-E/2}$ (i.e., $U$ is a diagonal matrix
with diagonal entries all equal to $e^{-E/2}$). Considering the action of the
operator $\mathbb L' = U^{-1}\, \mathbb L \, U$ on the vector $\vec f$ and by
consecutive applications of \ref{prop:Fourier1} we have
\begin{equation}
\begin{split}
\mathbb L'\vec f
&= e^{E/2} \tilde{\nabla} \cdot
\left( e^{-E} \tilde{\nabla}(e^{E/2} \vec f)\right) \\
&= e^{E/2} \tilde{\nabla} \cdot \left( e^{-E} \tilde{\nabla} e^{E/2}
\right) \, \vec f
+ \left( e^{-E/2} \tilde{\nabla} e^{E/2}
+ e^{E/2} \tilde{\nabla} e^{-E/2} \right)
\cdot \left(\tilde{\nabla} \vec f\right)
+ \tilde{\nabla}^2 \vec f \\
&= e^{E/2} \tilde{\nabla} \cdot \left( e^{-E} \tilde{\nabla} e^{E/2}
\right) \, \vec f
+ \tilde{\nabla}^2 \vec f.
\end{split}
\end{equation}
We now note that the first term above is symmetric since it is a diagonal
operator, and so is the operator in the second term, i.e.
$\tilde{\nabla}^2$ (due to \ref{prop:Fourier4}). Hence, $\mathbb L'$ is
symmetric. Note that this concludes $\mathbb L$ being diagonalizable, and
moreover, since $\kappa_U \leq e^{\Delta/2}$, we also have
$\kappa_V\leq e^{\Delta/2}$. Next, we show that $\mathbb L'\preccurlyeq 0$.
\begin{equation}
\begin{split}
\langle f, \mathbb L'\, f\rangle &= \sum_{n\in[-N..N]^d} e^{E[n]/2} f[n]
\,\tilde{\nabla}\cdot\left(
e^{-E} \, \tilde{\nabla}(e^{E/2} f) \right)_{[n]}\\
&\overset{(a)}{=} \sum_{n\in[-N..N]^d} \tilde{\nabla}
\cdot \left( e^{E/2} \, f \,  \tilde{\nabla}(e^{E/2} f) \right)_{[n]}
- \sum_{n\in[-N..N]^d} e^{-E[n]} \,
\norm{ \tilde{\nabla}e^{E/2} f}^2_{[n]}\\
&\overset{(b)}{=} - \sum_{n\in[-N..N]^d} e^{-E[n]} \,
\left(\norm{ \tilde{\nabla}e^{E/2} f}^2\right)_{[n]} \leq 0
\end{split}
\end{equation}
where (a) and (b) follow from \ref{prop:Fourier1} and \ref
{prop:Fourier3}, respectively. Also, note that the expression is zero if and
only if $f[n] \propto e^{-E[n]/2}$. Therefore, the only eigen-direction of
$\mathbb L'$ corresponding to the eigenvalue $0$ is that of $e^{-E/2}$. Using
the similarity transform between $\mathbb L$ and $\mathbb L'$, this consequently
implies that the kernel of $\mathbb L$ is the subspace spanned by the
discretized Gibbs distribution.

\textit{Claim (c):} We note that
$\mathbb L = \left(\tilde{\nabla} \cdot \right) \circ
e^{-E} \circ \tilde{\nabla} \circ e^{E}$,
where $\left(\tilde{\nabla} \cdot \right)$ is the discrete divergence
operator. We can now upper bound the spectral norm of $L$ by noting that
$\norm{e^{-E}} \, \norm{e^{E}} \leq e^{\Delta}$, and
$\norm{\tilde{\nabla}} \leq \sqrt{d} \, \frac{2\pi\, N}{l}$, and also
$\norm{\left(\tilde{\nabla} \cdot \right)} \leq \sqrt{d}\frac{2\pi\, N}{l}$.
Furthermore, using \eqref{eq:FP-equality}, the triangle inequality,
and \ref{prop:Fourier2}, we conclude that
$\norm{\mathbb L} \leq \frac{dN^2}{l^2}
\left( 4\pi^2 + 48^2 (\ln N)^2 \Delta + 96 \pi \Delta \ln N\right)$, for $N>3$.
\end{proof}

In the following lemma, we prove that evolution under $\mathbb L$, does not
dramatically change the $l_2$-norm of the state under evolution. We denote the
vector of all ones by $\mathbf{1} \in \mathcal{V}_N$.

\begin{lem}
\label{lem:Lnorm}
Consider the differential equation $\frac{d}{dt}\vec{u} = \mathbb L \, \vec{u}$,
with initial condition $\vec{u}(0)= \mathbf{1}$. We have
\begin{align}
\sup_{t \geq 0} \norm{ \vec u(t) }
&\leq e^{\Delta/2}\, \norm{\mathbf{1}}, \label{eq:maxnorm} \\
\inf_{t \geq 0} \norm{ \vec u(t) }
&=  \norm{\mathbf{1}} \label{eq:minnorm}
\end{align}
\end{lem}
\begin{proof}

We write the solution as $\vec u(t) = e^{\mathbb Lt} \vec u(0)$. From
\ref{lem:discL1} we have $\norm{e^{\mathbb Lt}} = \norm{V^{-1} \, e^{Dt} \, V }
\leq \norm{V} \, \norm{V^{-1}} \, \norm{e^{Dt}}$. Therefore
$\kappa_V = \norm{V} \, \norm{V^{-1}}$ and $\norm{e^{Dt}} \leq 1$ imply
that $\norm{e^{\mathbb Lt}} \leq \kappa_V \leq e^{\Delta/2}$. This proves
\eqref{eq:maxnorm}.

For \eqref{eq:minnorm} we use the fact that $\bra{\mathbf{1}} \mathbb L =0$
(which follows from \ref{prop:Fourier3}), to conclude that
$\langle \mathbf{1}, u(t) \rangle =  \langle \mathbf{1}, u(0) \rangle$.
Using the Cauchy-Schwartz inequality one has
\begin{align}
\norm{u(t)} \, \norm{\mathbf{1}} \geq  \langle \mathbf{1}, u(0) \rangle
\end{align}
and given $ \langle \mathbf{1}, u(0) \rangle = \norm{\mathbf{1}}^2$,
we have $\norm{u(t)} \geq \norm{\mathbf{1}}$. The result follows by noting
that $t=0$ this inequality is an equality.
\end{proof}

\subsection{Auxiliary lemmas}
\label{sec:FP-analytic}

In this section we upper bound the error in solving the discretization of the
Fokker--Planck equation. Here $\vec{u(t)}$ denotes the discretization of the
actual solution to the Fokker--Planck equation(i.e., the differential equation in
the continuous domain). We shorten our notation and denote this solution by
$\vec{u}$. In contrast, we denote the solution to the discretized Fokker--Planck
equation by $\vec{v}$, that is $\vec{v}$ satisfies the linear system $\frac
{d\vec v}{dt} = \mathbb L \vec v$.

\begin{lem}\label{lem:NS}
Let $u(x,t)$ ($\forall t\geq 0$) be a solution to the Fokker--Planck equation
\eqref{eq:FP}. Then $\max_{x} e^{E}\, u(x,t)$ is a non-increasing function of
time.
\end{lem}

\begin{proof}
Let $v(x,t) = e^{E} u(x,t)$. Using \eqref{eq:FP} $v$ satisfies
\begin{align}
\label{eq:FP-2}
\partial_t v= -\nabla E \cdot \nabla v + \nabla^2 v= \mathcal L^* v
\end{align}
which is the backward Kolmogorov equation with $\mathcal L^*$ the adjoint of
the operator $\mathcal L$ (see for instance \cite{pavliotis2014stochastic} or
\citep[Section 2.2]{vanhandel2016probability}). From this we can generate two
proofs to the lemma.

Let $x^\ast$ be a local maximum of $v(\cdot,t)$. Since $\nabla v(x^\ast,t) = 0$,
and $\nabla^2 v(x^\ast,t) \leq 0$, one concludes from \eqref{eq:FP-2} that the
value of any local maximum of $v$ can only decrease with time. Another argument
relies on observing that the solution to the backward Kolmogorov equation is
the expectation
\begin{align}
\label{eq:bk}
v(x,t+s) = \mathbb{E}\left[ v(X_{t+s},t) \big| X_t = x  \right]
\quad (\forall s, t\geq 0)
\end{align}
where $(X_t)_{t\geq 0}$ is a toroidal stochastic process. However, the
expectation of a function is at most its maximum, therefore
\begin{align}
\label{eq:ub-1}
v(x,t+s) \leq \max_{y\in \mathbb{T}} v(y,t) \quad (\forall x).
\end{align}
It now suffices to take the maximum of the left hand side of
\eqref{eq:ub-1} to prove the claim.
\end{proof}

From hereon we assume $E$ is a potential for which $e^{-E}$ is
$(C, a)$-semi-analytic. Note that for the shifted potential $-E + \delta$
the semi-analyticity parameter $C$ may be replaced with $C e^{\delta}$. Therefore
without loss of generality we assume $E$ attains its minimum value at zero.
Note also that in this case $\mathcal{U} \geq e^{-\Delta}$ as it pertains to
\cref{prop:upsampling}.

\begin{lem}
\label{lem:mainError}
Let $\vec u$ denote the discretization of the solution to the Fokker--Planck
equation (\eqref{eq:FP-2}), with the initial condition $u(0) = \mathbf{1}$.
Assuming $e^{-E}$ and $u$ are both $(C, a)$-semi-analytic, and given
$N\geq \max(4ad,4)$, we have
\begin{align}
\label{eq:mainError}
\norm{ \frac{d}{dt}\vec{u(t)} - \mathbb L\vec{u(t)}}
\leq 8\times 10^5 \pi e^3 \,
\frac{\sqrt{d}\,e^{\Delta} \, C \, (1+lL/48)
(a^3 + a^2)}{l^2} \, \left(2N+1\right)^{d/2} \, e^{-0.4N}
\end{align}
for every point $t \geq 0$ in time, where $L$ is the Lipschitz constant of $E$.
\footnote{More pedantically, we may let $L$ be
the maximum absolute value of the partial derivatives that $E$ attains on the
lattice $L = \max_{x\in\mathbb{T}} \max_{j\in[d]} \abs{\partial_j E}$.}
\end{lem}

\begin{proof}
We write
\begin{align}\label{eq:combinedError}
\mathbb L\vec u - \frac{d}{dt}\vec u
&= e^{E} \left( \norm{\tilde{\nabla} e^{-E} }^2
- \norm{\nabla e^{-E} }^2 \right) \, \vec u
+ e^{E} \left( \tilde{\nabla^2} e^{-E}
- \nabla^2 e^{-E}\right) \vec u  \notag \\
& + e^{E} \left( \nabla e^{-E} \cdot \nabla \vec u
- \tilde{\nabla} e^{-E} \cdot \tilde{\nabla}u \right)
+ \left(\tilde{\nabla^2} \vec u - \vec{\nabla^2 u}\right)
\end{align}
and bound every term on the right hand side as follows. For the first term
\begin{align}
\label{eq:1st}
\begin{split}
\sum_{n\in[-N..N]^d}
& \left|e^{E[n]} \, u[n] \, \left( \norm{\tilde{\nabla}
e^{-E} }^2 - \norm{\nabla
e^{-E} }^2 \right)_{[n]} \right|^2 \\
& \leq \left(\max_{x} e^{E(x)} u(x)\right)^2 \,
\sum_{n\in[-N..N]^d} \left|  \norm{\tilde{\nabla}
e^{-E} }^2 - \norm{\nabla e^{-E} }^2 \right|^2 \\
&\overset{(a)}\leq e^{2\, \max_x E(x)}
\bigg[ \sum_{n\in[-N..N]^d} \left( \norm{\tilde{\nabla}
e^{-E} } - \norm{\nabla e^{-E} } \right)^2_{[n]}
\left( \norm{\tilde{\nabla} e^{-E} }
+ \norm{\nabla e^{-E} } \right)^2_{[n]}\bigg] \\
&\overset{(b)}\leq d\,e^{2\Delta} \left(\frac{48}{l} N \ln N
+ L\right)^2\sum_{n\in[-N..N]^d}\norm{\tilde{\nabla}
e^{-E} - \nabla e^{-E} }^2_{[n]} \\
&\overset{(c)}\leq A_1 \frac{dC^2a^2\, e^{2\Delta}}{l^4}
\left( N \ln N + \frac{lL}{48} \right)^2
\left(2N+1\right)^d\, e^{- N/a}
\end{split}
\end{align}
where we have used \cref{lem:NS} in (a), a triangle inequality of
$l_2$ norms together with \ref{prop:Fourier2} in (b), and
inequality \eqref{eq:first} in (c). Here
$A_1 = 3200\times48^2 \pi^2\, e^6 < 8\times 10^6 \pi^2 e^6$ is a constant.

For the second term in \eqref{eq:combinedError}, we use inequality
\eqref{eq:second} and \cref{lem:NS} to write
\begin{align}
\label{eq:2nd}
\sum_{n \in [-N.N]^d } \left| e^{E[n]} \, u[n]
\left( \tilde{\nabla^2} e^{-E} - \nabla^2 e^{-E} \right)
\right|^2 &\leq A_2 \frac{C^2 a^4\, e^{2\Delta}}{l^4} \,
\left(2N+1\right)^d\, e^{ -0.8N/a}
\end{align}
with $A_2 = 8\times 10^4 \pi^4 e^6$ being a constant.

We rewrite the third term as
\begin{align}
e^{E[n]} \left( \nabla e^{-E} \cdot \nabla u
- \tilde{\nabla} e^{-E} \cdot \tilde{\nabla}u \right)_{[n]}
= e^{E[n]} \left[\left( \nabla e^{-E}
- \tilde{\nabla} e^{-E}\right) \cdot \tilde{\nabla} u
+ \nabla e^{-E} \cdot \left( \nabla u-\tilde{\nabla} u\right)\right]
\end{align}
which allows us to conclude that
\begin{equation}
\begin{split}
\left| e^{E[n]} \left( \nabla e^{-E} \cdot \nabla u
- \tilde{\nabla} e^{-E} \cdot \tilde{\nabla}u \right)_{[n]} \right|
&\leq e^{E[n]} \, \left| \left(\nabla e^{-E}
- \tilde{\nabla} e^{-E}\right) \cdot \tilde{\nabla} u \right|
+ \left| \nabla E[n] \cdot \left( \nabla u-\tilde{\nabla} u\right) \right|\\
&\leq e^{E[n]} \, \norm{ \nabla e^{-E} -  \tilde{\nabla} e^{-E} }
\norm{\tilde{\nabla} u}
+ \norm{\nabla E} \norm{ \nabla u-\tilde{\nabla} u }.
\end{split}
\end{equation}
Hence (by the elementary inequality $(a+b)^2 \leq 2a^2 + 2b^2$ and using
\ref{prop:Fourier2}) we have
\begin{equation}
\begin{split}
\norm{ e^{E} \left( \nabla e^{-E} \cdot \nabla u
- \tilde{\nabla} e^{-E} \cdot \tilde{\nabla}u \right) }^2
&\leq 2\frac{d\,e^{2 \max_x E(x)}}{l^4}
\left(48 N \log N\right)^2 \norm{ \nabla e^{-E}
- \tilde{\nabla} e^{-E} }^2 \\
&+ 2 d L^2 \norm{ \nabla u-\tilde{\nabla} u }^2
\end{split}
\end{equation}
which by applying \eqref{eq:first} implies
\begin{align}
\label{eq:3rd}
\norm{ e^{E} \left( \nabla e^{-E} \cdot \nabla u
- \tilde{\nabla} e^{-E} \cdot \tilde{\nabla}u \right) }^2
\leq 2A_1 \frac{d \, e^{2 \Delta} \, C^2 a^2}{l^4}
\left( N \ln N + \frac{lL}{48} \right)^2
\left(2N+1\right)^{d} \, e^{- N/a}.
\end{align}

Finally the last term of \eqref{eq:combinedError} is taken care of directly
using \eqref{eq:second}:
\begin{align}
\label{eq:4th}
\norm{ \tilde{\nabla^2} u - \nabla^2 u}
\leq \sqrt{A_2} \frac{C\, a^2}{l^2}\left(2N+1\right)^{d/2} e^{-0.4 N}
\end{align}

We now observe that $N\ln N + x \leq N\ln N (1+x)$ for all $x>0$ and all
$N \geq 3$. This together with $N\ln N \leq N^2 \leq 100a^2 e^{0.1 N/a}$,
and combined with \eqref{eq:1st}, \eqref{eq:2nd}, \eqref{eq:3rd},
\eqref{eq:4th}, and \eqref{eq:combinedError} yield
\begin{align}
\norm{ \frac{d}{dt}\vec{u(t)} - \mathbb L\vec{u(t)}}
\leq A \frac{\sqrt{d}e^{\Delta}
C\, (1+\frac{lL}{48}) (a^3 + a^2)}{l^2}
\left(2N+1\right)^{d/2} e^{-0.4N}
\end{align}
where $A = 4\times 100\sqrt{2A_1}
< 1.6\times 10^6 \pi e^3$.
\end{proof}

\begin{prop}
\label{prop:error-last}
Let $u(x,t)$ be the exact solution to the diffusion process and further let
$\vec{v}$ be the solution to the discretized (in space) differential
equation i.e., $\vec{v}$ satisfies $\frac{d\vec{v}}{dt} =
\mathbb L \vec{v}$. Assume $\{u(\cdot,t): t\in[0,T]\}$ consists of $
(C,a)$-semi-analytic functions, and further assume $N\geq \max(4da,4)$. Then,
\begin{align}
\label{eq:disc-FP-error}
\norm{\vec{u}(T) - \vec{v}(T)}
\leq 1.6\times 10^6 \pi e^3
\frac{T\,e^{3\Delta/2}\, C^2\, (a^3+a^2)\,
(1 + \frac{l L}{48})}{l^2} \left(2N+1\right)^{d/2}
e^{-0.4 \frac{N}{a}}.
\end{align}
\end{prop}

\begin{proof}
For convenience we will denote the right hand side of \eqref{eq:mainError}
as $f[N]$ (which defers from the right hand side of \eqref{eq:disc-FP-error}
above by a factor of $Te^{\Delta/2}$).

We now write $\frac{d}{dt} \vec{u}(t) = \mathbb L \vec{u}(t) + \vec{e}(t)$ where
$\norm{\vec{e}(t)}$ is upper bounded in \cref{lem:mainError}. Note also that
$\frac{d}{dt}\vec{v} = \mathbb L \vec{v}$ by definition. Hence, letting
$\vec{\epsilon} := \vec{u} - \vec{v}$ we get
\begin{align}\label{eq:ErrorF-2}
\frac{d}{dt}\vec{\epsilon}(t) = \mathbb L\vec{\epsilon}(t) + \vec{e}(t).
\end{align}
By \ref{lem:discL1}, $\mathbb L = V^{-1} \, D \, V$ where $D\leq 0$ and
$\kappa_V \leq e^{\Delta/2}$. It is left to upper bound the norm of
$\vec{\epsilon}(T)$. We multiply both sides of \eqref{eq:ErrorF-2} from left by
$V$, and let $\vec{\mathcal{E}}:=V\vec{\epsilon}$ and $\vec{b}:=V\vec{e}$ to get
\begin{align}
\frac{d}{dt}\vec{\mathcal{E}}(t) = D \, \vec{\mathcal{E}}(t) + \vec{b}(t).
\end{align}
Now, we take the inner product with respect to $\vec{\mathcal{E}}$:
\begin{align}
\langle\vec{\mathcal{E}}, \frac{d}{dt} \vec{\mathcal{E}}\rangle
= \langle\vec{\mathcal{E}}, D \, \vec{\mathcal{E}}\rangle
+ \langle \vec{\mathcal{E}}(t), \vec{b}(t)\rangle.
\end{align}
Since $D$ is negative semi-definite
\begin{align}
\Re\left( \langle\vec{\mathcal{E}}, \frac{d}{dt} \vec{\mathcal{E}}\rangle \right)
\leq \Re\left(\langle \vec{\mathcal{E}}(t), \vec{b}(t)\rangle\right)
\leq \norm{\vec{\mathcal{E}}}\, \norm{\vec{b}}.
\end{align}
Therefore since
$\frac{d}{dt} \norm{\vec{\mathcal{E}}}^2 = 2\Re
\left(\langle\vec{\mathcal{E}}, \frac{d}{dt} \vec{\mathcal{E}}\rangle\right)$,
we conclude
\begin{align}
\frac{d}{dt} \norm{\vec{\mathcal{E}}(t)} \leq \norm{\vec{b}(t)}.
\end{align}
Recalling $\vec{\mathcal{E}} = V \vec{\epsilon}$ and $\vec{b} = V \vec{e}$
we get
\begin{align}
\frac{d}{dt} \norm{V\, \vec{\epsilon}} \leq f[N] \norm{V \ket{e}}.
\end{align}
where $\ket{e(t)}$ is $\vec{e}(t)$ normalized. We use
$\norm{V\, \ket{e}}\leq \sigma_{\max}(V)$ to conclude that
\begin{align}
\norm{V\, \vec{\epsilon}(T)} \leq T f[N] \sigma_{\max}\left(V\right).
\end{align}
Finally, using $\norm{V \vec{\epsilon}} \geq \norm{\vec{\epsilon}}
\sigma_{\min}\left( V\right)$ and $\kappa_V \leq e^{\Delta/2}$ we complete
the proof.
\end{proof}

\section{Training energy-based models}
\label{sec:app-EBM}

In machine learning, energy-based models (EBM) are used to regress a Gibbs
distribution
\begin{equation}
\label{eq:Gibbs-distribution}
p_\theta (x) = \exp(-\beta E_\theta (x))/ Z_{\beta, \theta}
\end{equation}
known as the \emph{model distribution} from an unknown distribution $\ptarget$
represented by classical data. The EBM comprises a function approximator for
an \emph{energy potential} $E_\theta: \mathbb R^d \to \mathbb R$. Here
$\theta \in \mathbb R^m$ denotes a vector of $m$ model parameters (e.g., weights
and biases of a deep neural network). Given a set of i.i.d training samples
$D= \{x_1, \dots, x_N\} \subset \mathbb R^d$, the goal of the learning procedure
is to find a vector of model parameters $\theta^* \in \mathbb R^m$ that attain
optimal regression of $\ptarget$ via $p_{\theta^*}$ with respect to the
Kullback-Leibler (KL) distance between the two distributions. It is easy to
see that this is equivalent to maximizing the log-likelihood of the training
data:
\begin{equation}
\label{eq:data-model-kl}
\KL(p_\text{data}(x)||p_\theta(x))=
-\mathbb E_{x \sim \ptarget} [\log p_\theta(x)] + \text{constant}.
\end{equation}

However, we do not need access to the value of the likelihood directly but
rather the gradient of the log-probability of the model. We have
\begin{equation}
\label{eq:grad-log-prob}
\begin{split}
\mathbb E_{\ptarget}[\nabla_\theta \log p_\theta(x)]
= -\beta \mathbb E_{\ptarget} [\nabla_\theta E_\theta(x)]
+ \beta \mathbb E_{p_\theta}[\nabla_\theta E_\theta(x)].
\end{split}
\end{equation}
While the first term is easy to approximate using the data samples the second
term is approximated through costly Gibbs sampling. Indeed, if we can
efficiently draw samples from the model distribution $p_\theta$, we have access
to unbiased estimates of the log-likelihood gradient, which in turn can be used
to train the EBM via stochastic gradient descent \cite{song2021train}.

The energy function is a composition of linear functions (affine
transformations) with nonlinear ones such
as the sigmoid function \footnote{Although the
rectified linear unit (ReLU) is more typically used it does not fall under the
semi-analyticity assumptions of our paper.} for which we can build a quantum
oracle as in \cref{fig:EBM}.
Our algorithm works by queries to an oracle for the discretization of
the generator of the Fokker--Planck equation $L$ (see \eqref{eq:def-dis-L}).
In order to construct this oracle, we use the expression \eqref{eq:FP-equality}
restated here as
\begin{align}\label{eq:O4L}
\mathbb L(-) = \tilde{\nabla^2} E (-)
+ \tilde{\nabla} E \cdot \tilde{\nabla} (-)
+ \tilde{\nabla^2} (-)
\end{align}
constructed using $2d(2N+1)$ replicas of the energy oracle.

After iterative queries to the oracle of $\mathbb L$,
\cref{alg:gibbs-sampling} returns a sample from the model distribution of the
EBM. Repeated executions will then provide an approximation of the second term
in \eqref{eq:grad-log-prob}. This in turn allows for updating the model
parameters $\theta$ via stochastic gradient descent. And finally, repeated
descent steps will result in an approximation for trained parameters
$\theta^*$.

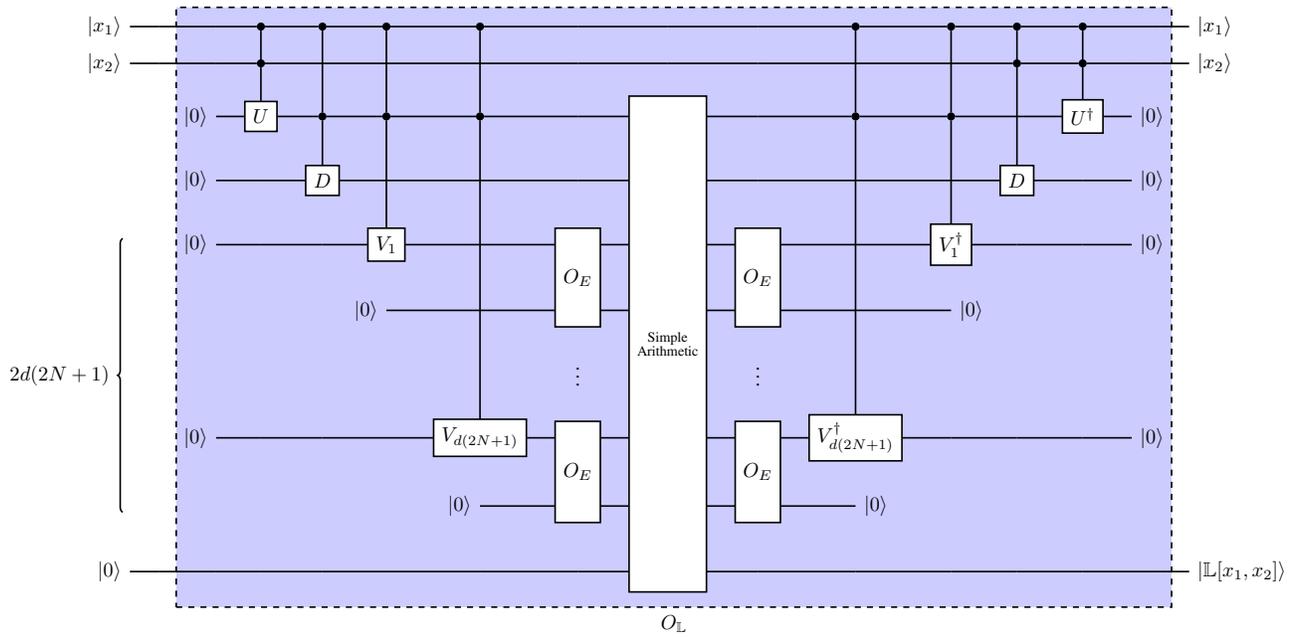
\begin{figure}
\centering
\begin{adjustbox}{max width = \textwidth}
\begin{quantikz}
\lstick{$\ket{x_1}$} & \qw & \qw\gategroup[10,steps=15,style={dashed,
fill=blue!20, inner xsep=2pt},
background,label style={label position=below,anchor=
north,yshift=-0.2cm}]{{\sc $O_\mathbb L$}}
& \qw & \ctrl{1} & \ctrl{2} & \ctrl{2} & \ctrl{2}
& \qw & \qw & \qw & \ctrl{2} & \ctrl{2} & \ctrl{1} & \ctrl{1} & \qw & \qw & \qw
\rstick{$\ket{x_1}$} \\
\lstick{$\ket{x_2}$} & \qw & \qw & \qw & \ctrl{1} & \qw & \qw & \qw & \qw & \qw
& \qw & \qw & \qw & \ctrl{2} & \ctrl{1} & \qw & \qw & \qw \rstick{$\ket{x_2}$}\\
&&& \lstick{$\ket 0$} & \gate{U} & \ctrl{1} & \ctrl{2} & \ctrl{5}
& \qw & \gate[wires = 8, nwires=5]{
\text{$\substack{\text{Simple}\\\text{Arithmetic}}$}} & \qw & \ctrl{5}
& \ctrl{2} & \qw & \gate{U^\dagger} & \qw\rstick{$\ket 0$} \\
&&& \lstick{$\ket 0$} & \qw & \gate{D} & \qw & \qw & \qw && \qw & \qw & \qw &
\gate{D} & \qw & \qw\rstick{$\ket 0$} & &&&& \\
\lstick[wires=5]{$2d(2N+1)$}
&&& \lstick{$\ket{0}$}  & \qw & \qw & \gate{V_1} & \qw & \gate[wires=2]{O_E} &
& \gate[wires=2]{O_E} & \qw & \gate{V_1^\dagger} & \qw & \qw &
\qw \rstick{$\ket 0$} \\
&&&&&& \lstick{$\ket{0}$} & \qw & & & & \qw & \qw\rstick{$\ket 0$} &&&& \\
&&&&&&& & \vdots& & \vdots & &&& \\
&&& \lstick{$\ket 0$} & \qw & \qw & \qw & \gate{V_{d(2N+1)}} &
\gate[wires=2]{O_E} & & \gate[wires=2]{O_E} & \gate{V_{d(2N+1)}^\dagger} & \qw
& \qw & \qw & \qw\rstick{$\ket 0$}\\
&&&&&&& \lstick{$\ket 0$}  & & & & \qw\rstick{$\ket 0$} & & \\
\lstick{$\ket{0}$} & \qw & \qw & \qw & \qw & \qw & \qw & \qw & \qw &
& \qw & \qw & \qw & \qw & \qw & \qw & \qw
& \qw \rstick{$\ket{\mathbb L[x_1,x_2]}$}
\end{quantikz}
\end{adjustbox}
\caption{The circuit for the oracle of discrete generator $\mathbb L$ comprising
$2d(2N+1)$ copies of the energy potential oracle, $O_E$. To query
$\mathbb L[x_1,x_2] = \bra{x_2}\mathbb L \ket{x_1}$, first the controlled-$U$
gate checks for the difference between $x_1$ and $x_2$: the third register is
set to $\ket{i}$ if $x_1$ and $x_2$ defer only on their $i$-th entry. The state
remains unchanged, if $x_1=x_2$, and it is set to a null state $\ket \perp$
otherwise. Conditioned on this third register being at state $\ket{i}$, another
register (the fourth register) computes the distance between $x_1$ and $x_2$
along the $i$-th axis on the lattice (the controlled-$D$ gate). Again,
conditioned on the state of the third register, we query the energy function at
specific lattice points to compute either $\partial_iE(x_1)$ (if the third
register is in $\ket i$), or $\nabla^2 E(x_1)$ (if the third register is in
$\ket{0}$) using the sequence of controlled-$V_j$ gates. The estimation of
these derivatives exploits Fourier spectral method (see \cref{sec:app}) and is
applied via a circuit performing simple arithmetic.}
\label{fig:Oracle4L}
\end{figure}

Alternatively, we may use the controlled variant of \cref{fig:Oracle4L}
in order to perform mean estimations of the components of the gradient
$\nabla_\theta E_\theta(x)$ as per \cref{cor:body-mean-estimation} quantumly,
as opposed to using samples provided by the quantum computer to perform
classical estimation of the expectation $\mathbb E_{p_\theta}
[\nabla_\theta E_\theta]$. The mean estimation algorithm queries this controlled
oracle of $\mathbb L$ and additionally the controlled oracles of the $m$ partial
derivatives of $E_\theta$. The construction of the latter oracles can be
automated in the same fashion as automatic differentiation in ML.

\section{Proofs of the results in \cref{sec:main-results}}

\subsection{Gibbs sampling}
\label{sec:results-sampling}

Our algorithm benefits from the use of the high precision quantum linear
differential equation solver developed by \citet
{berry2017quantum}. Krovi later proved in \yrcite{krovi2023improved} that this
solver is more efficient than initially shown in the original work. The solver
is designed to tackle the problem of interest, which is solving the following
ODE at time $T>0$:
\begin{align}
\frac{d \vec x}{dt} = A \vec x + \vec b, \; \vec x(0)=\vec x_{\mathrm{init}},
\end{align}
where $\vec x, \vec b\in\mathbb{C}^n$ and $A\in \mathbb{C}^{N\times N}$ is a
matrix whose all eigenvalues have non-positive real parts. The aim of `solving'
the ODE is to prepare a quantum state that encodes the entries of $x(T)$. The
main idea that is used in the construction of the algorithm is the truncation
of the Taylor expansion of the exponential function, as the $\vec x
(T)$ satisfies the following closed-form solution (\citep[Lemma 6]
{krovi2023improved}).
\begin{align}
\vec x(T) = e^{At} \vec x(0)
+ \left[ \int_{0}^T e^{As}\, \mathrm ds \right] \, \vec b
\end{align}
We directly apply their algorithm to solve \cref{eq:FP-2}. Hence, we restate
their complexity result.

\begin{thm}[Adoption of Theorem 7 of \cite{krovi2023improved}]
\label{thm:Berry-et-al}
Suppose $A=VDV^{-1}$ is an $N\times N$ diagonalizable matrix, where $D = diag
(\lambda_1,\cdots,\lambda_N)$ satisfies $\Re(\lambda_i) \leq 0$ for any
$i\in[N]$. In addition, suppose $A$ has at most $s$ nonzero entries in any row
and column, and we have an oracle $O_A$ that computes these entries. Suppose
$\vec x_{init}$ and $\vec b$ are $N$-dimensional vectors with known norms and
that we have two controlled oracles, $O_x$ and $O_b$ that prepare states
proportional to $\vec x_{init}$ and $b$, respectively. Let $\vec x$ evolve
according to the differential equation
\begin{align}
\frac{d}{dt}\vec x = A \vec x + \vec b
\end{align}
with the initial condition $\vec x(0) = \vec x_{init}$. Let $T>0$ and
\begin{align}
g = \max_{t\in [0,T]} \norm{\vec x(t)}/\norm{ \vec x(T)}.
\end{align}
Then there exists a quantum algorithm that produces a state $\epsilon$-close to
$\vec{x}(T)/\norm{\vec x(T)}$ in $l^2$-norm, succeeding with $\Omega(1)$
probability, with a flag indicating success, using
\begin{align}
O\Big(gT\|A\|\kappa_V\poly\Big(s,\log d,\log(1+\frac{Te^2\|b\|}{\|x_T\|}),\log(\frac{1}{\epsilon}),\log(T\|A\|\kappa_V)\Big)\Big)\,,
\end{align}
queries overall, where $\kappa_V = \norm{V} \norm{V^{-1}}$ is the condition number of $V$.
In addition, the gate complexity of the algorithm is larger than its query
complexity by a factor of
\begin{equation}
    O(\plog (1+\frac{Te^2\|b\|}{\|x_T\|}, 1/\epsilon, T\|A\|))\,,
\end{equation}.
\end{thm}
\begin{proof}
We can observe that the only difference in the complexity result between our theorem and the one presented in \citep[Theorem 7]{krovi2023improved} is the substitution of $C(A)$ with $\kappa_V$. It is worth recalling from \citep[Definition 5]{krovi2023improved} that $C(A)$ is defined as the maximum norm of $e^{At}$ over the interval $[0,T]$. By imposing the condition $A = V D V^{-1}$ with $D\leq 0$, we readily obtain $C(A)\leq \kappa_V$. It is worth noting that in the case where $A$ is not diagonalizable, the relation between $C(A)$ and $\kappa_V$ is presented in \citep[Lemma 4]{krovi2023improved}.
\end{proof}

We now provide the proof of main result.

\begin{thm}[\cref{thm:body-main-result} in the manuscript]
\label{thm:main-result}
Given an $L$-Lipschitz periodic potential $E$,
suppose that the one-parameter family of all probability measures
$\{e^{\mathcal{L}t} \rho_0: t\geq 0\}$ consists of semi-analytic functions with
parameters $(C,a)$. \cref{alg:gibbs-sampling} samples from
a distribution $\epsilon$-close to the Gibbs distribution (in total variation
distance), by making
\begin{align*}
\mathcal{O}\left( d^3\, \frac{\kappa_{E/2}}{l^2}\, e^{\frac{\Delta}2}
\max \Bigg\{ a^4 d^4, \log^4 \!\left(
\frac{\sqrt{d} e^{\frac{5\Delta}4} C a^3 (1+l\,L)}{\epsilon} \right)\!
\Bigg\}
\plog\left( \frac{ad e^{\Delta}\log(C(1+lL))}{\epsilon} \right)\! \right)
\end{align*}
queries to the oracle of the energy function. The algorithm succeeds with
bounded probability of failure and returns a flag indicating its success.
In addition, the gate complexity of the algorithm is larger only by a factor of
$\plog(Cad e^{\Delta} (1+lL))/\epsilon)$.
\end{thm}

\begin{proof}
We first note that the energy function we consider in the Fokker--Planck equation
is $E/2$. Here we provide the method to obtain a $6\epsilon$-approximate
sampler. Using \cref{prop:error-last} we have
\begin{align}
\norm{\ket{e^{\mathbb LT}\, \rho_0} - \ket{e^{\mathcal{L}T} \, \rho_0}}
\leq 3.2\times 10^6 \pi e^3
\frac{\sqrt{d} T e^{3\Delta/4} \, C \, (1+\frac{lL}{48})
(a^3 + a^2)}{l^2} e^{-0.4N}
\end{align}
where we have also used \cref{lem:unitnorm} and \cref{lem:Lnorm}. Hence, if we
let $A = 3.2\times 10^6 \pi e^3$, we may choose
\begin{align}\label{eq:choose N}
N = \left\lceil\max\Bigg\{0.4^{-1}
\log\left(\frac{A\,\sqrt{d} T\, e^{3\Delta/4} \, C\,(a^2+a^3)
(1+l\,L/48)}{l^2\, \epsilon} \right),
4ad, 4\Bigg\}\right\rceil
\end{align}
to guarantee an at most $\epsilon$ distance between
$\ket{e^{\mathbb LT}\, \rho_0}$ and
$\ket{\rho_T}= \ket{e^{\mathcal{L}T}\, \rho_0}$. Now we apply
\cref{thm:Berry-et-al} to obtain an output state $\ket{\mathcal{A}}$, for which
we have $\norm{ \ket{\mathcal{A}} - \ket{\rho_T} } \leq 2\epsilon$.
Hence, by \cref{thm:interpolation}, continuous sampling from the algorithm's
output will provide samples from a $5\epsilon$-close distribution to the
distribution proportional to $\rho_T^2$.

Now we need to set $T$. \cref{lem:T}, together with \cref{corol:PI}
implies that  choosing
\begin{align}
T = \kappa_{E/2} \log(2 e^{\Delta/2}/\epsilon)
\end{align}
guarantees that the distribution proportional to $\rho_T^2$ is $\epsilon$-close
(in total variation distance) to the Gibbs distribution. Overall, our sampling
procedure returns samples from a distribution $6\epsilon$-close to the Gibbs
distribution.

The complexity of the algorithm according to \cref{thm:Berry-et-al} is now
obtained by noting firstly that $\kappa_V \leq e^{\Delta/4}$ from
\ref{lem:discL1} for $E/2$. Next, we note that the sparsity of $\mathbb L$ is
$s= O(dN)$. Also $g= O(e^{\frac{\Delta}4})$ by \cref{lem:Lnorm}. The
norm of $\mathbb L$ is bounded by \ref{lem:discL3} as
$\norm{\mathbb L}= O(\Delta dN^2/ l^2 \plog N)$. Finally $\beta \leq 1$ for us
also using \cref{lem:Lnorm}. This provides the complexity of every term in
\cref{thm:Berry-et-al} with respect to $N$. We also note that
\begin{align}
N = \Theta^\ast
\left(\max \Bigg\{ \log \left( \sqrt{d} \frac{e^{5\Delta/4} \,a^3 C\,
(1+lL)}{\epsilon} \right), a d \Bigg\} \right)
\end{align}
by our choice of $T$. Finally, a query to $\mathbb L$ requires $\mathcal{O}(dN)$
queries to $\mathcal{O}_E$ and this completes the proof.
\end{proof}

\subsection{Mean estimation}
\label{sec:results-mean}

In this section, we delve into how the Gibbs sampler discussed earlier can be
employed to calculate the expected values of random variables with bounded
variance. Specifically, we consider a periodic function $f:[-\frac{l}{2},\frac
{l}{2}]^d\rightarrow \mathbb R$ that belongs to $L^2(\rho)$, and we aim at
estimating ${\EE}_{\rho} f(X)$, where $X$ is a random variable with
distribution $\rho$. We utilize the state-of-the-art estimation algorithm
presented in \cite{kothari2023mean} to compute the expected value of our
function.

The main problem of interest is that of the mean estimation of a classical
random variable, whose classical probability amplitudes are encoded in a
quantum state. \citet{brassard2002quantum} consider having
access to a unitary $U$, that acts as $U\ket 0 = \sqrt{p} \ket 0 + \sqrt
{1-p} \ket{ 0^\perp}$, where $\ket{ 0^\perp}$ is a vector orthogonal to $\ket
0$. They prove that $\mathcal O(\frac{1}{\epsilon})$ queries to controlled-$U$
is sufficient to estimate $p$ with precision $\epsilon$, and with high
probability. The proof is based on the fact that $\ket \psi :=U\ket{0}$ can be
viewed as
\begin{align}
\ket\psi = \frac{1}{2}
e^{i\theta}
\left(\ket 0 - i|0^\perp\rangle\right)
+ \frac{1}{2}
e^{-i\theta}
\left(\ket 0 + i|0^\perp\rangle\right)
\end{align}
where $\sin \theta = \sqrt{p}$. We then note that $\left(\ket 0 \pm
i|0^\perp\rangle\right)$ are eigenvectors of a rotation matrix with rotation
angle $\phi$ in the $\ket{0}, |0^\perp\rangle$ plane, with eigenvalues $e^
{\pm i\phi}$. As the Grover diffusion operator is itself a rotation with
angle $2\theta$, the phase estimation algorithm (with the unitary being
the Grover operator, and the input state being $\ket \psi$) will reveal
$\theta$, and consequently $p$. One can think of this algorithm as an
estimation algorithm for a binary random variable. Note that classically one
requires $\Omega(\frac{1}{\epsilon^2})$ samples in order to achieve an
$\epsilon$-accurate estimation of $p$. This quadratic speedup with respect to
the error parameter is sometimes referred to as the Heisenberg limit, and as we
discuss later, is not restricted to the case of binary random variables.

Subsequently, \citet{montanaro2015quantum} and \citet{li2018quantum} extended
the above algorithm and obtained mean estimation algorithms for more generic
cases. \citet{hamoudi2018quantum} combines the latter algorithms and
obtains a desired complexity of $\mathcal{O}^\ast(\frac{1}{\epsilon})$. The
recent work of \cite{kothari2023mean} is the state of the art and provides an
algorithm that we directly apply for our mean estimation tasks.

Assume we have access to controlled-$U$, and its inverse, such that $U \ket 0
= \sum_{x\in\Omega} \sqrt{p_x} \ket{x}$. Further, one can assume having access
to controlled versions of a unitary $F$ and its inverse ($F^\dagger$) that acts
as $F\ket{x}\ket{0}\ket{0} = \ket{x}\ket{f(x)}\ket{0}$, for some function $f$.
Note that $F$ is allowed to exploit auxiliary qubits for the evaluation of $f$.
Having access to such quantum circuits is phrased as `having the code' for the
random variable $f(X)$ in \cite{kothari2023mean}. We restate the following
theorem from their work.

\begin{thm}[Theorem 1.3 of \cite{kothari2023mean}]\label{thm:Robin}
There is a computationally efficient quantum algorithm with the following
properties: Given `the code' for a random variable $f(X)$, the algorithm makes
$O(n\log \frac{1}{\delta})$ queries to the oracles for the controlled unitaries
$U$, $U^\dagger$, $F$, and $F^\dagger$ to output an estimation $\widehat \mu$
such that
\begin{align}
\mathbb P\left[ |\widehat \mu - \EE [f(X)] |
\geq  \frac{\Var [f(X)]}{n} \right] \leq \delta.
\end{align}
\end{thm}

The algorithm they propose is again based on Gorver's diffusion operators.
However, they use unitaries with complex phases (as opposed to reflections).
Let us now state our mean estimation result.

\begin{corol}[\cref{cor:body-mean-estimation} in the manuscript]
\label{cor:mean-estimation}
Let $E$ be an energy function, satisfying the assumptions made in \cref
{thm:main-result}. Furthermore, let $f$ be an $L_f$-Lipschitz $l$-periodic
function with diameter $\Delta_f$. There is a quantum algorithm that returns an
estimate $\widehat \mu$ to $\mathbb E[f(X)]$, with additive error at most
$\epsilon >0$ and success probability at least $1-\delta$, making
\begin{align}
\mathcal O \left( d^7 a^4e^{\Delta/2} \frac{\kappa_{E/2}}{l^2}
\frac{\Delta_f}{\epsilon} \log(\frac{1}{\delta})
\plog\left( C,a,\frac{1}{\epsilon}, \Delta_f, L_fl, Ll \right)\right)
\end{align}
queries to the controlled and standalone oracles of the energy function $E$ and
the function $f$.
\end{corol}

\begin{proof}
Consider the quantum circuit that implements line \ref{alg-line:2} of \cref
{alg:gibbs-sampling}, and call it $U$. We can manipulate this circuit to obtain
a unitary $\tilde{U}$ such that $\norm{\ket{u(T)} - \tilde U \ket{ u
(0)}} \leq \epsilon_1$, by making $\mathcal O \left( \log \frac{1}
{\epsilon_1} \right)$ calls to $U$, $U^\dagger$, and additional gates. This is
achieved via fixed-point amplitude amplification algorithm\footnote{The $\frac
{\pi}{3}$-amplitude amplification algorithm of \cite{grover2005fixed} has a
dependence on the success probability of the algorithm, which is later improved
in the works of \citet{yoder2014fixed} and \citet{gilyen2019quantum}. However,
as the success probability of our circuit is $\Omega(1)$, we do not need to
utilize the more complex algorithms.} \cite{grover2005fixed}. Note that a
total-variation distance of $\epsilon_1$ between the two distributions, results
in at most a $M\epsilon_1$ distance between the expected values. Furthermore,
we note that expectation with respect to the algorithm's output would be far
from the actual value by at most
\begin{align}
\frac{d/2(lL_f + lL\Delta_f + 10\sqrt{2}/3 e^4 \Delta_f aC)}{M}
+ 2\Delta_f\epsilon_1,
\end{align}
which follows from the total variation bounds obtained above, and further, that
of \cref{lem:choose-M}. Hence, implementing line \ref{alg-line:3} of \cref
{alg:gibbs-sampling} with $M = \frac{\mathrm{poly}(C,a,\Delta_f,L_fl,Ll)}
{\epsilon}$, and $\epsilon_1 = \frac{\epsilon}{8\Delta_f}$, results in at most a
distance $\epsilon/2$ from the ideas expectation. Finally, setting $n = \frac
{\Delta_f}{\epsilon}$ and applying \cref{thm:Robin} concludes the result.
\end{proof}

\section{Lemmas used in \cref{sec:app-semianalytic}}

\begin{lem}
\label{lem:cool}
For any integer $m\geq 0$ and $z>0$
\begin{align}
\sum_{k=0}^{\infty} \frac{z^k\, k^m}{k!} \leq e^z \, \max\{z^m,1\}\, m!.
\end{align}
\end{lem}

\begin{proof}
We note that
\begin{equation}
\label{eq:cool3}
\begin{split}
\sum_{k=0}^{\infty} \frac{z^k\,k^m}{k!}
= \frac{\partial^m}{\partial x^m} \Bigg|_{x=0}
\sum_{k=0}^\infty \frac{z^k \,e^{kx}}{k!}
= \frac{\partial^m}{\partial x^m} \Bigg|_{x=0} e^{z\,e^x}.
\end{split}
\end{equation}
Defining $f(x) := e^{z\,e^x}$ and $g(x) := z\,e^{x}$ we observe that
$f'(x) = f(x)\, g(x)$ and further $g'(x) = g(x)$. Therefore, one can expand the
$s$-th derivative as follows
\begin{align}
\frac{\partial^s}{\partial x^s} f(x)
= \sum_{r=1}^{s} C_{s}[r] \, \left(g(x)\right)^r \, f(x)
\end{align}
Taking derivative of both sides yields the following recursive relations
\begin{equation}
\begin{split}
C_{s+1}[r] = \begin{cases}
C_{s}[1], \;\; &\text{if } r=1,\\
r\, C_s[r] + C_s[r-1], \;\; &\text{if } 2\leq r\leq s,\\
C_s[s], \;\; &\text{if } r = s+1.
\end{cases}
\end{split}
\end{equation}
Therefore
\begin{equation}
\begin{split}
\sum_{r=1}^{s+1} C_{s+1}[r] = \sum_{r=1}^s C_s[r]\, (r+1)
\leq (s+1) \, \sum_{r=1}^s C_s[r]
\end{split}
\end{equation}
which given $C_1[1] = 1$ implies that
\begin{align}
\label{eq:coollem}
\sum_{r=1}^m C_m[r] \leq m!.
\end{align}
Lastly, we note that
\begin{align}\label{eq:coollem-2}
\frac{\partial^m}{\partial x^m} \Bigg|_{x=0} e^{z\,e^x}
\leq e^z\,\max\{z^m,1\}\, \sum_{r=1}^m C_m[r] \leq e^z\,z^m\, m!
\end{align}
since $g(0) = z$, and $f(0) = e^z$, and the last inequality follows from
\eqref{eq:coollem}. Combined with \eqref{eq:cool3} the result follows.
\end{proof}

\begin{lem}\label{lem:example-3}
For any $z>1$ and $m\in \mathbb N$, the following holds
\begin{align}
\sum_{k=0}^\infty k^m \, z^{-k}
\leq \frac{1}{1-z^{-1}}\max\left( 2,\frac{2}{z-1} \right)^{m} \, m!.
\end{align}
\end{lem}
\begin{proof}

First, we note that the function we are upper bounding is a special case of the
Lerch transcendents \cite{gradshteyntables}, for which we also provide a lower
bound in \cref{ex:inv-cos}. In particular,
$\Phi(z^{-1},m,0) = \sum_{k=0}^\infty k^m z^{-k}$. We shall now prove the result
by setting $\alpha = \ln z$ and writing
\begin{align}
\sum_{k\geq 0} k^m z^{-k}
= \frac{\partial^m}{\partial \alpha ^m} \sum_{k\geq 0} e^{\alpha k}
= \frac{\partial^m}{\partial \alpha ^m} \frac{1}{1-e^{\alpha}}.
\end{align}
By a simple induction, we arrive at the following form
\begin{align}
\frac{\partial^m}{\partial \alpha ^m} \frac{1}{1-e^{\alpha}}
= \sum_{r=1}^{m} C_{m}[r] \frac{e^{r\alpha}}{(1-e^\alpha)^{r+1}},
\end{align}
with the following recursive relation for the coefficients
\begin{align}
C_{s+1}[r] = \begin{cases}
C_s[1], & \text{if } r=1,\\
r \left( C_s[r] + C_s[r-1] \right) & \text{if } 2\leq r\leq s,\\
r C_s[s], & \text{if } r=s+1.
\end{cases}
\end{align}
Hence, we have $\sum_{r=1}^m C_m[r] \leq 2m \sum_{r=1}^{m-1} C_{m-1}[r]$, and
consequently $\sum_{r=1}^m C_m[r] \leq 2^m m!$, as $C_1[1]=1$. Therefore,
\begin{align}
\frac{\partial^m}{\partial \alpha ^m} \frac{1}{1-e^{\alpha}}
\leq \frac{1}{1-z^{-1}} \, \max\{
 1,\frac{1}{z-1}\}^m \,2^m \, m!.
\end{align}
\end{proof}

\begin{lem}
\label{lem:fac}
For any $m\geq 1$
\begin{align}
\sum_{k=0}^m k! (m-k)! \leq 3\, m!.
\end{align}
\end{lem}

\begin{proof}
Note that the inequality could be checked by direct calculation for $m=1, 2, 3$.
We now consider $m \geq 4$ and note that
\begin{align}
\label{eq:fac2}
\sum_{k=0}^m k! (m-k)! = m! \, \sum_{k=0}^m \frac{1}{{m \choose k}}
\end{align}
However, we have $\min_{k = 2, \cdots, m-2} {m\choose k} \geq m $ and hence
\begin{equation}\label{eq:fac1}
\begin{split}
\sum_{k=0}^m \frac{1}{{m \choose k}}
= 2 + \frac{2}{m} + \sum_{k=2}^{m-2} \frac{1}{{m \choose k}}
\leq 2 + \frac{2}{m} + \frac{m-3}{m} \leq 3.
\end{split}
\end{equation}
\end{proof}

\begin{lem}\label{lem:fact2}
For all $d \geq 1$
\begin{align}
\sum_{\stackrel{i_1,i_2,\cdots,i_d\in\{0,\cdots,m\}}{i_1+i_2+\cdots+i_d=m}}
i_1! \,i_2!\, \cdots \, i_d! \leq 3^{d-1} \, m!.
\end{align}
\end{lem}

\begin{proof}
We prove this claim by induction. The base case ($d=1$) is trivially true.
Assuming the statement is correct for $d$, we have
\begin{equation}
\begin{split}
\sum_{i_1+i_2+\cdots+i_d=m} i_1! \,i_2!\, \cdots \, i_{d+1}!
&= \sum_{i_1=0}^m i_1!
\left( \sum_{i_2+\cdots+i_{d+1}=m-i_1} i_2!\, \cdots i_{d+1}! \right)\\
&\leq 3^{d-1} \sum_{i_1=0}^m i_1\, (m-i_1)!\\
&\leq 3^d \, m!
\end{split}
\end{equation}
where the last step is due to \cref{lem:fac}.
\end{proof}

\begin{lem}
\label{lem:useful} Let $m$ be an integer greater than $d$. We have
\begin{align}
\max_{x\in[-1,1]^d}
\sum_{p\in\mathbb{Z}^d\setminus \{0\}} \norm{x+2\,p}^{-2m} \leq 2^{d+1}
\end{align}
for a universal constant $\xi$.
\end{lem}

\begin{proof}
Firstly, we note that
\begin{align}
\max_{x\in[-1,1]^d} \sum_{p\in\mathbb{Z}^d\setminus \{0\}}
\norm{x+2\,p}^{-2m}
&\leq \sum_{j=1}^d \, {d\choose j} \, \max_{x\in[-1,1]^j}
\sum_{p\in(\mathbb{Z}\setminus\{0\})^j} \frac{1}{\norm{x + 2p}^{2m}}\\
&\leq \sum_{j=1}^d \, {d\choose j} \, \max_{x\in[-1,1]^j}
\sum_{p\in(\mathbb{Z}\setminus\{0\})^j} \frac{1}{\norm{x + 2p}^{2j}}.
\end{align}
We now note that
$\sum_{p\in(\mathbb{Z}\setminus\{0\})^j} \frac{1}{\norm{x + 2p}^{2m}}$ is
maximized over $[-1,1]^j$ for $x$ being one of the corner points. Hence, we can
upper bound the summation by the following integral with respect to the
volume form $dV_j$ of the $j$-dimensional ball in the $l_2$ norm:
\begin{equation}
\begin{split}
\max_{x\in[-1,1]^d} \sum_{p\in\mathbb{Z}^d\setminus \{0\}} \norm{x+2\,p}^{-2m}
&\leq \sum_{j=1}^d \, {d\choose j} \, \left( \frac{1}{\sqrt{j}}
+ \frac{1}{2^j} \int_{r = 1}^\infty \frac{dV_j}{r^{2j}} \right)\\
&\leq \sum_{j=1}^d \, {d\choose j} \, \left(
\frac{1}{\sqrt{j}} + \frac{\pi^{j/2}}{2^j \Gamma(j/2+1)} \,
\int_{r=1}^\infty r^{-1-j} \, dr\right)\\
&= \sum_{j=1}^d \, {d\choose j} \, \left( \frac{1}{\sqrt{j}}
+ \frac{\pi^{j/2}}{2^j \Gamma(j/2+1)} \, \frac{1}{j}  \right)\\
&\leq 2^d \, \sup_{j\in\mathbb{N}} \left(  \frac{1}{\sqrt{j}}
+ \frac{\pi^{j/2}}{2^j \Gamma(j/2+1)} \, \frac{1}{j} \right)
= 2^{d+1}
\end{split}
\end{equation}
\end{proof}

\begin{lem}[Lemma 13 of \cite{berry2017quantum}]
\label{lem:unitnorm}
Let $\overrightarrow{a}$ and $\overrightarrow{b}$ be two vectors of the same
vector space. It is the case that
\begin{align}
\norm{ \frac{\overrightarrow{a}}{\norm{\overrightarrow{a}}}
- \frac{\overrightarrow{b}}{\norm{\overrightarrow{b}}} } \leq
\frac{2\,\norm{\overrightarrow{a} - \overrightarrow{b} }}{\max\left[
\norm{\overrightarrow{a}}, \norm{\overrightarrow{b}} \right]}.
\end{align}
\end{lem}

\begin{proof}
Without loss of generality, we assume $\overrightarrow{a}$ has a larger norm. We
then write
\begin{equation}
\begin{split}
\norm{ \frac{\overrightarrow{a}}{\norm{\overrightarrow{a}}}
- \frac{\overrightarrow{b}}{\norm{\overrightarrow{b}}} }
&=  \norm{ \frac{\overrightarrow{a}}{\norm{\overrightarrow{a}}}
- \frac{\overrightarrow{b}}{\norm{\overrightarrow{a}}}
+ \frac{\overrightarrow{b}}{\norm{\overrightarrow{a}}}
- \frac{\overrightarrow{b}}{\norm{\overrightarrow{b}}} }\\
&\leq \norm{ \frac{\overrightarrow{a}}{\norm{\overrightarrow{a}}}
- \frac{\overrightarrow{b}}{\norm{\overrightarrow{a}}} }
+ \norm{ \frac{\overrightarrow{b}}{\norm{\overrightarrow{a}}}
- \frac{\overrightarrow{b}}{\norm{\overrightarrow{b}}} }\\
& \leq \frac{ 2 \norm{\overrightarrow{a}
- \overrightarrow{b} }}{ \norm{\overrightarrow{a}} }
\end{split}
\end{equation}
where in the last inequality we have used the triangle inequality
$\norm{\overrightarrow{a}} - \norm{\overrightarrow{b}}
\leq \norm{\overrightarrow{a} - \overrightarrow{b}}$.
\end{proof}

\begin{lem}
\label{lem:TV-bound}
Let $\ket{\psi}$ and $\ket{\phi}$ be two quantum states residing in a finite
dimensional Hilbert space. Let us denote the output measurement probabilities in
the computational basis by $P_\psi$ and $P_\phi$. Then we have
\begin{align}
\TV\left(P_\psi , P_\phi\right) \leq \norm{\ket{\psi} - \ket{\phi}}.
\end{align}
\end{lem}

\begin{proof}
We have
\begin{equation}
\begin{split}
\TV\left(P_\psi , P_\phi\right)
&= \frac{1}{2} \, \sum_{i\in I}
\left| \left|\psi(i)\right|^2 - \left|\phi(i)\right|^2 \right|\\
&= \frac{1}{2} \, \sum_{i\in I} \bigl| \,\left|\psi(i) \, \right|
- \left|\, \phi(i)\right| \bigr| \cdot \bigl| \left|\psi(i) \right|
+ \left|\phi(i)\right| \, \bigr| \\
&\leq \frac{1}{2} \, \sqrt{ \sum_{i\in I} \bigl| \,\left|\psi(i) \, \right|
- \left|\, \phi(i)\right| \bigr|^2} \, \sqrt{ \sum_{i\in I} \bigl|
\,\left|\psi(i) \, \right| + \left|\, \phi(i)\right| \bigr|^2 } \\
&\overset{(a)}{\leq} \frac{1}{\sqrt{2}} \, \sqrt{ \sum_{i\in I}  \bigl| \psi(i)
- \phi(i)\bigr|^2} \, \, \sqrt{ \sum_{i\in I} \left| \psi(i) \right|^2
+ \left| \phi(i) \right|^2 }\\
&= \norm{\ket{\psi} - \ket{\phi}}
\end{split}
\end{equation}
where (a) follows from the basic inequalities $(a+b)^2\leq 2a^2 + 2b^2$ and
$\big|\abs{a}-\abs{b} \big|\leq \abs{a-b}$.
\end{proof}

\begin{lem}
\label{lem:choose-M}
Let $u:\mathbb{R}^d \rightarrow \mathbb R$ be $l$-periodic along each dimension.
Further let $u$ be $(C,a)$-semi-analytic and $L$-Lipschitz.
Also, let $\mu$ be the probability
density proportional to $u^2$, and further, $\widehat{\mu}$ be the probability
density associated with the continuous sampling from $\ket{u_M}$. Choosing $M$
as
\begin{align}
\label{eq:choose-M}
M = \left\lceil\frac1{\delta}
\frac{L l d/2 + 10/3\sqrt 2\,ae^4 C}{\mathcal{U}}\right\rceil
\end{align}
where $\mathcal{U} = \sqrt{\EE[u^2(X)]}$, we are guaranteed to have
$\TV\left( \mu , \widehat{\mu} \right) \leq \delta$.
\end{lem}

\begin{proof}
We have
\begin{align}
\TV(\mu, \widehat{\mu})
= \frac12 \int dx \,\left| \frac{u^2}{l^d \mathcal{U}^2}
- \sum_{n\in[-M..M]^d} \mathbf{1}_{\{ x\in \mathbb{B}_n \}}
\frac{u^2_M[n]}{\norm{\vec{u_M}}^2} \left(\frac{2M+1}{l}\right)^d \right|.
\end{align}
Using the triangle inequality, we have
\begin{equation}
\begin{split}
\TV(\mu,\widehat{\mu}) &\leq \frac12 \int dx\,
\left| \frac{u^2}{l^d \mathcal{U}^2}
- \sum_{n\in[-M..M]^d} \mathbf{1}_{\{x\in\mathbb{B}_n\}}
\frac{u_M^2[n]}{\mathcal{U}^2 l^d} \right|\\
&\qquad +\frac12 \int \frac{dx}{l^d}\,
\sum_{n\in[-M..M]^d} \mathbf{1}_{\{ x\in\mathbb{B}_n \}}
u_M^2[n] \, \cdot \left| \frac{1}{\mathcal{U}^2}
- \frac{(2M+1)^d}{\norm{\vec{u_M}}^2} \right|\\
&= \frac{1}{2l^d \mathcal{U}^2} \int dx\, \sum_{n\in [-M..M]^d}
\mathbf{1}_{\{ x\in \mathbb B_n\}} \left| u^2 - u_M^2[n]\right|
+ \frac{1}{2\mathcal{U}^2} \left| \frac{\norm{\vec{u_M}}^2}{(2M+1)^d}
- \mathcal U^2 \right|\\
&\overset{(a)}{\leq} \frac{L l d}{\mathcal U (2M+1)}
+ \frac{2\sqrt 2e^3 C}{\mathcal {U}} e^{-3M/5a}
\end{split}
\end{equation}
where in (a) we use \cref{lem:distance-bound} with the choice of $M$ in
\eqref{eq:choose-M}. Using the inequality $e^{-x/e} \leq 1/x$, we obtain
\begin{align}
\TV(\mu,\widehat{\mu}) \leq
\frac{L l d/2 + 10/3\sqrt 2\,ae^4 C}{\mathcal U M}
\end{align}
which concludes the proof.
\end{proof}

\section{Lemmas used in \cref{sec:results-sampling}}

\begin{lem}
\label{lem:T}
Let $u(x)$ be an $l$-periodic real-valued function satisfying
\begin{align}
\sqrt{ \int_{\mathbb{T}} \rho_s \left(\frac{u}{\rho_s} - 1\right)^2 } \leq
\delta\, \sqrt{\int_{\mathbb{T}} \rho_s \left(\frac{1}{V\,\rho_s} - 1\right)^2}
\end{align}
for some $\delta >0$, with $V = l^d$ (the volume of the torus $\mathbb T$).
Then,
\begin{align}
\frac{1}{2}\int_{\mathbb{T}} \left| \frac{u^2}{\int_{\mathbb{T}} u^2}
- \frac{\rho_s^2}{\int_{\mathbb{T}}\rho_s^2} \right|
\leq 2\, \delta\, e^{\Delta/2}.
\end{align}
\end{lem}

\begin{proof}
Note that from the assumption
\begin{align}
\label{eq:normy1}
\sqrt{ \int_{\mathbb{T}} \left(u - \rho_s\right)^2 }
\leq e^{\Delta/2} \, \delta\, \sqrt{ \int_{\mathbb{T}}
\left(\frac{1}{V} - \rho_s\right)^2 }.
\end{align}
By a similar argument as in \cref{lem:TV-bound} we have
\begin{align}
\label{eq:normy0}
\frac{1}{2}\int_{\mathbb{T}} \left| \frac{u^2}{\int u^2}
- \frac{\rho_s^2}{\int_{\mathbb{T}}\rho_s^2} \right|
\leq \sqrt{ \int_{\mathbb{T}} \left( \frac{u}{\sqrt{\int_{\mathbb{T}} u^2}}
- \frac{\rho_s}{\sqrt{\int_{\mathbb{T}} \rho_s^2}} \right)^2 }.
\end{align}
Furthermore, using the triangle inequality (c.f., \cref{lem:unitnorm})
\begin{align}
\label{eq:normy2}
\sqrt{ \int_{\mathbb{T}} \left( \frac{u}{\sqrt{\int_{\mathbb{T}} u^2}}
- \frac{\rho_s}{\sqrt{\int_{\mathbb{T}} \rho_s^2}} \right)^2 } \leq 2\,
\frac{\sqrt{\int_{\mathbb{T}} (u-\rho_s)^2 }}{\sqrt{\int_{\mathbb{T}}\rho_s^2}}.
\end{align}
Now, putting equations \eqref{eq:normy1}, \eqref{eq:normy0}, and
\eqref{eq:normy2} together, we have
\begin{align}
\label{eq:normy3}
\frac{1}{2}\int_{\mathbb{T}} \left| \frac{u^2}{\int u^2}
- \frac{\rho_s^2}{\int_{\mathbb{T}}\rho_s^2} \right|
\leq 2\delta e^{\Delta/2} \frac{ \sqrt{ \int_{\mathbb{T}}
\left(\frac{1}{V} - \rho_s\right)^2}}{\sqrt{ \int_{\mathbb{T}} \rho_s^2 }}.
\end{align}
Now consider a random variable $X$ drawn uniformly at random from $\mathbb{T}$,
and define $Y := \rho_s(X)$. It is clear that
\begin{align}
\frac{ \sqrt{ \int_{\mathbb{T}} \left(\frac{1}{V}
- \rho_s\right)^2}}{\sqrt{ \int_{\mathbb{T}} \rho_s^2 }}
= \sqrt{\frac{ \mathbb{E}\left[ \left( Y
- \frac{1}{V}\right)^2\right] }{\mathbb{E}\left[Y^2\right]}}.
\end{align}
Furthermore, note that $\mathbb{E}[Y] = \frac{1}{V}$, which implies
\begin{align}
\frac{ \sqrt{ \int_{\mathbb{T}} \left(\frac{1}{V}
- \rho_s\right)^2}}{\sqrt{ \int_{\mathbb{T}} \rho_s^2 }}
= \sqrt{\frac{\mathrm{Var}\left[Y\right]}{\mathbb{E}\left[ Y^2\right]}} \leq 1.
\end{align}
Combining the latter equation with \eqref{eq:normy3} completes the proof.
\end{proof}


\end{document}